\documentclass[10pt,a4paper]{article}

\usepackage[margin=1.75cm]{geometry}

\usepackage{hyperref}
\usepackage{abstract}

\usepackage{fancyhdr}
\usepackage{amssymb, amsmath}
\usepackage{graphicx}
\usepackage{comment}
\usepackage{enumerate}
\usepackage{color}
\usepackage{tikz}

\title{
Stochastic models of the binding kinetics of VEGF-A to VEGFR1 and VEGFR2 in endothelial cells~\footnote{This work
 has been supported by The Leverhulme Trust
 (RPG-2012-772). M.~Nowicka is supported by EPSRC and AstraZeneca (CASE studentship).
 We also acknowledge the University of Leeds for the permission to use
 the High Performance Computing facilities ARC1 and ARC2.
$^\dagger$M.~L\'opez-Garc\'ia and M.~Nowicka have contributed equally to this work.
\newline
$^{\&}$Corresponding author. E-mails: m.lopezgarcia@leeds.ac.uk (M.~L\'opez-Garc\'ia),
mm11ms@leeds.ac.uk (M.~Nowicka), grant@maths.leeds.ac.uk (G.~Lythe),
s.ponnambalam@leeds.ac.uk (S.~Ponnambalam), carmen@maths.leeds.ac.uk (C.~Molina-Par\'{\i}s).
}
}

\author{
M.~L\'opez-Garc\'ia$^{1,\dagger,\&}$, M.~Nowicka$^{1,\dagger,\&}$,
C.~Bendtsen$^{2}$,
G.~Lythe$^{1,\&}$, 
\\
S.~Ponnambalam$^{3,\&}$,
C.~Molina-Par\'{\i}s$^{1,\&}$\\
\normalsize $^{1}$Department of Applied Mathematics, School of Mathematics, University of Leeds,
LS2 9JT Leeds, UK \\
\normalsize $^{2}$ 
Quantitative Biology, Discovery Sciences, IMED,
AstraZeneca,
Cambridge Science Park,
\\
\normalsize
 Milton Road,  CB4 0WG Cambridge, UK
\\
\normalsize $^{3}$Endothelial Cell Biology Unit, School of Molecular and Cellular Biology, 
\\
\normalsize
University of Leeds,
LS2 9JT Leeds, UK
}

\date{22nd of June  2016}

\begin{document}

\maketitle

\begin{abstract}

Vascular endothelial growth factor receptors (VEGFRs) are receptor
tyrosine kinases (RTKs) that regulate proliferation, migration,
angiogenesis and vascular permeability of endothelial cells. Vascular
endothelial growth factor receptor~1 (VEGFR1) and  receptor~2 (VEGFR2) bind
vascular endothelial growth factors (VEGFs), inducing receptor
dimerisation and activation, characterised by phosphorylation of
tyrosine residues in their cytoplasmic domain. Although recent
experimental evidence suggests that RTK signalling occurs both on the
plasma membrane and intra-cellularly, and reveals the role of
endocytosis in RTK signal transduction, we still lack knowledge of
VEGFR phosphorylation-site use and of the spatiotemporal regulation of
VEGFR signalling. In this paper, we introduce four stochastic
mathematical models to study the binding kinetics of vascular
endothelial growth factor VEGF-A to VEGFR1 and VEGFR2, and RTK
phosphorylation.  The formation of phosphorylated (homo- and hetero-)
dimers on the cell surface is a two-step process: diffusive transport and
binding.  The first two of our models (Model~1 and Model~2) only consider
VEGFR2. This simplification
allows us to introduce new stochastic descriptors making use of a
matrix-analytic approach. The two remaining models (Model~3.1 and Model~3.2)
describe the competition of VEGFR1 and VEGFR2 for ligand availability, and
are analysed making use of Gillespie (stochastic) simulations and the
van Kampen approximation. Under the hypothesis that bound
phosphorylated receptor dimers are the signalling units, we study the
time to reach a threshold number of such complexes. Our results
indicate that the presence of VEGFR1
does not only affect 
the timescale to
reach a given signalling threshold,
but it also affects the maximum
attainable threshold. This result is consistent with the conjectured
role of VEGFR1 as a decoy receptor, that prevents VEGF-A binding to
VEGFR2, and thus, VEGFR2 attaining suitable phosphorylation levels. We
identify an optimum range of ligand concentration for sustained dimer
phosphorylation.  Finally, sensitivity analysis identifies the rate of
monomer formation as the parameter that our descriptors depend most
strongly on.

\end{abstract}



\section{Introduction}
\label{Sect1}

Vascular endothelial growth factors (VEGFs) are a  family of bivalent ligands consisting of mammalian
and virus-encoded members. The first discovered member was VEGF-A~\cite{Alarcon07}. The ligand
occurs in different isoforms of varying lengths. Mounting evidence suggests that
the various isoforms are involved in diverse cellular responses~\cite{Olsson06}. VEGFs specifically bind to three
type-V receptor tyrosine kinases (RTKs), VEGFR1, VEGFR2 and VEGFR3, as well as co-receptors,
 such as neuropilins.
In physiological conditions, the vascular endothelium expresses VEGFR1 and VEGFR2, whereas the
lymphatic endothelium expresses VEGFR2 and VEGFR3~\cite{Cross03}. Each receptor has an extra-cellular
domain for binding ligand, a trans-membrane domain, and an intra-cellular or cytoplasmic
domain~\cite{Lauffenburger96}. Like many other RTKs, VEGFRs normally require dimerisation to become activated:
once VEGF binds to VEGFRs, the intra-cellular domains become activated through auto-phosphorylation
and start  cascades of intra-cellular enzymatic reactions~\cite{Olsson06}.

In order to model endothelial cell behaviour regulated by VEGFR/VEGF signalling,
initial cell surface binding events and
subsequent intra-cellular trafficking processes must be first quantified. Once this foundation is established,
cellular behaviour can more easily be analysed based on the number, state, and location of all molecules and
complexes involved. The receptor population is involved in binding to other receptors or membrane associated molecules,
internalisation, recycling, degradation and synthesis, broadly termed ``trafficking'' events. Both VEGFR monomers
and VEGFR dimers undergo internalisation by the same mechanism. The molecules are internalised and transferred
to the early endosome, in a process called endocytosis. After entering the early endosome, monomeric and dimeric
VEGFRs follow different pathways. The latter are transported to the late endosome and then to the lysosome for
degradation, whereas the former are rapidly recycled to the membrane~\cite{Teis03}.

VEGF-induced signalling cascades can cause diverse cellular responses such as cell motility,  division or
death ({\em i.e.,} apoptosis). Thus, a quantitative study of binding and phosphorylation kinetics is crucial to the understanding
of processes like angiogenesis and vasculogenesis. In Ref.~\cite{Alarcon07} a stochastic model is proposed
 which includes binding, dimerisation, endocytosis and early signalling events (activation of enzymes carrying an SH2
domain). The authors carry out an analysis of the master equation of the process, by a generalisation of the
Wentzel-Kramers-Brillouin method, to address the contribution of ligand-induced receptor dimerisation,
activation of src-homology-2 domain-carrying kinases and receptor internalisation in the behaviour of the
VEGF/VEGFR system, where only one receptor type is considered (VEGFR2).

In order to analyse in detail the dimerisation and phosphorylation kinetics on the cell membrane,
it is usual to consider mathematical models which neglect
internalisation events, and strictly focus on the biochemical reactions taking place on the cell surface.
In Ref.~\cite{Mac07} the authors introduce a comprehensive set of models with different dimerisation pathways:
 the first allowing pre-dimerisation without
ligand and the second considering only ligand-induced receptor dimerisation. In this way, the authors
can address  the role of
 pre-formed dimers in the binding process.
It is also worth mentioning Ref.~\cite{Olsson06}, where the authors reported that blood flow might activate
VEGFRs in a ligand-independent manner (promoting the activation of mechano-sensory complexes).
The consideration of more than one receptor type in Ref.~\cite{Mac07} is also essential, given that the authors
note that prostacyclin synthesis has been reported to be under the control of VEGFR heterodimers, which
suggests that the signalling of heterodimers is unique and significant for cellular responses. In most papers,
VEGFR1 is often neglected,  even when it might  be essential for the recruitment of
haematopoietic precursors and migration of monocytes and macrophages~\cite{Olsson06}.
Furthermore, in many biological responses to VEGF, the contribution
of both VEGFR1 and VEGFR2 might be required for a balanced signalling~\cite{Olsson06}.
 VEGFR signal transduction
models have to provide a context for potential communication between different VEGF receptors at the plasma
 membrane (through heterodimerisation). Therefore, the dynamics of competition for ligand availability between
 VEGFR1 and VEGFR2 needs still to be analysed in greater depth.

There is a wealth of previous studies that have developed mathematical models of RTKs
and their role in cellular responses. For example,
in Ref.~\cite{Starbuck90} the authors consider a different receptor tyrosine kinase, the
epithelial growth factor receptor (EGFR) to study the role of epithelial growth factor (EGF)
on B82 fibroblasts. They argue that the receptor signal is generated at a rate proportional to the number of
activated receptors present, so that the amount of phosphorylated dimers is directly related to the
initiation of signalling cascades. In Ref.~\cite{Tan13}, the authors consider a mathematical model of
pre-formed RTK dimers, with instantaneous phosphorylation of dimers upon ligand
binding. However, phosphorylation is in fact a multi-step process, in which 
the different tyrosine domains of each receptor transfer phosphate (from ATP) 
onto specific tyrosine residues of the partner receptor, {\em i.e.,} 
trans-autophosphorylation~\cite{Olsson06}.
In Ref.~\cite{Alarcon06}, stochastic models of receptor oligomerisation by a bivalent ligand are introduced to
study the role of ligand-induced receptor cross-linking in cell activation. A particular feature of this study is that a small number of receptors
is considered, making a stochastic approach more suitable than a deterministic one
(see Ref.~\cite{Mac05a} for a comparison between deterministic and stochastic approaches in VEGFR
models). In order to relate receptor-ligand dynamics on the cellular membrane to cell activation,  the authors
introduce a threshold number, $\theta$, of bound oligomers
that need to be formed before a cellular response can take place. Once the stochastic
 process reaches this threshold, they study
(by means of Gillespie simulations)  the probability of staying above this threshold for a given time,
$T=10 \; k_{\text{off}}^{-1}$, which is identified with the time required for the activation of kinases and
 for the signalling pathway to be initiated~\cite{Alarcon06}.

In this paper, we aim to study the dynamics of VEGFR1, VEGFR2 and VEGF-A. We first focus on the binding
kinetics of VEGFR2/VEGF-A on the cell surface, and introduce a mathematical model (Model~1),
in which monomeric receptors, VEGFR2, can bind a bivalent ligand, VEGF-A, and receptor dimerisation
is ligand induced. This model is similar to Model~1 of Ref.~\cite{Alarcon06}. However, in order to circumvent
the need to introduce a fixed time to stay above the threshold to lead to a cellular response,
we consider phosphorylation an intrinsic characteristic of the cross-linked VEGFR2 dimers. In Model~1,
dimers are considered to be instantaneously phosphorylated, so that the time to initiate the signalling cascade
 is identified with the time  to reach a given threshold number of phosphorylated dimers. In Model~2,
 phosphorylation of dimers is considered as a new reaction in the process,
and we equally consider the possibility of dimer de-phosphorylation. We then compute the time to reach
a given threshold number of phosphorylated dimers in Model~2.
Finally, and in order to study the role of VEGFR1 in the dynamics of  VEGFR2 and VEGF-A,
we introduce two stochastic models (Model~3.1 and  Model~3.2), which are extensions
of Model~1 and Model~2, respectively,  in the presence of VEGFR1.

As stated in Ref.~\cite{Alarcon06}, the analytical treatment of multi-variate stochastic processes is usually
extremely difficult, and numerical approaches, such as Gillespie simulations, are used instead. However,
it is still possible to carry out an analytical study of these processes without solving the master equation.
Here we make use of a matrix-analytical technique in order to consider  a number of stochastic descriptors,
 conveniently defined in the spirit of Ref.~\cite{Alarcon06}. This matrix-analytic approach, which has its origins
  in the seminal work by M.~Neuts~\cite{Neuts94}, allows us to study the stochastic descriptors
of interest for moderate concentrations of ligands and receptors, as discussed in Section~\ref{Sect2}.
Matrix-analytic techniques have historically been developed in the context of Queueing Theory~\cite{Latouche99}.
However, more recently, they have been applied in Mathematical Biology~\cite{GomezCorral12a,GomezCorral12b}
(competition model between two species of individuals).

The paper is organised as follows. In Section~\ref{Sect2}, four different stochastic models are introduced
to describe the binding dynamics of receptor monomers and dimers on the surface of endothelial cells.
The models include phosphorylation or competition for ligand availability. Matrix-analytic techniques
are applied, when possible, in order to study different stochastic descriptors of interest in the VEGF-A/VEGFR
system. One special property of this method is that a sensitivity analysis for the effect of binding,
dissociation and phosphorylation rates on the stochastic descriptors can be carried out.
In Section~\ref{Sect3}, parameter estimation is carried out following arguments first described in
Ref.~\cite{Lauffenburger96}. Finally, numerical results are described in Section~\ref{Sect4}, followed by
a discussion in Section~\ref{Sect5}. Standard notation used throughout the paper is introduced in Appendix~A, and
different matrices and algorithms defined in the paper are specified in Appendices~B and~C.
The application of the Van Kampen approximation~\cite{Van92}, when dealing with the master equation (to study
the transient behaviour of the Markov chains under consideration), is discussed in Section~\ref{Sect5},
and this method is shortly reviewed in Appendix~D.


\section{Stochastic models}
\label{Sect2}

In this Section, we introduce three different stochastic models for the binding kinetics of two different VEGFRs,
VEGFR1 and VEGFR2, to the bivalent ligand VEGF-A, taking place on the membrane
of a vascular endothelial cell. We consider a bivalent ligand that can bind to plasma membrane receptors,
creating receptor-with-ligand monomers.
The free pole of the ligand in a monomer can then bind to free receptors during the diffusion
of these receptors on the cell surface,
creating dimers consisting of two receptors bound to the ligand. Dimerisation of receptors is only induced
by ligands in our model and two free receptors
are not able to create a {\it pre-dimer} without ligand (ligand-induced dimerisation or LID). 
This is consistent with Ref.~\cite{Grunewald10}
and is the main assumption in
 Ref.~\cite[LID Model]{Mac07}. 
 There is also experimental support
 for this hypothesis: free VEGFR2 is observed (electron microscopy) in monomeric 
 form on the cell surface~\cite{ruch2007structure}.
 We note that, for low to moderately high ligand concentrations, as
  discussed in Section~\ref{Sect4}, no
significant differences are expected in our results by including pre-dimerisation~\cite{Mac07}. 
Yet, this would significantly
increase our model complexity. The
relevance in the system dynamics of pre-dimers is only expected under highly saturated situations; see, for
example, Ref.~\cite[Figures 2 and 3]{Mac07} for details, and Figure~\ref{FigLast}, of this paper for further
details 
related to this matter.

In SubSection~\ref{Subsect21} and SubSection~\ref{Subsect22}, we consider the processes where only VEGFR2
is expressed on the cell surface. In particular, in SubSection~\ref{Subsect21}, we propose a stochastic model
in which dimers are assumed to instantaneously phosphorylate, so that these dimers amount to the
signalling complexes on the plasma membrane, and the number of these complexes becomes a
 direct measure of signalling. In SubSection~\ref{Subsect22}, an extended model is considered,
 where phosphorylation (and de-phosphorylation) are included as new reactions in the process: 
  non-phosphorylated dimers may eventually become phosphorylated,
and these phosphorylated dimers are the complexes
initiating the signalling cascade. In any model considered in this paper, phosphorylation of a cross-linked
dimer is considered to be the synchronised activation of all
the tyrosine kinase residues in the intra-cellular tail of the receptors.
However, phosphorylation is a multi-step process, in which 
the different tyrosine domains of each receptor transfer phosphate (from ATP) 
onto specific tyrosine residues of the partner receptor~\cite{Olsson06}.
Our approach should be then viewed as an intermediate alternative
between modelling phosphorylation in an instantaneous way (Model~1 in Subsection~\ref{Subsect21},
as well as most of the published models, such  as Refs.~\cite{Alarcon06,Alarcon07}), and
considering dimer phosphorylation as a multi-step process, where each tyrosine residue becomes phosphorylated
upon ligand
stimulation, which is out of the scope of this paper. 
Thus, the mathematical models developed here will allow us to consider the phosphorylation
of bound and cross-linked VEGFR dimers as a separate reaction and how it affects the dynamics
of the VEGF-VEGFR association and dissociation. This reaction was discussed
 in Ref.~\cite{Mac07} but not included in the mathematical model.
In SubSection~\ref{Subsect23}, two different variants
of a third model are considered to analyse the competition dynamics between
VEGFR1 and VEGFR2, when both are expressed on the cell surface. Finally, a sensitivity analysis is developed in
SubSection~\ref{Subsect24}
in order to understand how the binding, dissociation and phosphorylation rates affect the dynamics of the
ligand/receptor system.

The study of the number of monomer, non-phosphorylated and phosphorylated dimer molecules
on the cell surface  over time can be viewed as
the analysis of the transient behaviour of a specific Markov chain, a problem which, in general,
 is not solvable in closed form~\cite{Kulkarni95}.
Therefore, one typically carries out Gillespie simulations~\cite{Gillespie77}, or applies moment-closure
techniques~\cite{Gillespie09,Hespanha08} to deal  with the master equation of
the Markov process under study. In this paper, we follow both approaches to study the
multi-variate competition models introduced in SubSection~\ref{Subsect23}.
However, for models in SubSection~\ref{Subsect21} and SubSection~\ref{Subsect22},
 it is possible to apply alternative procedures in order to analyse, in an exact way,
different characteristics of the processes under consideration. In particular, the matrix-analytic approach is
described in SubSection~\ref{Subsect21} and  SubSection~\ref{Subsect22},
by studying the Laplace-Stieltjes transforms of particular random variables of interest, and making use of first-step arguments
and auxiliary absorbing Markov chains conveniently defined. Moreover, a novel local sensitivity 
analysis for the Markov chains under study is adapted
and applied in SubSection~\ref{Subsect24}, by generalising arguments from Ref.~\cite{Caswell11}.
This analysis allows us
 to identify how the stochastic descriptors (considered in SubSection~\ref{Subsect21} and SubSection~\ref{Subsect22}),
are affected by the binding, dissociation and phosphorylation rates.


\subsection{Model~1: instantaneous phosphorylation}
\label{Subsect21}

\begin{figure}[htp!]
    \centering
        \includegraphics[scale=0.30]{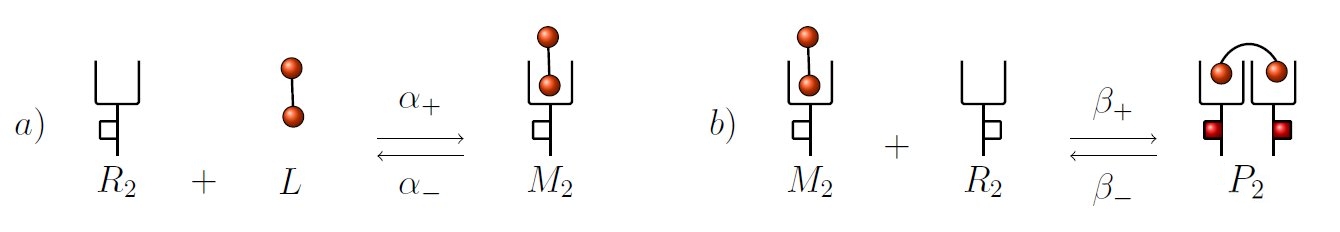}
    \caption{Reactions in Model~1. $a$) Formation and dissociation of bound monomers ($M_2$).
    $b$) Formation and dissociation of bound dimers ($P_2$), which instantaneously phosphorylate
 (represented by red phosphorylated residues in the intra-cellular tail of the receptors).}
    \label{Fig1}
\end{figure}

In this Section, we consider a model of the bivalent ligand VEGF-A that can bind VEGFR2
to form
{\it $M_2$} complexes. Free receptors can diffuse on the cell surface, so that eventually
they react with bound monomers $M_2$, to form ligand-bound and cross-linked receptor dimers, termed
{\it $P_2$} complexes. Once a $P_2$ complex is formed, it is instantaneously phosphorylated,
so that the $P_2$ complexes on the plasma membrane initiate signalling,
 in the spirit of Refs.~\cite{Alarcon06,Starbuck90}.
Finally, both bound monomers and dimers can dissociate. We 
assume that  de-phosphorylation of $P_2$ complexes 
is fast (see Table~\ref{tab: biochem parameters}), and that it takes place when cross-linked receptor dimers dissociate.
In this scenario
 four possible reactions
can occur with different binding and dissociation rates as
shown in Figure~\ref{Fig1}.

\begin{figure}[h]
    \centering
        \includegraphics[scale=0.85]{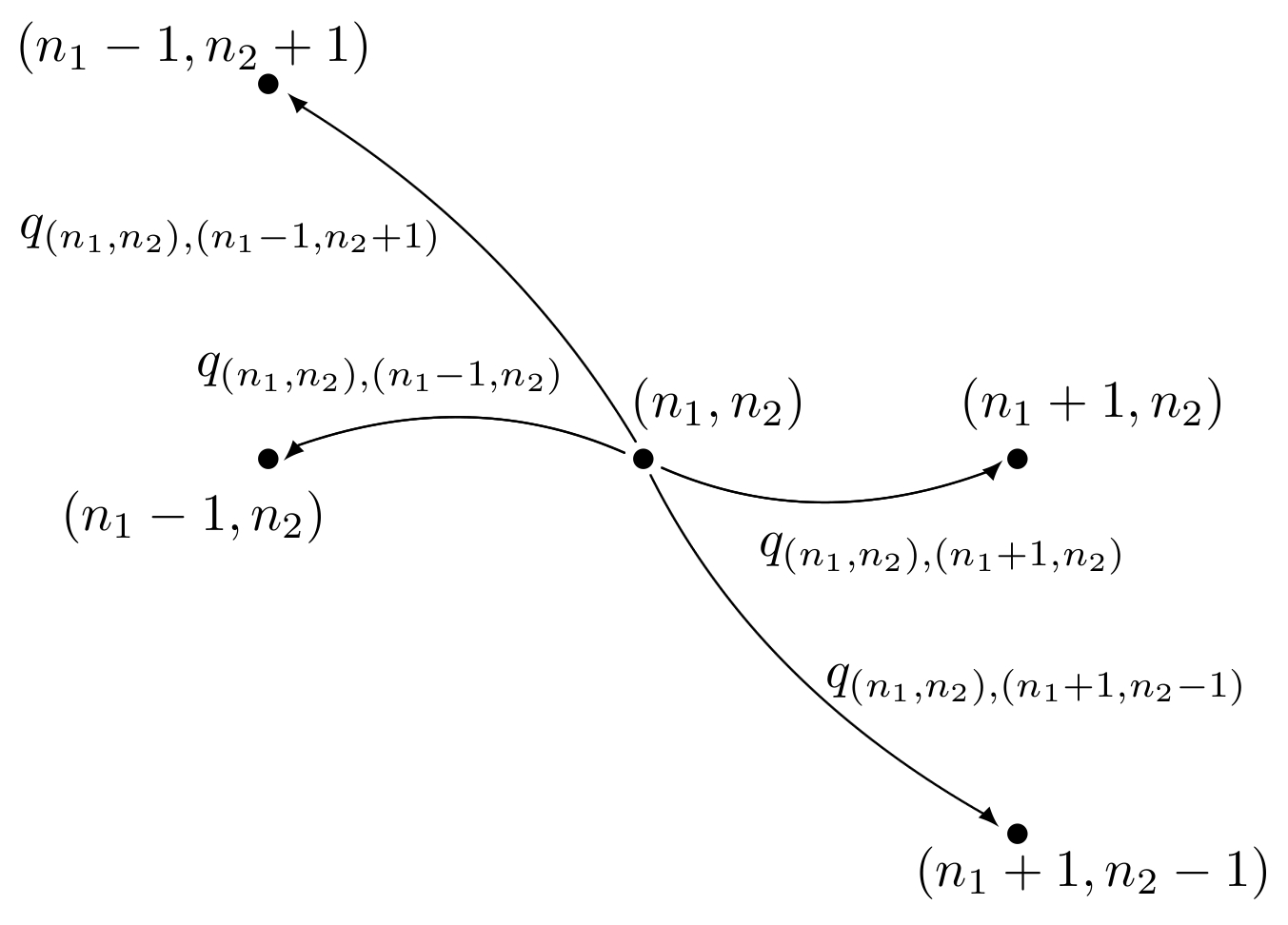}
    \caption{Transition diagram for model~1.}
    \label{Fig2}
\end{figure}

In what follows, we consider an environment with constant numbers, $n_{R_2}$ and $n_L$,
of receptors and ligands spatially well-mixed on the cell surface or in the extra-cellular space, respectively.
We are interested in the number of $M_2$ and $P_2$ complexes on the cell surface as a function of time,
which we
model using a stochastic approach: as a continuous time Markov chain (CTMC)
${\cal X}=\{{\bf X}(t)=(M_2(t),P_2(t)):~ t\geq0\}$, where $M_2(t)$ and $P_2(t)$ represent the number of
$M_2$ and $P_2$ complexes, respectively, at time $t$.
We note that, if we define the random variables $R_2(t)$ and $L(t)$ as the numbers of
free receptors and ligands, respectively, at time $t\geq0$,
it is clear that $R_2(t)=n_{R_2}-M_2(t)-2P_2(t)$ and $L(t)=n_{L}-M_2(t)-P_2(t)$ for all
$t\geq0$. Then, $R_2(t)$ and $L(t)$ are implicitly analysed in ${\cal X}$
and do not need to be explicitly considered in the CTMC. Moreover, it is
clear that $M_2(t),P_2(t)\geq0$ and, from the previous comments
\begin{eqnarray}
 L(t)\geq0 &\Rightarrow& M_2(t)+P_2(t) \ \leq \ n_L,
 \nonumber
 \\
 R_2(t)\geq0 &\Rightarrow& M_2(t)+2P_2(t) \ \leq \ n_{R_2},\nonumber
\end{eqnarray}
for all $t\geq0$, which specifies the space of states ${\cal S}$ of ${\cal X}$. 
Specifically, we note that given $(M_2(t),P_2(t))=(n_1,n_2)$ at some time instant $t\geq0$, then
\begin{itemize}
 \item if $2n_L\leq n_{R_2}$: $n_1+n_2 \ \leq \ n_L \ \Rightarrow \ n_1+2n_2 \ \leq \ n_{R_2}$, and
 \item if $n_{R_2}\leq n_{L}$: $n_1+2n_2 \ \leq \ n_{R_2} \ \Rightarrow \ n_1+n_2 \ \leq \ n_{L}$,
\end{itemize}
so that three different specifications of the space of states ${\cal S}$ are obtained,
 depending on the particular values of $n_{R_2}$ and $n_{L}$. In particular:

\begin{itemize}

  \item if $2n_L\leq n_{R_2}$, then ${\cal S}=\{(n_1,n_2)\in(\mathbb{N}\cup\{0\})^{2}:~ n_1+n_2\leq n_L\}$,

  \item if $n_{R_2}<2n_L<2n_{R_2}$, then ${\cal S}=\{(n_1,n_2)\in(\mathbb{N}\cup\{0\})^{2}:~ n_1+n_2\leq n_L,~ n_1+2n_2\leq n_{R_2}\}$, and

  \item if $n_{R_2}\leq n_L$, then ${\cal S}=\{(n_1,n_2)\in(\mathbb{N}\cup\{0\})^{2}:~ n_1+2n_2\leq n_{R_2}\}$.
\end{itemize}
Although we can deal with each of these cases in a similar manner, we only consider the case
$2n_L\leq n_{R_2}$, since this is the case under
physiological conditions of VEGFR2, see {\em e.g.,} Ref.~\cite{Kut07}.
In this case, ${\cal X}$ is defined over ${\cal S}=\{(n_1,n_2)\in(\mathbb{N}\cup\{0\})^{2}:~ n_1+n_2\leq n_L\}$.
From Figure~\ref{Fig1}, it is clear that transitions from states in the interior of ${\cal S}$,
that is, from states $(n_1,n_2)\in\mathbb{N}^{2}$
with $n_1+n_2< n_L$, can be to four adjacent states as shown in Figure~\ref{Fig2}.
Transitions for states within the boundary of ${\cal S}$ are obtained in a
similar way by discarding those transitions that leave ${\cal S}$.

Transitions between states in our CTMC are governed by the infinitesimal transition rates
$q_{(n_1,n_2),(n_1',n_2')}$, with $(n_1,n_2),(n_1',n_2')\in{\cal S}$.
These infinitesimal transition rates are  obtained by
mass action kinetics, and by the fact that
 if the process is in state $(n_1,n_2)$ at a given time, there are $(n_L-n_1-n_2)$ free ligands and
$(n_{R_2}-n_1-2n_2)$ free receptors available. Formation of $M_2$ complexes
directly depends on the number of free receptors and ligands, while their
dissociation only depends on the number of $M_2$ complexes.
Similar comments can be made for $P_2$ complexes. Finally, we note that formation of $M_2$
complexes and dissociation of $P_2$ complexes can take place with one of the two available
poles of the ligand involved in the reaction. Then, the specific values
of the non-null infinitesimal transition rates are given by
\begin{eqnarray}\label{Eqn2}
 q_{(n_1,n_2),(n_1',n_2')} &=& \left\{\begin{array}{ll}
2\alpha_{+}(n_{R_2}-n_1-2n_2)(n_L-n_1-n_2), & \hbox{if $(n_1',n_2')=(n_1+1,n_2)$,}\\
\alpha_{-}n_1, & \hbox{if $(n_1',n_2')=(n_1-1,n_2)$,}\\
\beta_{+}n_1(n_{R_2}-n_1-2n_2), & \hbox{if $(n_1',n_2')=(n_1-1,n_2+1)$,}\\
2\beta_{-}n_2, & \hbox{if $(n_1',n_2')=(n_1+1,n_2-1)$,}
\end{array}\right.
\end{eqnarray}
where $\alpha_{+}, \alpha_{-}, \beta_{+}$ and $\beta_{-}$ are positive constants representing
binding and dissociation rates for $M_2$ and $P_2$ complexes, respectively.

Our objectives in this Section amount to study two descriptors of interest:

\begin{enumerate}

  \item Starting from any state $(n_1,n_2)\in{\cal S}$, the time to reach a number $N_2>n_2$ of $P_2$ complexes.

  \item Starting from any state $(n_1,n_2)\in{\cal S}$, the stationary distribution of the process.

\end{enumerate}

Descriptor~2 allows us to obtain the theoretical state of the system at steady state,
thus enabling us to analyse the long-term dynamics,
for example, for different initial numbers of ligand and receptor. Descriptor~1 is the most relevant
and useful,
 since it is the time to reach
some pre-defined threshold number of signalling complexes, or equivalently  a pre-defined threshold for cell activation
(see Refs.~\cite{Alarcon06,Starbuck90}).
One of our aims is to analyse how this time depends on the binding and dissociation
rates, as well as on the  numbers of ligand and receptor on the cell surface. Numerical results obtained in
Section~\ref{Sect4} show, by means of two competition models
introduced in SubSection~\ref{Subsect23}, how the presence of a VEGFR2 competitor, VEGFR1,
induces a delay in the time to reach a given threshold of $P_2$ complexes.

The analysis carried out in this Section is based in the use of levels for the organisation of state space,
Laplace-Stieltjes transforms,
first-step arguments and auxiliary absorbing Markov chains.
Although the theoretical results and algorithms presented in this Section
are valid regardless of the exact values of $n_L$ and $n_{R_2}$ (with $2n_L\leq n_{R_2}$),
the dimensionality of the state space, ${\cal S}$, can become computationally intractable,
 even
for moderate values of the total number of ligands $n_L$, since
\begin{eqnarray*}
 \#{\cal S} &=& \frac{(n_L+1)(n_L+2)}{2}.
\end{eqnarray*}
Thus, in order to minimise the computational effort in our procedures, a strong focus on algorithmic issues is necessary throughout the paper.
We first organise the space of states ${\cal S}$ by levels (groups of states) as
\begin{eqnarray*}
 {\cal S} &=& \bigcup\limits^{n_L}_{k=0}L(k),
\end{eqnarray*}
where $L(k)=\{(n_1,n_2):~ n_2=k\}$, $0\leq k\leq n_L$, so that $J(k)=\# L(k)=n_L-k+1$. That is, a level $L(k)$
comprises all the possible states
$(n_1,n_2)$ of the process with a total number of $P_2$ complexes equal to $k$. Moreover, we order these levels as
\begin{eqnarray*}
 L(0)\prec L(1)\prec\dots\prec L(n_L),
\end{eqnarray*}
and states inside a level, $L(k)=\{(0,k),(1,k),\dots,(n_L-k,k)\}$, $0\leq k\leq n_L$, are ordered as
\begin{eqnarray*}
 (0,k)\prec (1,k)\prec\dots\prec (n_L-k,k).
\end{eqnarray*}

{Given} the transitions of Figure~\ref{Fig2}, it is clear that from a state $(n_1,n_2)$ at level $L(n_2)$,
the process can only
move to states in the same level, $L(n_2)$, and to states at adjacent levels, $L(n_2-1)$ and $L(n_2+1)$.
That is, if the state of the system is $(n_1,n_2)$ (and then,
the process is in level $L(n_2)$), the only possible transitions are to $(n_1-1,n_2)$
 (if a monomer dissociates, in which case the process
remains in level $L(n_2)$), to $(n_1+1,n_2)$
(if a monomer is formed, leaving the process in level $L(n_2)$),
to $(n_1+1,n_2-1)$ (if a dimer dissociates, and the process then decreases to level $L(n_2-1)$),
or to
$(n_1-1,n_2+1)$ (if a dimer is created, increasing the level of the process to $L(n_2+1)$).

The organisation of ${\cal S}$, previously proposed, becomes crucial in order to obtain a
convenient structure for the infinitesimal generator ${\bf Q}$ of ${\cal X}$, the matrix containing the
transition rates in the Markov chain. Specifically, the above yields an infinitesimal generator which has
the following tridiagonal-by-block structure
 \begin{eqnarray}\label{Eqn3}
 {\bf Q} &=& \left(\begin{array}{cccccc}
{\bf A}_{0,0} & {\bf A}_{0,1} & {\bf 0}_{J(0)\times J(2)} & \dots & {\bf 0}_{J(0)\times J(n_L-1)} & {\bf 0}_{J(0)\times J(n_L)} \\
{\bf A}_{1,0} & {\bf A}_{1,1} & {\bf A}_{1,2} & \dots & {\bf 0}_{J(1)\times J(n_L-1)} & {\bf 0}_{J(1)\times J(n_L)} \\
{\bf 0}_{J(2)\times J(0)} & {\bf A}_{2,1} & {\bf A}_{2,2} & \dots & {\bf 0}_{J(2)\times J(n_L-1)} & {\bf 0}_{J(2)\times J(n_L)} \\
\vdots & \vdots & \vdots & \ddots & \vdots & \vdots \\
{\bf 0}_{J(n_L-1)\times J(0)} & {\bf 0}_{J(n_L-1)\times J(1)} & {\bf 0}_{J(n_L-1)\times J(2)} & \dots & {\bf A}_{n_L-1,n_L-1} & {\bf A}_{n_L-1,n_L} \\
{\bf 0}_{J(n_L)\times J(0)} & {\bf 0}_{J(n_L)\times J(1)} & {\bf 0}_{J(n_L)\times J(2)} & \dots & {\bf A}_{n_L,n_L-1} & {\bf A}_{n_L,n_L}
                   \end{array}\right),\nonumber\\
& &
\end{eqnarray}
where sub-matrices ${\bf A}_{k,k'}$ contain the infinitesimal transition rates
of the transitions from states at level $L(k)$ to states at level $L(k')$,
with $k'\in\{k-1,k,k+1\}$. Expressions for sub-matrices ${\bf A}_{k,k'}$ are directly derived from~\eqref{Eqn2},
and are specified in Appendix~B.1.

Firstly, we consider the time to obtain a number $N_2>0$ of $P_2$ complexes.
In particular, given an initial state of the process $(n_1,n_2)$, and a certain threshold $N_2>0$,
we consider the random variable
\begin{eqnarray*}
 T_{(n_1,n_2)}(N_2) &=& \hbox{\it ``Time to reach a number of $P_2$ complexes equal to $N_2$ in ${\cal X}$,}\\
&& \hbox{\it if the process starts at $(n_1,n_2)\in{\cal S}$''}.
\end{eqnarray*}
We observe that this time is $0$ for $N_2\leq n_2$. In order to study this descriptor for $N_2>n_2$,
we make use of an auxiliary CTMC, ${\cal X}(N_2)$, which depends on the threshold value $N_2$. We define
${\cal X}(N_2)$ over ${\cal S}(N_2)$ with
\begin{eqnarray*}
 {\cal S}(N_2) &=& {\cal C}(N_2)\cup\{\bar N_2\},
\end{eqnarray*}
where we denote ${\cal C}(N_2)=\cup_{k=0}^{N_2-1}L(k)$, and where ${\bar N_2}$ is
a macro-state obtained by lumping together all states in the set $\cup_{k=N_2}^{n_L}L(k)$.
Regarding the transition rates of this auxiliary CTMC,
we retain those transitions of ${\cal X}$ between states in ${\cal C}(N_2)$,
and we consider ${\bar N_2}$ as an absorbing macro-state, so that once
${\cal X}(N_2)$ enters  ${\bar N_2}$, it does not leave this state.
Transitions from states in level $L(N_2-1)$ to states in $L(N_2)$ of the original process ${\cal X}$,
 become transitions from
states in level $L(N_2-1)$ to the macro-state ${\bar N_2}$ in ${\cal X}(N_2)$,
where their infinitesimal transition rates are directly obtained
from the original ones as follows:
\begin{eqnarray*}
 q_{(n_1,n_2),{\bar N_2}} &=& \sum\limits_{(n_1',n_2')\in L(N_2)}q_{(n_1,n_2),(n_1',n_2')},\quad \forall (n_1,n_2)\in L(N_2-1).
\end{eqnarray*}
The process ${\cal X}(N_2)$ can be seen as the process ${\cal X}$ until $N_2$ of $P_2$ complexes are
formed. Then, ${\cal X}(N_2)$ ends
since ${\bar N_2}$ is an absorbing state for this auxiliary process.
With ${\cal X}(N_2)$ so defined, it is clear that the time
taken to obtain a number $N_2$ of $P_2$ complexes in the original process ${\cal X}$ is equal
to the time until absorption at ${\bar N_2}$ in the absorbing process ${\cal X}(N_2)$,
 which is known to follow a continuous
phase-type (PH) distribution, see {\em e.g.,} Refs.~\cite{Kulkarni95,Latouche99}.
Analysing the exact distribution of a continuous PH distribution,  in general, is a (well known) difficult
problem. In our case, it would imply obtaining the exponential matrix
$\exp({\bf T}(N_2))=\sum_{n=0}^{+\infty}\frac{{\bf T}(N_2)^{n}}{n!}$,
where ${\bf T}(N_2)$ is a specific sub-matrix of the infinitesimal generator of ${\cal X}(N_2)$.
Here, we instead make use of the Laplace-Stieltjes transform of $T_{(n_1,n_2)}(N_2)$,
which completely determines its distribution, and which
allows us to obtain any $l$-th order moment $E[T_{(n_1,n_2)}(N_2)^l]$.
Moreover, we can efficiently calculate the $l$-th order moment by using
the $(l-1)$-th order moment, proceeding recursively, with the computational effort devoted to
 obtaining inverses of square blocks ${\bf A}_{k,k}$, which have
dimension $J(k)=n_L-k+1$. Again, the proposed organisation of states is crucial for the construction
of an efficient algorithm. If we define the
Laplace-Stieltjes transform of $T_{(n_1,n_2)}(N_2)$ as
\begin{eqnarray*}
 \phi^{N_2}_{(n_1,n_2)}(z) &=& E\left[e^{-zT_{(n_1,n_2)}(N_2)}\right],\quad \Re(z)\geq0,
\end{eqnarray*}
\noindent then, the different $l$-th order moments of $T_{(n_1,n_2)}(N_2)$ can be obtained as
\begin{eqnarray*}
E\left[T_{(n_1,n_2)}(N_2)^l\right] &=& \left.(-1)^l\frac{d^l}{dz^l}\phi^{N_2}_{(n_1,n_2)}(z)\right|_{z=0},\quad \forall l\geq1.
\end{eqnarray*}
We can apply a first-step argument in order to obtain a system of linear equations
for the Laplace-Stieltjes transforms $\phi^{N_2}_{(n_1,n_2)}(z)$, given a
state $(n_1,n_2)\in{\cal{S}}(N_2)$. We can write down the equation
\begin{eqnarray}\label{Eqn4}
 \phi^{N_2}_{(n_1,n_2)}(z) &=& (1-\delta_{n_1+n_2,n_L})
 \frac{2\alpha_{+}(n_{R_2}-n_1-2n_2)(n_L-n_1-n_2)}{z+A_{(n_1,n_2)}}\phi^{N_2}_{(n_1+1,n_2)}(z)+(1-\delta_{n_1,0})\nonumber\\
&&\times\frac{\alpha_{-}n_1}{z+A_{(n_1,n_2)}}\phi^{N_2}_{(n_1-1,n_2)}(z)+(1-\delta_{n_1,0})
\frac{\beta_{+}n_1(n_{R_2}-n_1-2n_2)}{z+A_{(n_1,n_2)}}\big(\delta_{n_2,N_2-1}\\
&&+(1-\delta_{n_2,N_2-1})\phi^{N_2}_{(n_1-1,n_2+1)}(z)\big)+(1-\delta_{n_2,0})
\frac{2\beta_{-}n_2}{z+A_{(n_1,n_2)}}\phi^{N_2}_{(n_1+1,n_2-1)}(z),
\nonumber
\end{eqnarray}
where from now on $A_{(n_1,n_2)}
=2\alpha_+(n_{R_2}-n_1-2n_2)(n_L-n_1-n_2)+\alpha_{-}n_1+\beta_{+}n_1(n_{R_2}-n_1-2n_2)+2\beta_{-}n_2$. Eq.~\eqref{Eqn4}
relates the Laplace-Stieltjes transforms corresponding to all the states of ${\cal S}(N_2)$, 
so that a system of linear equations is obtained. If we organise the Laplace-Stieltjes
transforms in vectors by levels as follows
\[
{\bf g}^{N_2}(z)=({\bf g}^{N_2}_{0}(z)^T,{\bf g}^{N_2}_{1}(z)^T,{\bf g}^{N_2}_{2}(z)^T,\dots,{\bf g}^{N_2}_{N_2-1}(z)^T)^T,
\]
with
${\bf g}^{N_2}_{k}(z)=(\phi^{N_2}_{(0,k)}(z),\phi^{N_2}_{(1,k)}(z),\phi^{N_2}_{(2,k)}(z),\dots,\phi^{N_2}_{(n_L-k,k)}(z))^T$, for $0\leq k\leq N_2-1$,
then the system given in~\eqref{Eqn4} can be expressed in matrix form as
\begin{eqnarray}\label{Eqn5}
 {\bf g}^{N_2}(z) &=& {\bf A}^{N_2}(z) \; {\bf g}^{N_2}(z)+{\bf a}^{N_2}(z),
\end{eqnarray}
with the matrix ${\bf A}^{N_2}(z)$ given by
\begin{eqnarray*}
\left(\begin{array}{ccccccc}
{\bf A}_{0,0}(z) & {\bf A}_{0,1}(z) & {\bf 0}_{J(0)\times J(2)} & \dots & {\bf 0}_{J(0)\times J(N_2-2)} & {\bf 0}_{J(0)\times J(N_2-1)} \\
{\bf A}_{1,0}(z) & {\bf A}_{1,1}(z) & {\bf A}_{1,2}(z) & \dots & {\bf 0}_{J(1)\times J(N_2-2)} & {\bf 0}_{J(1)\times J(N_2-1)} \\
{\bf 0}_{J(2)\times J(0)} & {\bf A}_{2,1}(z) & {\bf A}_{2,2}(z) & \dots & {\bf 0}_{J(2)\times J(N_2-2)} & {\bf 0}_{J(2)\times J(N_2-1)} \\
\vdots & \vdots & \vdots & \ddots & \vdots & \vdots \\
{\bf 0}_{J(N_2-2)\times J(0)} & {\bf 0}_{J(N_2-2)\times J(1)} & {\bf 0}_{J(N_2-2)\times J(2)} & 
\dots & {\bf A}_{N_2-2,N_2-2}(z) & {\bf A}_{N_2-2,N_2-1}(z) \\
{\bf 0}_{J(N_2-1)\times J(0)} & {\bf 0}_{J(N_2-1)\times J(1)} & {\bf 0}_{J(N_2-1)\times J(2)} & 
\dots & {\bf A}_{N_2-1,N_2-2}(z) & {\bf A}_{N_2-1,N_2-1}(z)
\end{array}\right),
\end{eqnarray*}
and the vector
\begin{eqnarray*}
{\bf a}^{N_2}(z) &=& \left(\begin{array}{c}
{\bf 0}_{J(0)}\\
{\bf 0}_{J(1)}\\
\vdots\\
{\bf 0}_{J(N_2-2)}\\
{\bf a}_{N_2-1}(z)
\end{array}\right).
\end{eqnarray*}
Sub-matrices ${\bf A}_{k,k'}(z)$ and the sub-vector ${\bf a}_{N_2-1}(z)$ are given in Appendix~B.2. 
Exploiting the special block structure of ${\bf A}^{N_2}(z)$,
allows for an efficient solution of the system in~\eqref{Eqn5}, in a recursive manner through a specialised block-Gaussian elimination process. 
This gives us Algorithm~1 (Part~1)
listed in Appendix~C. The calculation of the Laplace-Stieltjes transforms in Algorithm~1 (Part~1) has its
own merit, since it determines the distribution of the random variable under consideration. Moreover, the calculation of the distribution function of
$T_{(n_1,n_2)}(N_2)$ by numerical inversion of the transform is possible, although computationally expensive, 
and is not developed here (see {\em e.g.,} Ref.~\cite{Abbate92}).

Once the Laplace-Stieltjes transforms are in hand, we can obtain the different $l$-th order moments 
by successive differentiation of the system in~\eqref{Eqn5}. In particular, we can write
\begin{eqnarray}\label{Eqn6}
 {\bf m}^{N_2,(l)} &=& \sum\limits_{p=0}^{l}(-1)^p\binom{l}{p}\left.
 \frac{d^p}{dz^p}{\bf A}^{N_2}(z)\right|_{z=0}{\bf m}^{N_2,(l-p)}+(-1)^l\left.\frac{d^{l}}{dz^{l}}{\bf a}^{N_2}(z)\right|_{z=0},
\end{eqnarray}
where ${\bf m}^{N_2,(l)}$ is the column vector containing the 
desired moments $E[T_{(n_1,n_2)}(N_2)^l]$, for $(n_1,n_2)\in{\cal C}(N_2)$. We organise these
moments in sub-vectors by levels as
\[
{\bf m}^{N_2,(l)}=({\bf m}^{N_2,(l)T}_{0},{\bf m}^{N_2,(l)T}_{1},{\bf m}^{N_2,(l)T}_{2},\dots,\linebreak {\bf m}^{N_2,(l)T}_{N_2-1})^T,
\]
 with ${\bf m}^{N_2,(l)}_{k}=(E[T_{(0,k)}(N_2)^l],E[T_{(1,k)}(N_2)^l],E[T_{(2,k)}(N_2)^l],\ldots,E[T_{(n_L-k,k)}(N_2)^l])^T$, for $0\leq k\leq N_2-1$.
Note that the notation ${\bf m}^{N_2,(0)}={\bf g}^{N_2}(0)={\bf e}_{\# {\cal C}(N_2)}$ is implicit in~\eqref{Eqn6}. That is, the moment of order
$l=0$ is the Laplace-Stieltjes transform for $z=0$. Finally, the system in~\eqref{Eqn6}
 is rewritten following the calculus notation presented in Appendix~A,
as follows:
\begin{eqnarray}\label{Eqn7}
{\bf m}^{N_2,(l)} &=& {\bf A}^{N_2}(0){\bf m}^{N_2,(l)}
+\sum\limits_{p=1}^{l}\binom{l}{p}(-1)^{p}{\bf A}^{N_2,(p)}(0){\bf m}^{N_2,(l-p)}+(-1)^{l}{\bf a}^{N_2,(l)}(0).
\end{eqnarray}
It is clear that the direct calculation of the inverse $({\bf I}_{\#{\cal C}(N_2)}-{\bf A}^{N_2}(0))^{-1}$
 involved in the solution of~\eqref{Eqn7} can be
avoided by working by levels and solving~\eqref{Eqn7} in a similar way to Algorithm~1 (Part~1).
By starting with the known moment of order $p=0$, we proceed recursively by calculating
${\bf m}^{N_2,(p)}$ from ${\bf m}^{N_2,(p-1)}$, until the desired order $p=l$ is reached. This gives us Algorithm~1
(Part~2) listed in Appendix~C.
Expressions for the derivative matrices ${\bf A}^{N_2,(p)}(0)$ and ${\bf a}^{N_2,(p)}(0)$ are given in Appendix~B.3.

Finally, the long term behaviour
 of the process is given by the stationary distribution of the CTMC; that is, the probabilities
\begin{eqnarray*}
 \pi_{(n_1,n_2)} &=& \lim\limits_{t\rightarrow +\infty}\mathbb{P}((M_2(t),P_2(t))=(n_1,n_2)),\quad \forall(n_1,n_2)\in{\cal S},
\end{eqnarray*}
which do not depend on the initial state. We can store this distribution in a row vector
 ${\boldsymbol \pi}=({\boldsymbol \pi}_{0},{\boldsymbol \pi}_{2},\dots,{\boldsymbol \pi}_{n_L})$,
where the row sub-vector ${\boldsymbol \pi}_{k}$ contains the ordered probabilities $\pi_{(n_1,n_2)}$ for states at level $L(k)$. Solving the system
\begin{eqnarray*}
 {\boldsymbol \pi}{\bf Q} = {\bf 0}^{T}_{\#{\cal S}}
 \quad
 {\rm and}
 \quad
 {\boldsymbol \pi}{\bf e}_{\#{\cal S}} = 1,
\end{eqnarray*}
and adapting the arguments in Ref.~\cite[Chapter 10]{Latouche99}, we obtain Algorithm~2 listed in Appendix~C.
With ${\boldsymbol \pi}$ in hand,
the long term mean number of $M_2$ and $P_2$ complexes can be obtained as
\begin{eqnarray*}
 m_1 &=& \hbox{\it ``Mean number of $M_2$ complexes in steady state''}
  ~=~ \sum\limits_{k=0}^{n_L}k\left(\sum\limits_{j=0}^{n_L}({\boldsymbol \pi}_{j})_k\right),\\
 m_2 &=& \hbox{\it ``Mean number of $P_2$ complexes in steady state''} ~=~ \sum\limits_{k=0}^{n_L}k\left({\boldsymbol \pi}_{k}{\bf e}_{J(k)}\right).
\end{eqnarray*}


\subsection{Model~2: delayed phosphorylation}
\label{Subsect22}

\begin{figure}[htp!]
    \centering
      \includegraphics[scale=0.25]{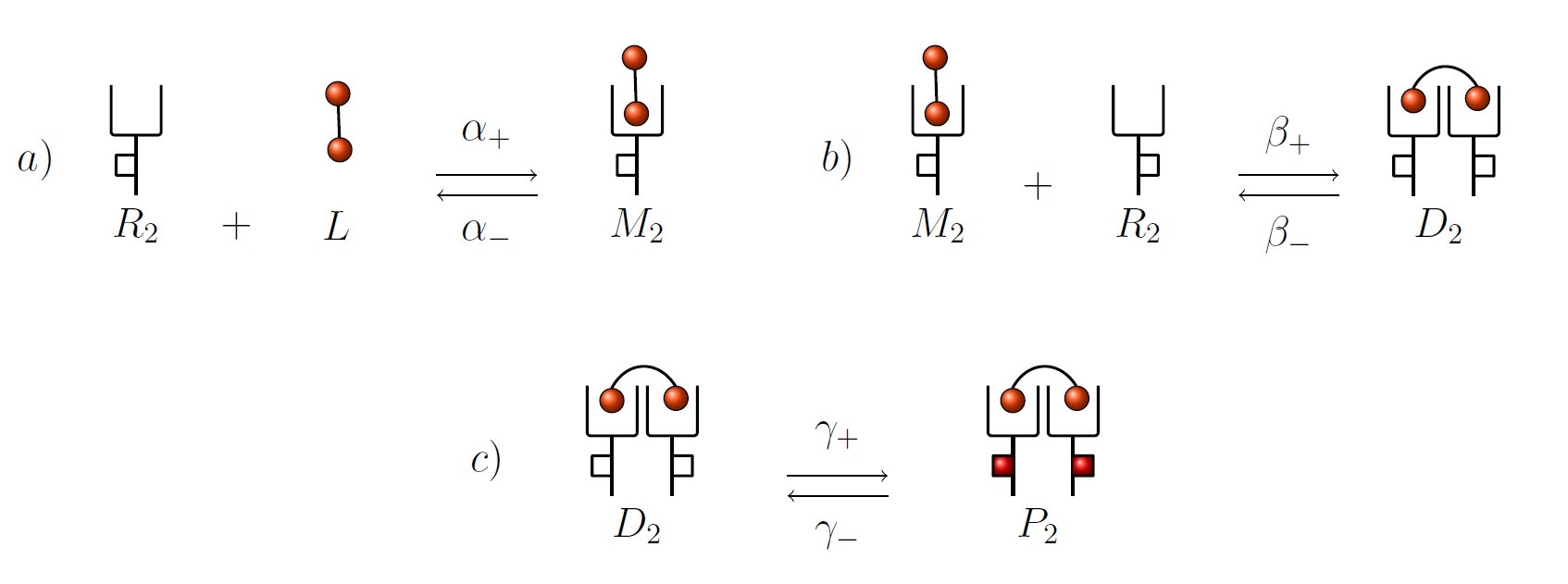}
    \caption{Reactions in Model~2. $a$) Formation and dissociation of bound monomers ($M_2$).
    $b$) Formation and dissociation of non-phosphorylated dimers ($D_2$).
$c$) Formation and de-phosphorylation of phosphorylated dimers ($P_2$).}
    \label{Fig3}
\end{figure}

In the previous Section, the $P_2$ complexes were instantaneously phosphorylated.
Here we relax this requirement and include phosphorylation as
an additional reaction (see Figure~\ref{Fig3}).
In what follows, we adapt the arguments of the previous Section to this model.
This allows us not only to evaluate the relevance of considering
phosphorylation as an independent reaction (with numerical results presented  in Section~\ref{Sect4}), but
also serves as an example of how to include new reactions in this type of models,
 while adapting the matrix-analytic arguments.

In brief, we consider the CTMC ${\cal \hat X}=\{{\bf \hat X}(t)=(\hat M_2(t),\hat D_2(t),\hat P_2(t)):~ t\geq0\}$, where
\begin{eqnarray*}
 \hat M_2(t) &=& \hbox{\it ``Number of $M_2$ complexes at time $t$''},\\
 \hat D_2(t) &=& \hbox{\it ``Number of $D_2$ complexes at time $t$''},\\
 \hat P_2(t) &=& \hbox{\it ``Number of $P_2$ complexes at time $t$''},
\end{eqnarray*}
for all $t\geq0$, where $D_2$ complexes refer to non-phosphorylated dimers
and $P_2$ to phosphorylated ones. From the reactions in Figure~\ref{Fig3}, it is clear that for all $t\geq0$
\begin{eqnarray*}
 \hat M_2(t)+\hat D_2(t)+\hat P_2(t) &\leq& n_L,\\
 \hat M_2(t)+2\hat D_2(t)+2\hat P_2(t) &\leq& n_{R_2},
\end{eqnarray*}
and, by assuming as previously that $2n_L\leq n_{R_2}$, it is straightforward to show that
\begin{eqnarray*}
 {\hat M}_2(t)+{\hat D}_2(t)+{\hat P}_2(t) \leq n_L,\quad \forall t\geq0 &\Rightarrow
 & {\hat M}_2(t)+2{\hat D}_2(t)+2{\hat P}_2(t) \leq n_{R_2},\quad \forall t\geq0,
\end{eqnarray*}
so that ${\cal \hat X}$ is defined over ${\cal \hat S}=\{(n_1,n_2,n_3)\in(\mathbb{N}\cup\{0\})^3:~ n_1+n_2+n_3\leq n_L\}$. We are thus, interested 
in the following descriptors, analogous to those of the previous Section:

\begin{enumerate}

  \item Starting from any state $(n_1,n_2,n_3)\in{\cal \hat S}$, the time to reach a number $N_3>n_3$ of $P_2$ complexes.

  \item Starting from any state $(n_1,n_2,n_3)\in{\cal \hat S}$, the stationary distribution of the system.

\end{enumerate}

\begin{figure}[htp!]
    \centering
    \includegraphics[scale=0.70]{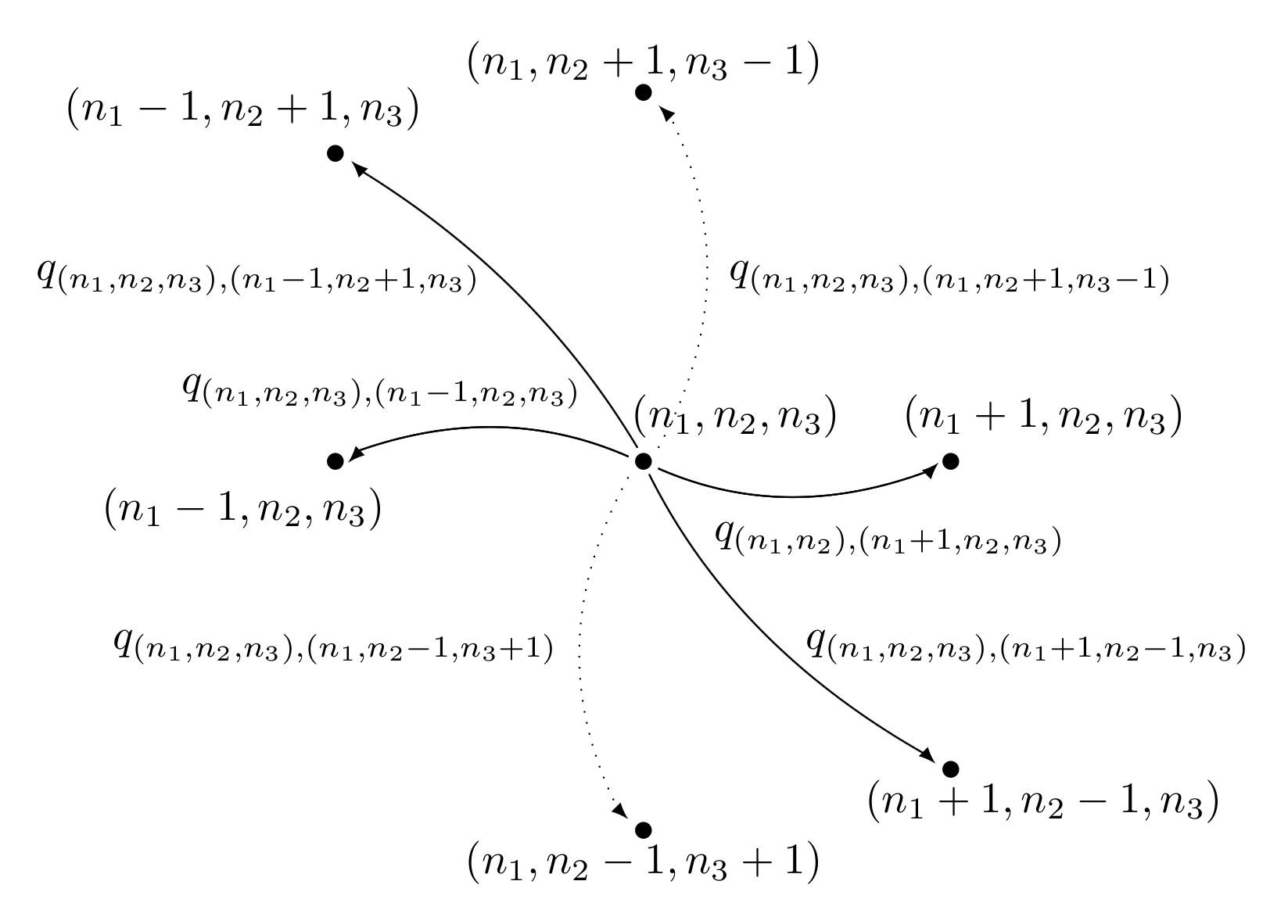}
    \caption{Transition diagram for model~2.}
    \label{Fig4}
\end{figure}

To study these descriptors, we again use level structures for the state space, and split ${\cal \hat S}$ in levels
as follows:
\begin{eqnarray*}
 {\cal \hat S} &=& \bigcup\limits_{k=0}^{n_L}{\hat L}(k),
\end{eqnarray*}
where ${\hat L}(k)=\{(n_1,n_2,n_3)\in{\cal \hat S}:~ n_3=k\}$, for $0\leq k\leq n_L$, so that
\begin{eqnarray*}
{\hat J}(k) &=& \# {\hat L}(k) ~=~ \frac{(n_L-k+1)(n_L-k+2)}{2}.
\end{eqnarray*}
The three-dimensionality of our process implies that each level ${\hat L}(k)$ may be split into
different sub-levels, as follows:
\begin{eqnarray*}
 {\hat L}(k) &=& \bigcup\limits_{r=0}^{n_L-k}l(k;r),
\end{eqnarray*}
with $l(k;r)=\{(n_1,n_2,n_3)\in{\cal \hat S}:~ n_2=r,~ n_3=k\}$, for $0\leq r\leq n_L-k$, $0\leq k\leq n_L$, and $J(k;r) = \# l(k;r)=n_L-r-k+1$. That is,
\begin{eqnarray*}
 l(k;r) &=& \{(0,r,k),(1,r,k),\dots,(n_L-r-k,r,k)\},\quad 0\leq r\leq n_L-k,~ 0\leq k\leq n_L,
\end{eqnarray*}
and states in $l(k;r)$ are ordered as indicated above. {From} Figure~\ref{Fig3}
the transition diagram can be obtained
(Figure~\ref{Fig4}), where non-null infinitesimal transition rates are obtained in a manner analogously to~\eqref{Eqn2}.

The given order of states and the organisation by levels and sub-levels, thus, yield an infinitesimal
generator similar to~\eqref{Eqn3}, where quantities $J(k)$
and matrices ${\bf A}_{k,k'}$ are replaced by ${\hat J}(k)$ and ${\bf \hat A}_{k,k'}$, respectively.
A matrix ${\bf \hat A}_{k,k'}$ contains the ordered infinitesimal transition rates corresponding to transitions from states at level
${\hat L}(k)$ to states at level ${\hat L}(k')$. Each matrix ${\bf \hat A}_{k,k'}$ is formed by sub-blocks ${\bf B}_{r,r'}^{k,k'}$ which contain the
infinitesimal transition rates corresponding to transitions from states at sub-level $l(k;r)\subset {\hat L}(k)$ to
 states at sub-level $l(k';r')\subset {\hat L}(k')$.
We observe that the dimension of the matrix ${\bf \hat A}_{k,k'}$ 
is $\hat{J}(k)\times \hat{J}(k')=\frac{(n_L-k+1)(n_L-k+2)}{2}\times\frac{(n_L-k'+1)(n_L-k'+2)}{2}$,
while the dimension of the sub-block 
${\bf B}_{r,r'}^{k,k'}$ inside ${\bf \hat A}_{k,k'}$ is $J(k;r)\times J(k';r')=(n_L-r-k+1)\times(n_L-r'-k'+1)$. 
Expressions for these matrices are given in Appendix ~B.4.

For an initial state $(n_1,n_2,n_3)\in{\cal \hat S}$ and a number $N_3>0$, we are now interested in the random variable
\begin{eqnarray*}
 T_{(n_1,n_2,n_3)}(N_3) &=& \hbox{\it ``Time to reach a number $N_3$ of $P_2$ complexes if the process}\\
&& \hbox{\it starts at $(n_1,n_2,n_3)$''}.
\end{eqnarray*}
We
omit $N_3$ in the notation for convenience, and
 denote the random variable under study $T_{(n_1,n_2,n_3)}$. Again, this time
is  $0$ for $N_3\leq n_3$. For $N_3>n_3$, we follow an argument similar to that of SubSection~\ref{Subsect21},
 so that the analysis of an
auxiliary absorbing CTMC requires the study  of $T_{(n_1,n_2,n_3)}$ as an absorption time in the auxiliary process.

 In order to obtain the different $l$-th order moments in an efficient way, we define the Laplace-Stieltjes transform of $T_{(n_1,n_2,n_3)}$ as
\begin{eqnarray*}
 \xi_{(n_1,n_2,n_3)}(z) &=& E\left[e^{-zT_{(n_1,n_2,n_3)}}\right],\quad \Re(z)\geq0,
\end{eqnarray*}
and the different $l$-th order moments of $T_{(n_1,n_2,n_3)}$ can be obtained as
\begin{eqnarray*}
E\left[T_{(n_1,n_2,n_3)}^l\right] &=& \left.(-1)^l\frac{d^l}{dz^l}\xi_{(n_1,n_2,n_3)}(z)\right|_{z=0},\quad \forall l\geq1.
\end{eqnarray*}
By a first-step argument (omitted here since it is analogous to~\eqref{Eqn4}), we obtain the system
\begin{eqnarray}\label{Eqn9}
 {\bf \hat g}(z) &=& {\bf \hat A}(z) \; {\bf \hat g}(z)+{\bf \hat a}(z),
\end{eqnarray}
where the Laplace-Stieltjes transforms are stored in vectors ${\bf \hat g}(z)$,
 following the order given by the levels and sub-levels,
and where the expressions for matrices ${\bf \hat A}(z)$ and ${\bf \hat a}(z)$ are omitted for brevity. 
By successive differentiation of the system in~\eqref{Eqn9}, we obtain the different $l$-th order moments 
$E[T^{l}_{(n_1,n_2,n_3)}]$ through an adapted version of Algorithm~1, with $N_2$ replaced by $N_3$, and with the $l$-th 
order moments stored in the vectors ${\bf \hat m}^{(l)}$.
We note that in the adapted version of Algorithm~1 to solve~\eqref{Eqn9}, which is omitted,
we need to deal with inverses of matrices with
dimension ${\hat J}(k)=\# {\hat L}(k)$. The complexity of transitions between states does not seem to
allow us to gain further efficiency in
our algorithms by working with inverses of matrices with the dimensions of the given sub-levels. 
However, in the special case
$\gamma_{-}=0$, that is, when
de-phosphorylation is neglected, it is possible to
improve the procedures so that the highest computational effort
is placed on inverting matrices with the dimensions of sub-levels instead of
levels, which would yield an Algorithm~3, that is  not described here.

 Finally, we focus on the stationary distribution of the process, that is, the probabilities
\begin{eqnarray*}
 \hat \pi_{(n_1,n_2,n_3)} &=& \lim\limits_{t\rightarrow + \infty}\mathbb{P}(({\hat M}_2(t),{\hat D}_2(t),
 {\hat P}_2(t))=(n_1,n_2,n_3)),\quad \forall(n_1,n_2,n_3)\in{\cal \hat S},
\end{eqnarray*}
which do not depend on the initial state. Similar arguments to those considered in SubSection~\ref{Subsect21} allow
us to obtain the stationary
distribution in a row vector ${\boldsymbol{\hat{\pi}}}=\left({\boldsymbol{\hat{\pi}}}_{0},{\boldsymbol{\hat{\pi}}}_{2},
\dots,{\boldsymbol{\hat{\pi}}}_{n_L}\right)$, where
${\boldsymbol{\hat{\pi}}}_{k}=\left({\boldsymbol{\hat{\pi}}}^{k}_{0},{\boldsymbol{\hat{\pi}}}^{k}_{2},\dots,{\boldsymbol{\hat{\pi}}}^{k}_{n_L-k}\right)$,
and where row sub-vectors ${\boldsymbol{\hat{\pi}}}^{k}_{r}$ contain, in an ordered manner, steady state probabilities
of states at sub-levels $l(k;r)$.
An adapted version of Algorithm~2 can be obtained, where the matrices ${\bf A}_{j,j'}$, in~\eqref{Eqn3},
 would be now replaced by the
matrices ${\bf \hat A}_{k,k'}$ previously defined. Once these vectors are in hand, it is clear that
\begin{eqnarray*}
 {\hat m}_1 &=& \hbox{\it ``Mean number of $M_2$ complexes in steady state''} ~=~ \sum\limits_{i=0}^{n_L}i\left(\sum_{k=0}^{n_L-i}\sum\limits_{r=0}^{n_L-i-k}({\boldsymbol{\hat{\pi}}}^{k}_{r})_i\right),\\
 {\hat m}_2 &=& \hbox{\it ``Mean number of $D_2$ complexes in steady state''} ~=~ \sum\limits_{r=0}^{n_L}r\left(\sum_{k=0}^{n_L-r}\sum\limits_{i=0}^{n_L-r-k}({\boldsymbol{\hat{\pi}}}^{k}_{r})_i\right),\\
 {\hat m}_3 &=& \hbox{\it ``Mean number of $P_2$ complexes in steady state''} ~=~ \sum\limits_{k=0}^{n_L}k\left(\sum_{r=0}^{n_L-k}\sum\limits_{i=0}^{n_L-r-k}({\boldsymbol{\hat{\pi}}}^{k}_{r})_i\right).
\end{eqnarray*}

\subsection{Model~3:  competition between VEGFR1 and VEGFR2}
\label{Subsect23}

In previous Sections, we have analysed the interaction between the bivalent ligand VEGF-A and the VEGFR2
on the cell surface. However, both VEGFR1 and VEGFR2 are
expressed on endothelial cells and can bind VEGF-A~\cite{mac2004model}.
VEGFR1 has a greater binding affinity  to VEGF-A than VEGFR2,
but as reported in Ref.~\cite{park1993vascular}, it may not 
be a ``receptor transmitting a mitogenic signal'', but a decoy receptor that prevents VEGF-A binding to VEGFR2.
On the other hand,
VEGFR2 phosphorylation and signalling is required
for the homeostasis of normal endothelial cells~\cite{Alarcon07}.
We, therefore, do not consider VEGFR1 phosphorylation in the models that follow~\cite{casaletto2012spatial}.
Then, VEGFR1 competes with VEGFR2 for ligand, and
these receptors will induce different signalling pathways. Finally, VEGFR1 and
VEGFR2 are found at different copy numbers in a variety of cell lines~\cite{Imoukhuede11,Imoukhuede12}.
Previous studies show that the heterogeneity in these two receptor
numbers contributes to a major complexity of the VEGF-A signal
transduction process, and should be studied further~\cite{Mac07}.
In this Section, we develop two competition models
 which include VEGFR1, VEGFR2 and VEGF-A.

\begin{figure}[htop]
\centering
\includegraphics[scale=0.25]
{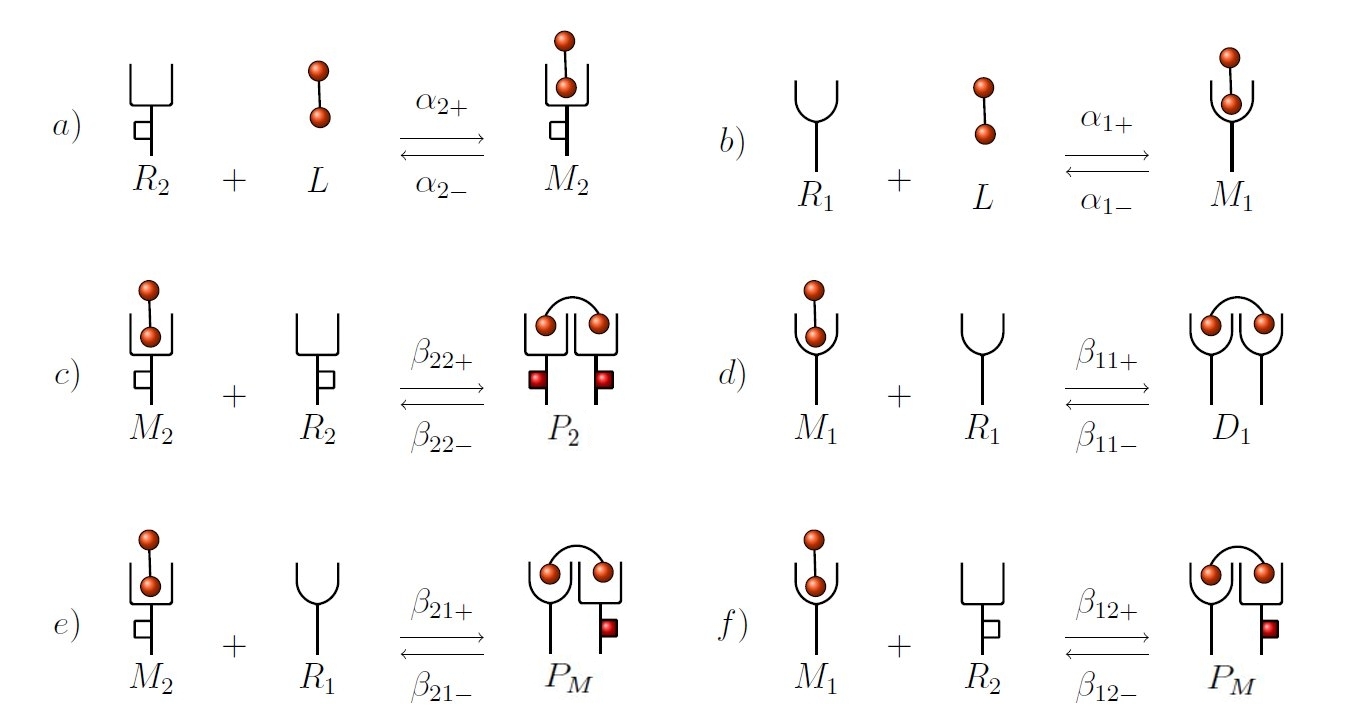}
\caption{Reactions of Model~3.1.
$a)$ Formation or dissociation of a bound monomer ($M_2$).
$b)$ Formation and dissociation of a bound  monomer ($M_1$).
$c)$ Formation or dissociation of bound homodimers ($P_2$).
$d)-f)$ Analogous reactions for homodimers $D_1$ and heterodimers $P_M$. VEGFR2 involved in a
bound dimer becomes instantaneously phosphorylated.}
\label{Fig5a}
\end{figure}

We first consider the case of instantaneous phosphorylation of  bound receptor dimers (Model~3.1).
In this model
 two types of bound monomers can be formed,
$M_1$ and $M_2$, as a result of the ligand binding to VEGFR1 and VEGFR2, respectively. Then,
ligand-induced receptor dimerisation leads to the formation of homodimers, $D_1$
and $P_2$, or heterodimers $P_M$.
VEGFR2 homodimers are instantaneously phosphorylated.
The complete set of reactions of Model~3.1 is given in Figure~\ref{Fig5a}.

We consider a CTMC ${\cal \tilde X}=\{{\bf \tilde X}(t):\ t\geq0\}$ where the state vector
${\bf \tilde X}(t) \in {\cal \tilde S}\subset\left(\mathbb{N}\cup\{0\}\right)^5$ is a collection of discrete 
random variables representing the number of each type of complex at time $t$:
\begin{eqnarray*}
{\bf \tilde X}(t) = ({\tilde M}_1(t), {\tilde M}_2(t), {\tilde D}_1(t), {\tilde P}_2(t), {\tilde P}_M(t)),
\end{eqnarray*}
and its joint probability distribution is given as
\begin{eqnarray*}
 \mathbb{P}_{\bf n}(t) =  \mathbb{P}({\bf \tilde X} (t) = {\bf n}),
\end{eqnarray*}
where ${\bf n} = (n_1, n_2, n_3, n_4, n_5)\in{\cal \tilde S}$. The space of states ${\cal \tilde S}$ can be identified by the implicit restrictions
imposed  by the reactions described in Figure~\ref{Fig5a}, and the consideration of 
fixed numbers $(n_{R_1},n_{R_2},n_{L})$ of receptors and ligands.
We may write the non-null infinitesimal transition rates, based on the reactions shown in Figure~\ref{Fig5a}, as follow:

\begin{eqnarray}\label{model31}
\begin{array}{ll}
q_{({\bf n},{\bf n'})} = &  \left\lbrace
\begin{array}{l}
2\alpha_{1+}(n_L-n_1-n_2-n_3-n_4-n_5)(n_{R_1}-n_1-2n_3-n_5), \\
\quad\quad\quad\quad\quad\quad\quad\quad\quad\quad\quad\quad\quad\quad\quad\quad\quad\quad\quad\quad\quad \ if \ {\bf n'} = (n_1+1,n_2,n_3,n_4,n_5), \\
\alpha_{1-}n_1, \quad\quad\quad\quad\quad\quad\quad\quad\quad\quad\quad\quad\quad\quad\quad\quad\quad\quad if \ {\bf n'} = (n_1-1,n_2,n_3,n_4,n_5), \\
 \beta_{11+}n_1 (n_{R_1}-n_1-2n_3-n_5), \quad\quad\quad\quad\quad\quad\quad\quad if \ {\bf n'} = (n_1-1,n_2,n_3+1,n_4,n_5), \\
 2\beta_{11-}n_3 ,\quad\quad\quad\quad\quad\quad\quad\quad\quad\quad\quad\quad\quad\quad\quad\quad\quad \ if \ {\bf n'} = (n_1+1,n_2,n_3-1,n_4,n_5), \\
 \beta_{12+}n_1(n_{R_2}-n_2-2n_4-n_5), \quad\quad\quad\quad\quad\quad\quad\quad if \ {\bf n'} = (n_1-1,n_2,n_3,n_4,n_5+1), \\
 \beta_{12-}n_5, \quad\quad\quad\quad\quad\quad\quad\quad\quad\quad\quad\quad\quad\quad\quad\quad\quad \ \ if \ {\bf n'} = (n_1+1,n_2,n_3,n_4,n_5-1), \\
 2 \alpha_{2+}(n_L-n_1-n_2-n_3-n_4-n_5)(n_{R_2}-n_2-2n_4-n_5),\\
\quad\quad\quad\quad\quad\quad\quad\quad\quad\quad\quad\quad\quad\quad\quad\quad\quad\quad\quad\quad\quad \ if \ {\bf n'} = (n_1,n_2+1,n_3,n_4,n_5), \\
 \alpha_{2-} n_2,\quad\quad\quad\quad\quad\quad\quad\quad\quad\quad\quad\quad\quad\quad\quad\quad\quad \quad if \ {\bf n'} = (n_1,n_2-1,n_3,n_4,n_5), \\
 \beta_{22+}n_2 (n_{R_2}-n_2-2n_4-n_5),\quad\quad\quad\quad\quad\quad\quad\quad if \ {\bf n'} = (n_1,n_2-1,n_3,n_4+1,n_5), \\
 2\beta_{22-}n_4, \quad\quad\quad\quad\quad\quad\quad\quad\quad\quad\quad\quad\quad\quad\quad\quad\quad \ if \ {\bf n'} = (n_1,n_2+1,n_3,n_4-1,n_5), \\
 \beta_{21+}n_2 (n_{R_1}-n_1-2n_3-n_5), \quad\quad\quad\quad\quad\quad\quad\quad if \ {\bf n'} = (n_1,n_2-1,n_3,n_4,n_5+1), \\
 \beta_{21-}n_5, \quad\quad\quad\quad\quad\quad\quad\quad\quad\quad\quad\quad\quad\quad\quad\quad\quad \ \ if \ {\bf n'} = (n_1,n_2+1,n_3,n_4,n_5-1),
\end{array}
\right.
\end{array}
\end{eqnarray}
and where $q_{(\bf n,n)} = - \sum_{\bf n \neq n'} q_{(\bf n, n')}$. The dynamics of the model can be described by the master equation
\begin{eqnarray}
\frac{d\mathbb{P}_{\bf n}(t)}{dt} &=&  \sum_{{\bf n'} \in S \; , {\bf n'} \neq {\bf n}} 
\; q_{({\bf n', n})} \; \mathbb{P}_{\bf n'}(t)
-\sum_{{\bf n'} \in S \; , {\bf n'} \neq {\bf n}} 
\; q_{({\bf n, n'})} \; \mathbb{P}_{\bf n}(t)
\; ,
\quad \forall {\bf n}\in{\cal S},
\label{eq: master R1-R2-L}
\end{eqnarray}
with the initial condition $\mathbb{P}_{(0,0,0,0,0)}(0) = 1$.

\begin{figure}[htp!]
\centering
\includegraphics[scale=0.3]{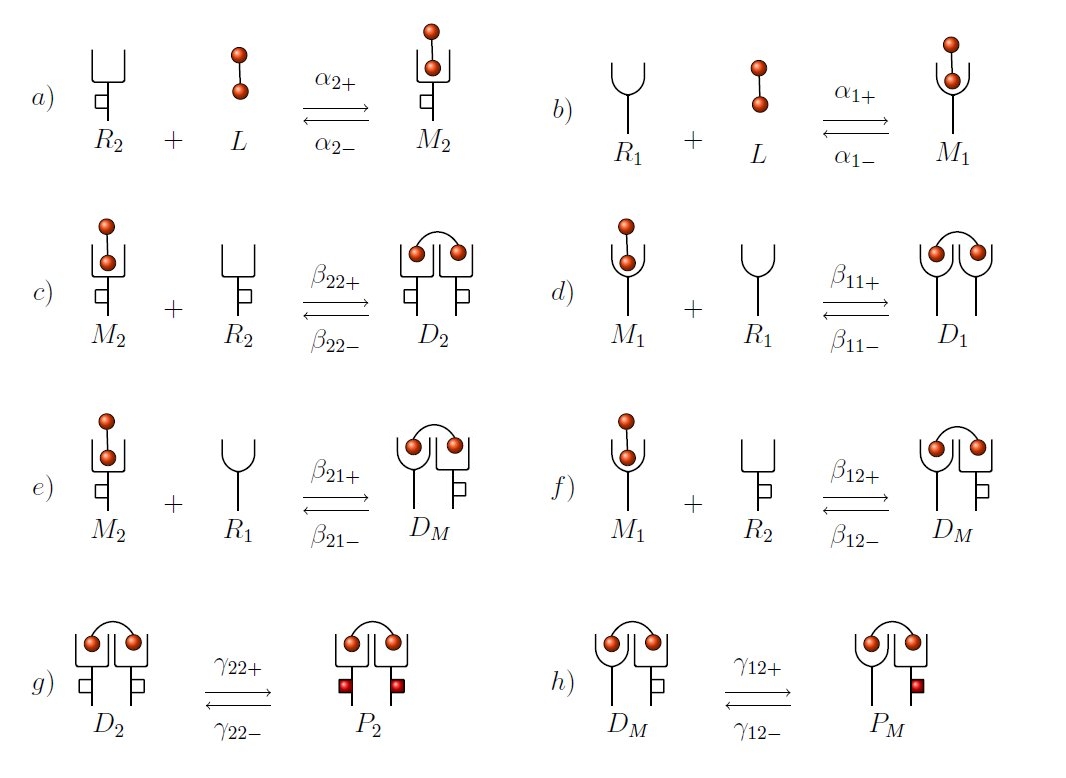}
\caption{Reactions in Model~3.2. Reactions $a)-b)$ are those of Figure~\ref{Fig5a}. Reactions $c)-f)$ describe
the formation of non-phosphorylated ligand cross-linked dimers. Reactions $g)$ and $h)$ represent,
 respectively, phosphorylation of homodimers $D_2$ and heterodimers $D_M$.}
\label{Fig5b}
\end{figure}

We can now introduce a variant of Model~3.1, denoted Model~3.2, as done in Section~\ref{Subsect22},
in which phosphorylation is not assumed to be instantaneous. In this case, the dimeric bound complexes,
$D_2$ and $D_M$, can become phosphorylated, $P_2$ and $P_M$, complexes, respectively.
The complete set of reactions of Model~3.2 is given in Figure~\ref{Fig5b}.
 Non-null
infinitesimal transition rates
can be obtained in a similar way to Model~3.1 (see~Eq.~\eqref{model31}), and an analogous master equation
to that in~\eqref{eq: master R1-R2-L}
can be written.
An analogous procedure to the one described in SubSections~\ref{Subsect21} and~\ref{Subsect22}
can be followed to study
stochastic descriptors in these competition models.
However, the dimensionality of these processes  makes them  intractable from a computational
point of view. Analytical methods, such as moment-closure techniques,
 may be used, within the validity of these approximations, to study
 the dynamics given in~\eqref{eq: master R1-R2-L}. We discuss the
application of the Van Kampen approximation  in Section~\ref{Sect5}, with a short explanation of its implementation in Appendix~D. 
However, given the restricted validity of the
 Van Kampen approximation in our competition models, Gillespie simulations are also carried out
  in Section~\ref{Sect4}.

\subsection{Local sensitivity analysis for stochastic descriptors and kinetic rates}
\label{Subsect24}

The objective of this Section
 is to develop a local sensitivity analysis to understand the
 effect that each of the  (binding, dissociation or phosphorylation) rates
($\alpha_{+}, \alpha_{-}, \beta_{+}, \beta_{-}, \gamma_{+}$ and $\gamma_{-}$)
has on the stochastic descriptors introduced in SubSections~\ref{Subsect21} and~\ref{Subsect22},
in a given neighbourhood of parameter space.
This selected  neighbourhood of parameter space may be obtained from a parameter estimation
of  in vitro and  in silico experiments, as shown in Section~\ref{Sect3}.
Our aim then is to obtain the partial derivatives of our descriptors with respect to each parameter,
so that these derivatives provide a measure of the effect of a
perturbation of the parameters on the descriptors.

Sensitivity analysis for CTMC with absorbing states  has been recently developed in Ref.~\cite{Caswell11}.
Although the Markov chains considered in this paper are, in general, non-absorbing,
the arguments in Ref.~\cite{Caswell11} can be clearly generalised to the CTMCs considered here. 
We adapt them in what follows,
 while keeping the spirit of the matrix-analytic approach.

For that aim, we consider a given matrix ${\bf A}_{m\times n}({\boldsymbol \theta})$, that depends on
${\boldsymbol \theta}=(\alpha_+,
\alpha_-,\beta_+,\beta_-,\gamma_+,\gamma_-)$,
the parameter vector,
and its element by element derivative with respect to $\theta_i\in\{\alpha_+,\alpha_-,\beta_+,\beta_-,
\gamma_+,\gamma_-\}$, ${\bf A}^{(\theta_i)}({\boldsymbol \theta})$.
It is then
possible to calculate the derivative of
${\bf A}^{-1}({\boldsymbol \theta})$ with respect to $\theta_i$
from ${\bf A}^{(\theta_i)}({\boldsymbol \theta})$ as (see Refs.~\cite{Magnus85,Magnus88})
\begin{eqnarray*}
 ({\bf A}^{-1})^{(\theta_i)}({\boldsymbol \theta}) &=& -{\bf A}^{-1}({\boldsymbol \theta}){\bf A}^{(\theta_i)}
 ({\boldsymbol \theta}){\bf A}^{-1}({\boldsymbol \theta}).
\end{eqnarray*}
We have made use of
this and other basic matrix calculus properties, as discussed
in Ref.~\cite{Caswell11}, to obtain Algorithm~1S and Algorithm~2S, which are given in Appendix~C,
 and can be obtained
 by sequentially differentiating all matrices in Algorithm~1 and Algorithm~2, respectively.
Finally,
the explicit details of the element by element
partial derivative of the matrices defined in Appendix~B, with respect
to any parameter, $\theta_i\in\{\alpha_{+},\alpha_{-},\beta_{+},\beta_{-},\gamma_{+},\gamma_{-}\}$,
have not been included in Appendix~B.

 It is clear that, since our descriptors are stored in the vectors ${\bf m}^{N_2,(l)}$, ${\bf \hat m}^{(l)}$
 (time to reach a
threshold number of $P_2$ complexes in Model~1 and Model~2, respectively)
and quantities $m_j$ and ${\hat m}_j$ (mean number of complexes
in steady state in Model~1 ($j\in\{1,2\}$) and  Model~2
($j\in\{1,2,3\}$), respectively), the objective in Algorithm~1S and
Algorithm~2S is to obtain the derivative vectors ${\bf m}^{N_2,(l,\theta_i)}$,
${\bf \hat m}^{(l,\theta_i)}$, ${\boldsymbol \pi}^{(\theta_i)}$ and ${\boldsymbol{\hat{\pi}}}^{(\theta_i)}$.
The first two vectors contain the derivatives of
the $l$-th order moments of the time  to reach a given  threshold  number of $P_2$ complexes, and the
last two yield the derivatives of quantities
$m_j$ and ${\hat m}_j$, with respect to each rate $\theta_i\in\{\alpha_{+},\alpha_{-},\beta_{+},\beta_{-},\gamma_{+},\gamma_{-}\}$.


\section{Parameter estimation}
\label{Sect3}

In this Section we show how the parameters
of the models introduced
  in Section~\ref{Sect2} can be estimated, based on the methods proposed by Lauffenburger
and Linderman in Ref.~\cite{Lauffenburger96}.
The transport mechanism of free ligand or free receptor is modelled by
molecular diffusion, since diffusive transport dominates convective transport
caused by fluid motion at cellular and sub-cellular length scales~\cite{Lauffenburger96,Weisz73}.

\subsection{Model~1 and Model~2:  single receptor (VEGFR1 or VEFGR2)}

\begin{figure}[h]
\centering
\includegraphics[scale=0.25]{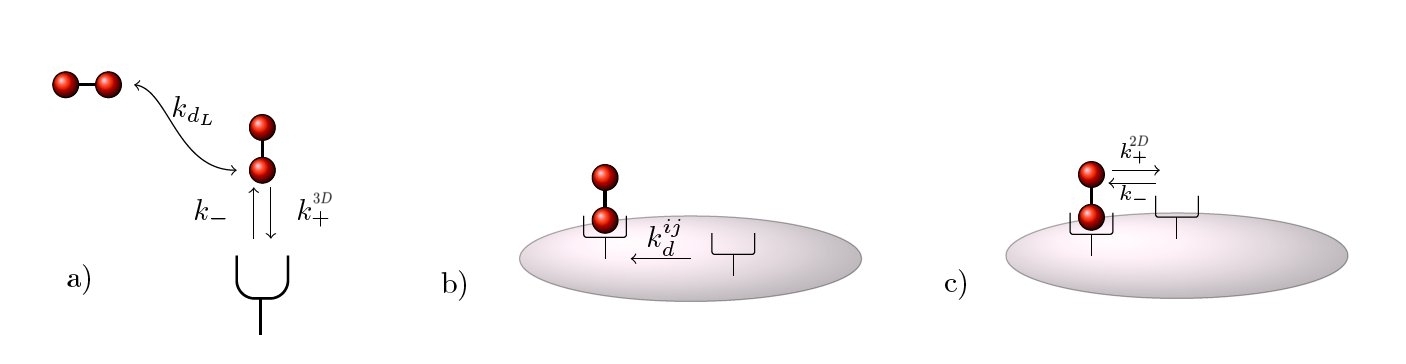}
\caption{$a$) Two-step binding and unbinding  of receptor and ligand. $k_{d_L}$ is the ligand transport rate, $k^{3D}_+/k_-$ are the intrinsic binding/unbinding rates.
$b$)  Diffusive transport of surface receptor. $k_d^{ij}$ is the transport rate for receptor $R_j$ diffusing towards bound monomer $M_i$ (if $i=j$,
then $k_d^{ij} = k_{d_R}$). $c$) Once in the reaction zone of $M_i$, $R_j$ can bind with rate $k_+^{2D}$ (which is a 2D version of $k_+^{3D}$) or unbind
with rate $k_-$.}
\label{Fig2Step}
\end{figure}

In this Section, we  estimate the parameters for the 
binding and unbinding of a single type of receptor (VEGFR1 or VEGFR2) to VEGF-A. We denote the
receptor molecule by $R$ and the ligand by $L$. The binding process between the receptor and the ligand, such as reaction $a)$ in
Figure~\ref{Fig1}, can be considered as a one-step process, with $k_{\rm on}$ the association constant and $k_{\rm off}$ the dissociation constant.
Constants $k_{\rm on}$ and $k_{\rm off}$ will be later identified with or directly related to the rates $\alpha_+$ and $\alpha_-$, respectively,
corresponding to the CTMCs under consideration in Section~\ref{Sect2} (Figures~\ref{Fig1}  and~\ref{Fig2}). However, chemical reactions such as
binding and unbinding events, are in fact two-step processes, requiring the ligand to move first into the neighbourhood of the receptor, with some
diffusion rate $k_{d_L}$, and then interacting with it with intrinsic rate $k^{3D}_+$~\cite{Lauffenburger96}. 
The mechanism of the reverse process is similar,
so that the separation of the receptor and the ligand occurs with intrinsic dissociation rate $k_-$ and the outward diffusion with transport rate
$k_{d_L}$; see Figure~\ref{Fig2Step} $a)$.

If we focus on a particular fraction $0<f<1$ of the cell, as
we will do in Section~\ref{Sect4}, the radius of this target surface is given by
\begin{eqnarray*}
 r &=& \sqrt{\dfrac{n_R s_c}{n_R^T \pi}},
\end{eqnarray*}
\par\noindent where $s_c$ is the total area of the cell surface, $n_R^T$ is the total number of receptors 
on the cell surface, and $n_R=fn_R^T$ is the
number of receptors present on the target surface, which amounts to the assumption of an homogeneous spatial distribution of VEGFR1 and VEGFR2 on the
cell surface~\cite{Ewan06,Mittar09}, neglecting receptor clustering,
 which might be initiated upon ligand simulation~\cite{Almqvist04}. Then, the contributions of rates $k_{d_L}$, $k^{3D}_+$ and $k_-$
to the overall association and dissociation rates, $k_{\rm on}$ and $k_{\rm off}$, respectively,
 are given by
\begin{eqnarray}
k_{\rm on} &=& \left( \frac{1}{k_{d_L}}+ \frac{1}{k^{3D}_+}\right)^{-1}, \quad k_{\rm off} \ = \ \left( \frac{k_{d_L}}{k_{d_L}+k^{3D}_+} \right) k_-,
\label{eq: alpha+, alpha-}
\end{eqnarray}
\par\noindent where the transport rate of the ligand is given by $k_{d_L}=\frac{4 \pi D_L r}{n_R}$, with $D_L$ the diffusion
coefficient of the ligand, so that
\begin{eqnarray*}
k_{\rm on} &=& \dfrac{4 \pi D_L r k^{3D}_+}{4 \pi D_L r + n_R k^{3D}_+}, \quad k_{\rm off} \ = \ \frac{4\pi D_L r k_-}{4 \pi D_L r + n_R k^{3D}_+}.
\end{eqnarray*}
A similar argument (Figure~\ref{Fig2Step} $b)$ and $c)$) applies when computing the overall rate of a receptor binding or unbinding to a second
receptor on the cell membrane~\cite{Lauffenburger96}, which occurs with rates
\begin{eqnarray}
k_{c} &=& \left( \frac{1}{k_{d_R}}+\frac{1}{k^{2D}_+}\right)^{-1}, \quad k_{u} \ = \ \left(  \frac{k_{d_R}}{k_{d_R} + k^{2D}_+}\right) k_-.
\label{eq: beta+, beta-}
\end{eqnarray}
\par\noindent Rate constants $k_{c}$ and $k_{u}$ will be later identified with or directly related to rates $\beta_+$ and $\beta_-$, respectively, for the CTMCs
considered in Section~\ref{Sect2}. In Eq.~\eqref{eq: beta+, beta-}, we have
\begin{eqnarray*}
k_{d_R} &=& \frac{2 \pi D_R}{\log \frac{w}{b}}
\end{eqnarray*}
\par\noindent is the transport rate of the free receptor (Figure~\ref{Fig2Step} $b)$), $D_R$ is the diffusion coefficient of the receptor on the
cell membrane, $b$ is the average radius of the receptor, and $w$ is the average distance between the target receptor and the diffusive free one, given by
\begin{eqnarray*}
w &=& 2 \sqrt{\dfrac{ s_c}{\pi \; n_R^T}}.
\end{eqnarray*}

We set the dissociation rate $k_{\rm off} = 1.32 \cdot 10^{-3} s^{-1}$ as reported in Ref.~\cite{Mac07}, 
to be the same for VEGFR1 and VEGFR2. From the
equilibrium dissociation rate
 $K_d=k_{\rm off}/(k_{\rm on}N_A)$ of the receptor and ligand under consideration, where $N_A$ is 
 Avogadro's number, it is possible then to
obtain $k_{\rm on}$. Given $k_{\rm on}$, transition rates $\alpha_+$ and $\alpha_-$ in Section~\ref{Sect2} are given by
\begin{eqnarray*}
\alpha_+ &=& \frac{k_{\rm on}}{f \; h \;s_c},\quad \alpha_- \ = \ k_{\rm off},
\end{eqnarray*}
where $h$ is the {\it height} of the experimental volume. Moreover, we can obtain $k^{3D}_+$ and $k_-$ 
from~\eqref{eq: alpha+, alpha-}, which allows us
to compute $k_c$ and $k_u$ in~\eqref{eq: beta+, beta-}. In particular, the intrinsic binding rate, $k^{3D}_+$,
 (with units of $s^{-1} \; volume^{-1}$) is
obtained from~\eqref{eq: alpha+, alpha-} in a 3D version, and needs to be transformed into its 2D version, $k^{2D}_+$, 
so that it can be used in the expression~\eqref{eq: beta+, beta-}. To that aim, we divide $k^{3D}_+$ by the average cell membrane thickness $h_m$ \cite{Lauffenburger96}. Once $k^{2D}_+$ and $k_-$ have been computed,
 rates $k_{c}$ and $k_u$ are derived from~\eqref{eq: beta+, beta-} and
\begin{eqnarray*}
\beta_+ &=& \frac{k_c}{f \; s_c},\quad \beta_-=k_u.
\end{eqnarray*}


\subsection{Model~3.1 and Model~3.2: two receptor types (VEGFR1 and VEFGR2)}

In this Section we establish the value of the parameters for the competition Model~3.1 and Model~3.2.
 In this case, there are two types of receptors (VEGFR1
and VEGFR2) in the system, $R_1$ and $R_2$, with the same diffusion coefficient $D_R$. Since the amount of each type of receptor on the cell surface is significantly
different (see Section~\ref{Sect4}), the average distance $w$ between two given receptors, considered in the previous Section, will depend on the particular
pair of receptors under consideration. This 
 changes the diffusion rate $k_{d_R}$ of each possible reaction in Figures~\ref{Fig5a} and~\ref{Fig5b}.
 
Let $k_d^{ij}$ be the transport rate for receptor $R_j$ diffusing towards monomer $M_i$ 
(Figure~\ref{Fig2Step} $b)$), with $i,j\in\{1,2\}$. The probability of monomer $M_i$ meeting
receptor $R_j$ can be approximated by
\begin{eqnarray*}
p_j &=&  \frac{n_{R_j}^T}{n_{R_1}^T+n_{R_2}^T},
\end{eqnarray*}
where $n_{R_j}^T$  is the total number of receptors $R_j$ per cell. In the same way,
the average distance between receptors $R_i$ and $R_j$ can be written as
\begin{eqnarray*}
w_{ij} &=& \left\{\begin{array}{ll}
2\sqrt{\frac{ s_c}{\pi \; n_{R_i}^T}}, & \hbox{\it if \; $i=j$},\\
2\sqrt{\frac{s_c}{\pi \; (n_{R_1}^T + n_{R_2}^T)}}, & \hbox{\it if \; $i\neq j$}.
\end{array}\right.
\end{eqnarray*}
Finally, the diffusion rates are then given by
\begin{eqnarray*}
k_d^{ij} &=& \left\{\begin{array}{ll}
\frac{2 \pi D_R}{\log \frac{w_{ii}}{b}}, & \hbox{\it if \; $i=j$},\\
\frac{2 \pi D_R}{\log \frac{w_{ij}}{b}} p_i, & \hbox{\it if \; $i\neq j$}.
\end{array}\right.
\end{eqnarray*}


\subsection{Sensitivity analysis for physiological parameters and kinetic rates}

 In this Section
 we are interested in
 studying how the key rates
 $k^{3D}_+$, $k_-$, $k_{c}$, and $k_{u}$ depend on some of the other parameters of the model.
 We note that these four constants are essential to describe the rates of the CTMCs
 considered in this paper (see Figures~\ref{Fig1}, \ref{Fig3}, \ref{Fig5a} and~\ref{Fig5b}).
 In particular, we analyse in what follows how these four kinetic rates depend on
{\it physiological} parameters such as the area $s_c$ of the cell, the total number $n_R^T$ of receptors, the ligand diffusion coefficient
$D_L$, the receptor diffusion coefficient $D_R$, and the receptor radius $b$.
 We carry out a sensitivity analysis which allows us to identify the most relevant parameter(s) of the model.

We obtain the partial derivatives of these four rates with respect  to the physiological parameters.
The effect of a given parameter on a kinetic rate is determined by the sign of the corresponding partial derivative, which 
is reported in Table~\ref{tab: sensitivity coefficients}.
As the cell surface increases, it becomes more difficult to find
nearby receptors, thus  binding/unbinding rates (intrinsic and overall)
decrease.
On the other hand, when the number of receptors increases, it is easier
to find nearby receptors,
so that the association and dissociation rates
are larger for increasing receptor numbers. If the diffusion coefficient of the ligand increases,
$k_{d_L}$ also increases and thus,
the intrinsic binding rate must decrease. Yet, if the diffusion coefficient of the ligand increases,
 the probability of dimerisation is greater, and overall rates grow. Finally, by increasing
 the receptor length, the average time to find a free receptor decreases.

\begin{table}[h]
\centering
\begin{tabular}{|l |l|}
\hline
Physiological parameter & Sign of partial derivatives\\
\hline
 Area of the cell surface, $s_c$  &  $
\frac{\partial k^{3D}_+}{\partial s_c}, \frac{\partial k_-}{\partial s_c}, \frac{\partial k_{c}}{\partial s_c}, \frac{\partial k_{u}}{\partial s_c} < 0 $\\ & \\
Total number of receptors, $n_R^T$ &
$
\frac{\partial k^{3D}_+}{\partial n_R^T}, \frac{\partial k_-}{\partial n_R^T}, \frac{\partial k_{c}}{\partial n_R^T}, \frac{\partial k_{u}}{\partial n_R^T} >0
$
\\ & \\
Diffusion coefficients, $D_L$ and $D_R$ & $
\frac{\partial k^{3D}_+}{\partial D_L}, \frac{\partial k_-}{\partial D_L} < 0, \ \ \ \frac{\partial k_{c}}{\partial D_R}, \frac{\partial k_{u}}{\partial D_R} > 0 $\\ & \\
Receptor radius, $b$ &
$\frac{\partial k_{c}}{\partial b}, \frac{\partial k_{u}}{\partial b} > 0$ \\ & \\
\hline
\end{tabular}
\caption{Signs of partial derivatives.}
\label{tab: sensitivity coefficients}
\end{table}

In order to compare the magnitudes of the different partial derivatives in Table~\ref{tab: sensitivity coefficients},
they need to be  normalised
by  the introduction of
 {\it sensitivity coefficients}. The sensitivity coefficient of a
 given dependent parameter with respect to an independent one
can be calculated from the corresponding partial derivative. Specifically, if a parameter $y$ depends on the parameter $z$ as $y=f(z)$, where $f(\cdot)$ is a
certain function, we can define their associated sensitivity coefficient as $\frac{\partial y}{\partial z} \; \frac{z^*}{y^*}$,
where $z^*$ is the actual value of the
parameter $z$ and $y^*=f(z^*)$. The quotient $\frac{z^*}{y^*}$ is then introduced to normalise
the partial derivative.  For example, let us
 focus on the partial derivatives of the binding rate $k^{3D}_+$. Given the definition of 
 $k^{3D}_+$, we have
\begin{eqnarray*}
 k^{3D}_+ &=& \dfrac{k_{\rm on}\; 4\pi D_L \sqrt{\frac{n_R \; s_c}{n_R^T \; \pi}}}{4\pi D_L \sqrt{\frac{n_R \; s_c}{n_R^T \; \pi}} - k_{\rm on} \;n_R},
\end{eqnarray*}
and the following partial derivatives can be computed:
\begin{eqnarray*}
\begin{array}{c c c}
\frac{\partial k^{3D}_+}{\partial D_L} &=& - \frac{k_{\rm on}^2 4\pi n_R  \sqrt{\frac{n_R \; s_c}{n_R^T \; \pi}}}{(4 \pi D_L \sqrt{\frac{n_R \; s_c}{n_R^T \; \pi}} - k_{\rm on} n_R)^2},\quad
\frac{\partial k^{3D}_+}{\partial n_R^T} \ = \ \frac{k_{\rm on}^2 4\pi D_L \sqrt{\frac{n_R s_c}{n_R^T \pi}} n_R }{2 n_R^T (4\pi D_L \sqrt{\frac{n_R s_c}{n_R^T \pi}} - k_{\rm on}n_R)^2},\quad
\frac{\partial k^{3D}_+}{\partial s_c} \ = \ - \frac{k_{\rm on}^2 n_R 4 \pi D_L \frac{n_R}{n_R^T \pi}}{2 \sqrt{\frac{n_R s_c}{n_R^T \pi}} (4\pi D_L \sqrt{\frac{n_R s_c}{n_R^T \pi}} - k_{\rm on}n_R)^2}.
\end{array}
\end{eqnarray*}
\par\noindent Then, regardless of the particular values of the parameters, it can be shown that
\begin{eqnarray*}
\bigg| \frac{\partial k^{3D}_{+}}{\partial D_L} \frac{D_L}{k^{3D}_{+}} \bigg|  &>& \bigg| \frac{\partial k^{3D}_{+}}{\partial n_R^T} \frac{n_R^T}{k^{3D}_{+}} \bigg| \ = \  \bigg| \frac{\partial k^{3D}_{+}}{\partial s_c} \frac{s_c}{k^{3D}_{+}}\bigg|.
\end{eqnarray*}
Similar arguments to the previous ones yield the following inequalities:
\begin{eqnarray*}
\bigg| \frac{\partial k_{-}}{\partial D_L} \frac{D_L}{k_{-}} \bigg|
&>& \bigg| \frac{\partial k_{-}}{\partial n_R^T} \frac{n_R^T}{k_{-}} \bigg| \ = \  \bigg| \frac{\partial k_{-}}{\partial s_c} \frac{s_c}{k_{-}}
\bigg|, \\
\bigg| \frac{\partial k_{c/u}}{\partial D_R} \frac{D_R}{k_{c/u}} \bigg| &>& \bigg| \frac{\partial k_{c/u}}{\partial n_R^T} \frac{n_R^T}{k_{c/u}} \bigg| \ = \ \bigg| \frac{\partial k_{c/u}}{\partial b} \frac{b}{k_{c/u}} \bigg| >  \bigg| \frac{\partial k_{c/u}}{\partial s_c} \frac{s_c}{k_{c/u}} \bigg|,
\end{eqnarray*}
so that the diffusion coefficients, $D_L$ and $D_R$, are the most sensitive physiological parameters in the binding and dissociation rates, while the specific value of the area of the cell surface is the
least sensitive one. We note that the previous inequalities are obtained under the following assumptions:

\begin{itemize}
 \item 
the binding rate is 
 much smaller than the diffusion rate of the ligand
 
\begin{eqnarray*}
k_{\rm on} n_R << 4 \pi D_L \sqrt{\frac{n_R \;s_c}{n_R^T \;\pi }},
\end{eqnarray*}

 \item surface receptor density is low, $b^2 \pi n_R << s_c$, which also
 implies that the average distance between receptors is
larger than the length of the receptor, and

 \item the intrinsic binding rate is greater than the diffusion rate of the receptor, $k^{2D}_+>2 \pi D_R$.
 
\end{itemize}


\section{Results}
\label{Sect4}

First, we note that all the rates involved in Models~1, 2, 3.1 and~3.2 (Figures~\ref{Fig1}, \ref{Fig3}, \ref{Fig5a} 
and~\ref{Fig5b},
respectively) and used in this Section, have been obtained by following the approach described in Section~\ref{Sect3}, 
with physiological parameters taken from the
literature. In particular, physiological parameters are given in Table~\ref{tab: biochem parameters}, 
and the computed rates corresponding to Models~1, 2, 3.1 and~3.2 are given in Table~\ref{tab: rates1} and Table~\ref{tab: rates2}. 
The equilibrium dissociation rate for VEGF-A and VEGFR1, and for VEGF-A and VEGFR2, is equal
to $K_d=30$pM and $K_d=150$pM, respectively, as reported in Ref.~\cite{Mac07}. These rates
are consistent with previously reported values for in silico experiements~\cite{mac2004model},
and agree with experimentally 
determined 
values~\cite{bikfalvi1991interaction,ewan2006intrinsic,huang1998expression,waltenberger1994different}.
Finally, the phosphorylation rate of the $D_M$ complexes in Model~3.2 is taken
to be $\gamma_{21+}=0.5\gamma_{22+}$, since only VEGFR2 is
assumed to become
   phosphorylated (we are neglecting VEGFR1 phosphorylation~\cite{casaletto2012spatial}), and de-phosphorylation
rate of $P_M$ complexes is taken to be $\gamma_{21-}=\gamma_{22-}$.

\begin{table}[h!]
\centering
\begin{tabular}{|l |c |c|}
\hline
Physiological parameter & Value & Reference\\
\hline
Endothelial cell surface area, $s_c$ & $1000 \; \mu m^2$ &~\cite{Mac07} \\
VEGF-A diffusion coefficient at $4\,^{\circ}\mathrm{C}$, $D_L$ & $5.8 \cdot 10^{-7} cm^2 s^{-1}$& ~\cite{Mac05b}\\
VEGFR1 and VEGFR2 diffusion coefficient, $D_R$ & $10^{-10} cm^2 s^{-1}$&~\cite{Linderman89}\\
VEGFR1 and VEGFR2 radius, $b$ &$0.5 \;nm$ &~\cite{Alarcon06}\\
Average membrane thickness of ECs, $h_m$ & $0.1\; \mu m $&~\cite{Aird07} \\
Height of the experimental volume, $h$ & $1 \;mm$ &~\cite{Mac07}\\
Dissociation rate, $k_{\rm off}$ & $1.32 \cdot 10^{-3} s^{-1}$ &~\cite{Mac07}\\
Equilibrium dissociation rate, $K_d$ for VEGFR1 & $30 \;pM$ &~\cite{Mac07} \\
Equilibrium dissociation rate, $K_d$ for VEGFR2 & $150 \;pM$ &~\cite{Mac07}\\
Phosphorylation rate for $D_2$ complexes, $\gamma_{22+}$ & $0.22 \;min^{-1}$&~\cite{Lauffenburger96} \\
De-phosphorylation rate for $P_2$ complexes, $\gamma_{22-}$ & $0.055 \;min^{-1}$ &~\cite{Lauffenburger96} \\
\hline
\end{tabular}
\caption{Physiological parameters}
\label{tab: biochem parameters}
\end{table}

\begin{table}[h!]
\centering
\begin{tabular}{|c |c| c| c| c| c |}
\hline
$\alpha_+$ & $\alpha_-$ & $\beta_+$ & $\beta_-$ & $\gamma_+$ & $\gamma_-$\\
\hline
 $3.653 \cdot 10^{-7}$ &   $ 1.320 \cdot 10^{-3}$ &  $  2.160 \cdot 10 ^{-4} $ & $  7.804 \cdot 10^{-5} $ &  $  3.667 \cdot 10^{-3}$ &  $9.167 \cdot 10^{-4} $\\
\hline
\end{tabular}
\caption{Kinetic rates (in $s^{-1}$) for Model~1 and Model~2, where $n_{R_1} = 0$, considering $4 \%$ of the cell. $\gamma_+$ and $\gamma_-$ are not considered in Model~1.}
\label{tab: rates1}
\end{table}

\begin{table}[h!]
\centering
\begin{tabular}{| c | c | c | c |}
\hline
 & $n_{R_1} = 64$ & $n_{R_1} = 72$ &  $n_{R_1} = 80$\\
 \hline
 $\alpha_{1+}$ & $ 1.827 \cdot 10^{-6}$ & $ 1.827 \cdot 10^{-6}$ & $ 1.827 \cdot 10^{-6}$ \\
$\alpha_{1-} $ &  $1.320 \cdot 10^{-3}$ & $1.320 \cdot 10^{-3}$ & $1.320 \cdot  10^{-3}$\\

$\alpha_{2+}$ & $ 3.653 \cdot 10^{-7}$ & $ 3.653 \cdot 10^{-7}$ & $ 3.653 \cdot 10^{-7}$ \\

$\alpha_{2-}$ & $1.320 \cdot 10^{-3}$ & $1.320 \cdot 10^{-3}$ & $1.320 \cdot  10^{-3}$\\

$\beta_{11+}$ & $2.074 \cdot 10^{-4}$ & $2.091 \cdot 10^{-4}$ & $ 2.105 \cdot 10^{-4}$\\

$\beta_{11-}$ & $1.499 \cdot 10^{-5}$ & $1.511 \cdot 10^{-5}$ & $ 1.521 \cdot 10^{-5}$ \\

$\beta_{12+}$ & $ 1.744 \cdot 10^{-4} $ &  $1.704 \cdot 10^{-4}$ & $1.665 \cdot 10^{-4}$\\

$\beta_{12-}$ & $6.303 \cdot 10^{-5}$ &  $6.156 \cdot 10^{-5}$ & $6.017 \cdot 10^{-5}$ \\

$\beta_{21+}$ & $5.039 \cdot 10^{-5} $ &  $5.530 \cdot 10^{-5}$ &
$5.997 \cdot 10^{-5}$\\

$\beta_{21-}$ & $3.642 \cdot 10^{-6}$ & $3.996 \cdot 10^{-6}$ & $4.333 \cdot 10^{-6}$ \\

$\beta_{22+}$ & $2.160 \cdot 10^{-4}$ & $2.160 \cdot 10^{-4}$ & $2.160 \cdot 10^{-4}$\\

$\beta_{22-}$ & $7.804 \cdot 10^{-5}$ & $7.804 \cdot 10^{-5}$ & $7.804 \cdot 10^{-5}$\\
\hline
$\gamma_{21+}$ & $1.833 \cdot 10^{-3}$ & $1.833 \cdot 10^{-3}$ & $1.833 \cdot 10^{-3}$ \\
$\gamma_{21-}$ & $9.167 \cdot 10^{-4} $ & $9.167 \cdot 10^{-4} $ & $9.167 \cdot 10^{-4} $\\
$\gamma_{22+}$ & $  3.667 \cdot 10^{-3}$ & $  3.667 \cdot 10^{-3}$ & $  3.667 \cdot 10^{-3}$ \\
$\gamma_{22-}$ & $9.167 \cdot 10^{-4} $ & $9.167 \cdot 10^{-4} $ & $9.167 \cdot 10^{-4} $\\
\hline
\end{tabular}
\caption{Kinetic rates (in $s^{-1}$) for Model~3.1 and Model~3.2 considering $4\%$ of the cell. $\gamma_{21+}$, $\gamma_{21-}$, $\gamma_{22+}$ and $\gamma_{22-}$ are not considered in Model~3.1.}
\label{tab: rates2}
\end{table}

We consider in this Section
the subset of endothelial cells, called
human umbilical vein endothelial cells  (HUVECs),
which have been characterised to express (on average)
  $5800$ VEGFR2s per cell~\cite{Imoukhuede12}. We focus on a $4\%$ of the cell surface ($f=0.04$) for computational reasons,
so that in this area the total number  of VEGFR2s is $n_{R_2}=232$. We note
that the size of the area under study is chosen so that the
algorithms in Appendix~C can be used in a reasonable running time. However, computational demand of the
Algorithms~1, 2, 1S and~2S for Model~1, and their
respective versions for Model~2, significantly differ from each other, so that the area under consideration could be eventually increased depending on the
particular descriptor under study. In SubSection~\ref{Subsect41}, our results correspond to Models~1 and~3.1, 
where immediate phosphorylation is assumed. In SubSection~\ref{Subsect42}
we develop  an analogous analysis for Models~2 and~3.2, where delayed phosphorylation is considered.
For these models, our objective is to analyse the 
dynamics of the different receptors and complexes for different VEGF-A ligand concentrations, 
and to study the competition effect that the
presence of VEGR1 has in the dynamics of VEGFR2. Finally, the sensitivity analysis of the descriptors with respect 
to the kinetic rates is carried out in SubSection~\ref{Subsect43}, following the arguments provided
in SubSection~\ref{Subsect24}.

\subsection{Immediate phosphorylation: Model~1 and Model~3.1}
\label{Subsect41}

\begin{figure}[h!]
    \centering
    \includegraphics[width=\textwidth]{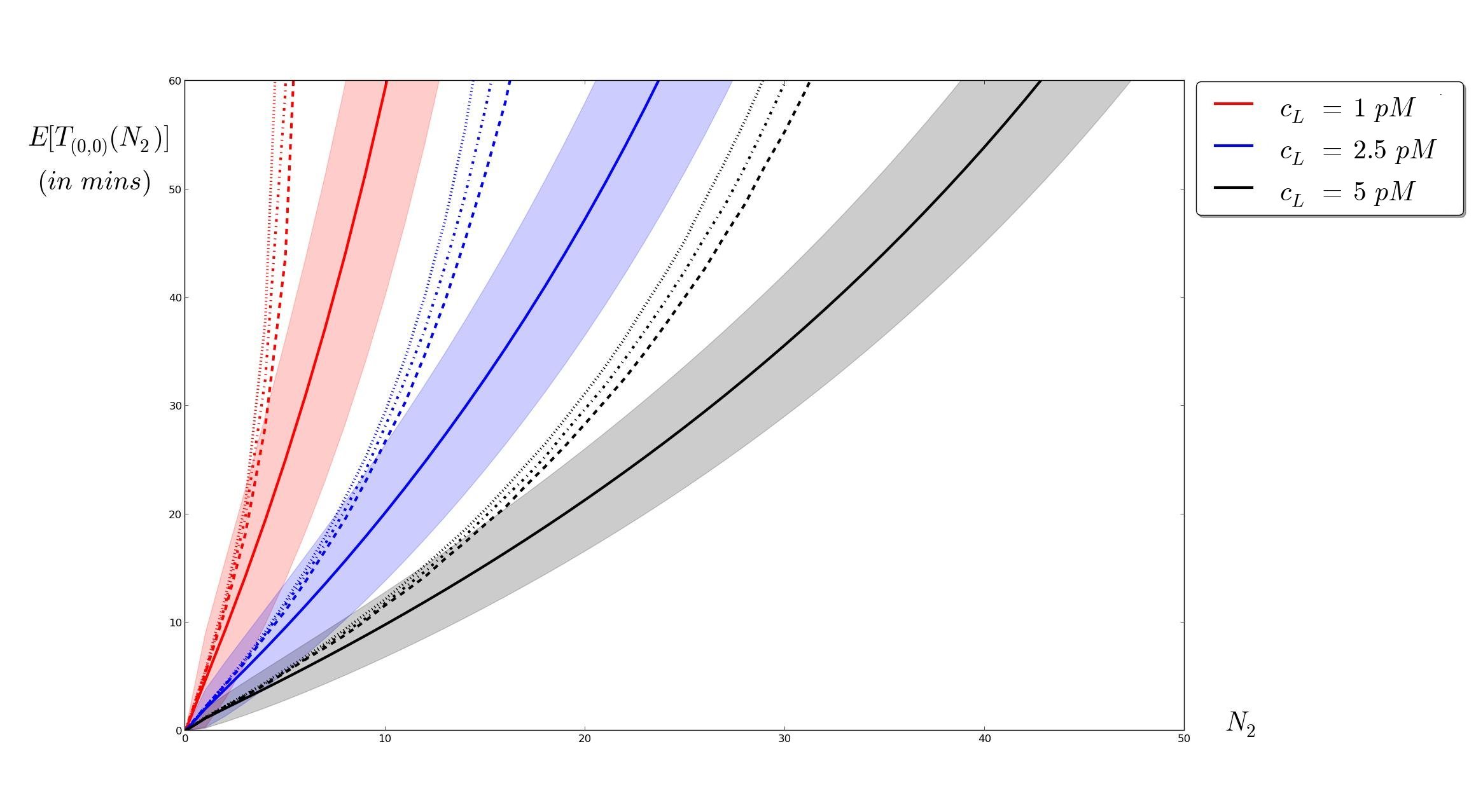}\newline
    \caption{$E[T_{(0,0)}(N_2)]$ for ({from {\it left} to {\it right}})  ligand concentrations $c_L \in \{1 pM, 2.5 pM, 5 pM  \}$. 
    {\it Solid} curves
represent the descriptor in the absence of VEGFR1. {\it Dashed}, {\it dash-dotted} and {\it dotted} curves represent the descriptor with $64$, $72$ and
$80$ numbers of VEGFR1s in the cell area under study, respectively.}
    \label{fig:1}
\end{figure}

In Figure \ref{fig:1}, we plot $E[T_{(0,0)}(N_2)]$ for values $0\leq N_2\leq n_L$, where $n_L\in\{23,58,116\}$ is the number
 of ligands considered, which corresponds to $10\%$, $25\%$ and $50\%$
of the total number of VEGFR2, respectively, and to the following ligand concentrations, $c_L \in \{1 pM, 2.5 pM, 5 pM  \}$.
The number of ligands considered in these three cases
verifies the condition $2n_L\leq n_{R_2}$, assumed in the analysis of $T_{(0,0)}(N_2)$, as discussed
 in SubSection~\ref{Subsect21}. 
 $T_{(0,0)}(N_2)$  is the continuous random variable that represents the time to reach a total number, $N_2$,
  of phosphorylated dimers $P_2$, given the initial
state $(0,0)$, in Model~1 with instantaneous phosphorylation (for details, see SubSection~\ref{Subsect21}).
The figures in this Section have been restricted to times up to $60$ min, to describe 
the early time 
dynamics of
the cell surface. The long-term behaviour of the system can be analysed by means of
the steady state distribution. In Figure~\ref{fig:1}, {\it solid} curves represent values of $E[T_{(0,0)}(N_2)]$ 
in the absence of VEGFR1; that is, these
quantities have been obtained in an exact way, making use of Algorithm~1 in Appendix~C. 
 Shaded areas have been obtained for 
 Model~1 (in the absence
of VEGFR1) by considering $E[T_{(0,0)}(N_2)]\pm SD[T_{(0,0)}(N_2)]$, 
where $SD[X]$ represents the standard deviation of the random variable $X$, which in the case
of $T_{(0,0)}(N_2)$ can be obtained from Algorithm~1 in Appendix~C.
On the other hand, {\it dashed}, {\it dash-dotted} and {\it dotted} curves correspond to values
of $E[T_{(0,0)}(N_2)]$ when $64$, $72$ and $80$ VEGFR1s are
present in the area under study of the cell surface, respectively. 
These numbers correspond to a total of $1600$, $1800$ and $2000$ VEGFR1s on the cell surface.
This choice is based on the fact that 
HUVECs express a total number of 
 $1800\pm 100$ VEGFR1 receptors per cell~\cite{Imoukhuede12}.
 In this case, these
quantities have been obtained making use of Gillespie
simulations of Model~3.1. 

In Figure~\ref{fig:1}, a monotonic behaviour can be easily identified. 
For a fixed value of $N_2$, $E[T_{(0,0)}(N_2)]$ is always smaller for
larger ligand concentrations, $c_L$. Indeed, an increase in the amount of ligand available to bind receptors 
will imply reaching the given
 signal threshold (encoded 
by the value of $N_2$) in a shorter time. 
Our results indicate that the presence of VEGFR1 
changes the time to reach this threshold:
 absence
of VEGFR1 ({\it solid} curves) and different concentrations of VEGFR1 ({\it dashed}, {\it dash-dotted} and {\it dotted} lines). 
In particular, VEGFR1 is known
to sequester ligands from VEGFR2 so that a delay in the formation of $P_2$ complexes is observed. 
However, the effect of VEGFR1 on
the formation of phosphorylated dimers, $P_2$, significantly differs for different ligand concentrations. 
As can be seen in Figure~\ref{fig:1}, the time delay due to the presence of VEGFR1 
 increases as the amount of ligand decreases. 
  For example, when the ligand concentration is given by
$c_L=1 pM $, the mean time $E[T_{(0,0)}(N_2)]$ to reach a threshold $N_2=5$ ($20\%$ of $n_L$) of phosphorylated dimers 
is $24.99$ min in the absence of
VEGFR1. On the other hand, these times approximately amount to $43.63$, $58.97$ and $83.36$ min when the number
 of VEGFR1s is $64$, $72$ and $80$, respectively.
Thus, the time delay caused by the presence of VEGFR1, relative to the time in the absence of
VEGFR1 is $74.59\%$, $135.97\%$ and $233.57\%$, respectively. 
If the ligand concentration is
 $c_L=5 pM$, these percentages are reduced to $40.56\%$, $48.74\%$ and $58.19\%$, respectively. 
 That is, the delay effect of VEGFR1 on the
formation of $P_2$ complexes increases 
as the amount of ligand decreases, as expected, given that 
the competition between VEGFR1 and
VEGFR2 to bind ligand, slowly vanishes as the concentration of ligand increases. This behaviour is still observed 
in  Gillespie simulations
when larger values of $c_L$ than those studied in Figure~\ref{fig:1}, are considered.

The presence of VEGFR1 does not only affect the time scales for obtaining a given signal threshold, 
but it also affects the {\it maximum}
threshold $N_2$ that is reached, which is 
 the asymptotic behaviour  observed 
in Figure~\ref{fig:1}. We note that, although in principle 
any threshold $N_2$ is achieved with probability $1$ in our models in the 
long-term~\footnote{
We are dealing with
irreducible CTMCs defined over a finite number of states, 
so that the stochastic process allows to
 visit  any state in the CTMC, and thus to reach any signal threshold.}, 
 Gillespie simulations show that this asymptotic behaviour,
  which approaches a value, $m_2$, corresponds to the steady state
value of the number of $P_2$ complexes. Thresholds above the steady state value do not seem to be reached
 in the time scale analysed in our simulations,
so that the steady state value should become a representative value of the signal
threshold that can be obtained in a biologically reasonable time scale. 
This steady state distribution is also significantly affected by the
presence of VEGFR1. This distribution can be obtained in an exact way in the 
absence of VEGFR1, by means of Algorithm~2 in Appendix~C. 
In this case, with no VEGFR1, the mean number $m_2$ of $P2$ complexes in steady state
is $22.10$, $53.32$, and $85.16$ for ligand concentrations corresponding to values $c_L=1 pM$, 
$2.5 pM$ and $5 pM$, respectively.

\begin{figure}[h!]
    \centering
    \includegraphics[width=0.75\textwidth]{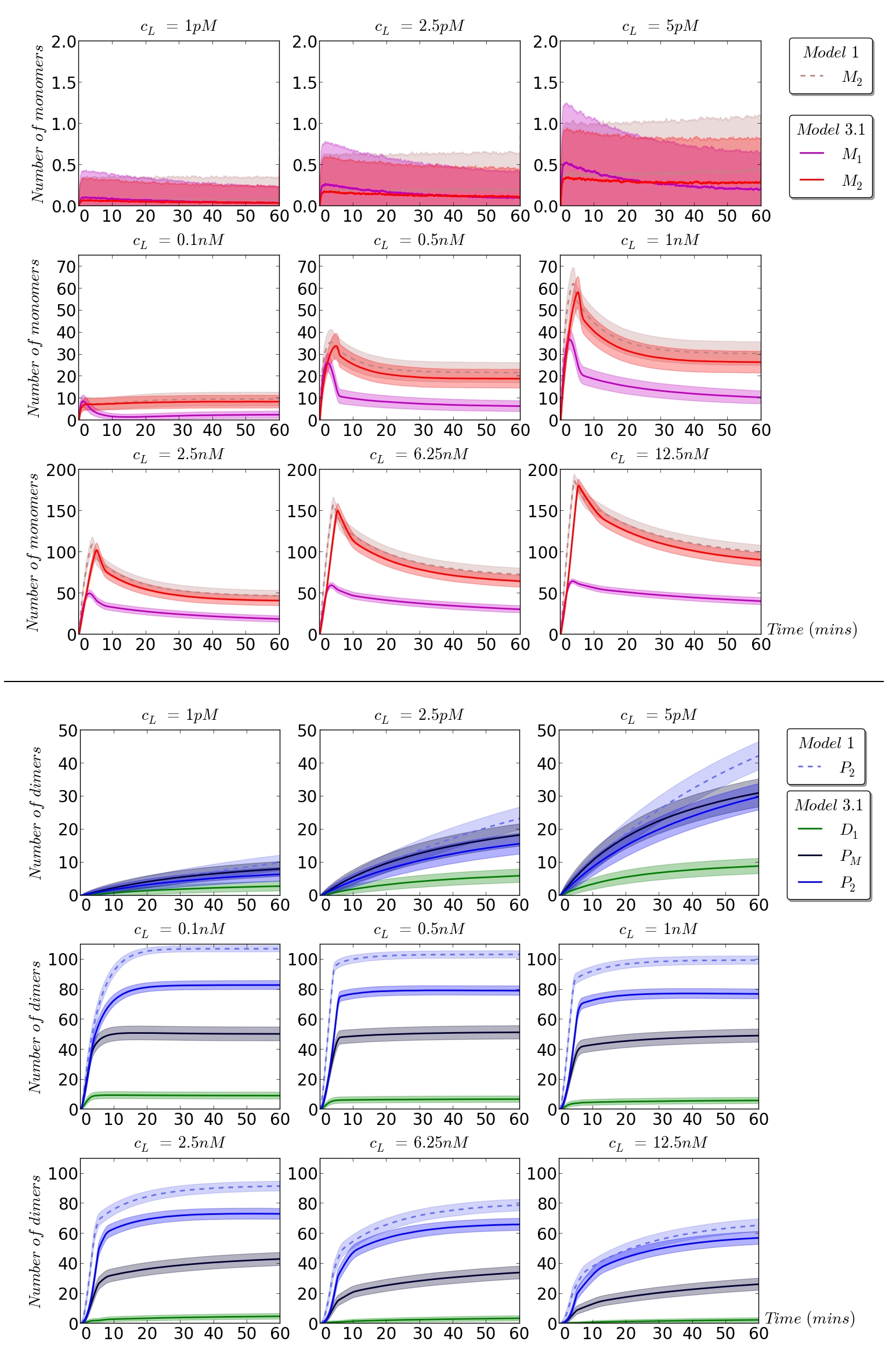}\newline
    \caption{Gillespie simulations of the process for different initial ligand concentrations $ c_L \in \{1pM, 2.5pM,$ $5pM, 0.1nM, 0.5nM, 1nM,2.5nM,6.25nM,12.5nM\}$.
{\it Dashed} lines correspond to Model~1
and  {\it solid} lines correspond to Model~3.1. Time course for monomers ({\it top}) and
dimers ({\it bottom}).}
    \label{fig v-k instantaneous phospho}
\end{figure}

In order to analyse the dynamics of the stochastic process in the presence of VEGFR1, in
Figure~\ref{fig v-k instantaneous phospho} we plot
 the means and standard deviations of the
random variables in Model~1 and Model~3.1 as a function of time. 
The time course has been generated by
means of Gillespie simulations, where we have broaden the VEGF-A
concentration range by considering $n_L \in\{0.1 n_{R_2}, 0.25 n_{R_2}, 0.5 n_{R_2},10 n_{R_2}, 50 n_{R_2}, 100n_{R_2},250n_{R_2},625n_{R_2},1250n_{R_2}\}$, which approximately
corresponds to concentrations $ c_L \in \{1pM, 2.5pM,$ $5pM, 0.1nM, 0.5nM, 1nM,2.5nM,6.25nM,12.5nM\}$. For low ligand concentrations the number of dimers grows as 
the VEGF-A concentration is increased.
For these concentrations the steady state has not been reached in the first
 $60$ min of the numerical simulation.
However, higher concentrations result in saturated situations,
 where the difference between the number of $P_2$ complexes with and without VEGFR1 becomes stable. In fact,
this saturation  results in lower numbers of $P_2$ complexes for 
 ligand concentrations higher than $c_L\sim2.5nM$. Thus, concentrations
around $0.1nM-2.5nM$ may be considered as optimum ones. 
As mentioned above, for ligand concentrations of order $c_L=\{6.25nM,12.5nM\}$, the
system exhibits a reduction in the number of
dimers, which is caused by the formation of monomeric bound complexes (see 
Figure~\ref{fig v-k instantaneous phospho}).
In fact, by analysing the formation of monomers 
as a function of  time, we observe, under optimum ligand
concentrations,  a peak of monomeric complexes in the first $5$ minutes, which is followed by a decrease to
the steady state values. 
For high ligand concentrations, the steady state value for 
monomeric complexes
increases,
 so that formation of dimers is effectively blocked.
 The inhibition of dimer formation at high ligand concentrations is intrinsically related
 to the LID assumption, where the formation of free receptor pre-dimers is not allowed. If free receptor dimers
 were to be considered, their effect would be negligible for ligand concentrations 
 below $1nM$~\cite{Mac07}.
 
\subsection{Delayed phosphorylation: Model~2 and Model~3.2}
\label{Subsect42}

\begin{figure}[h!]
    \centering
    \includegraphics[width=\textwidth]{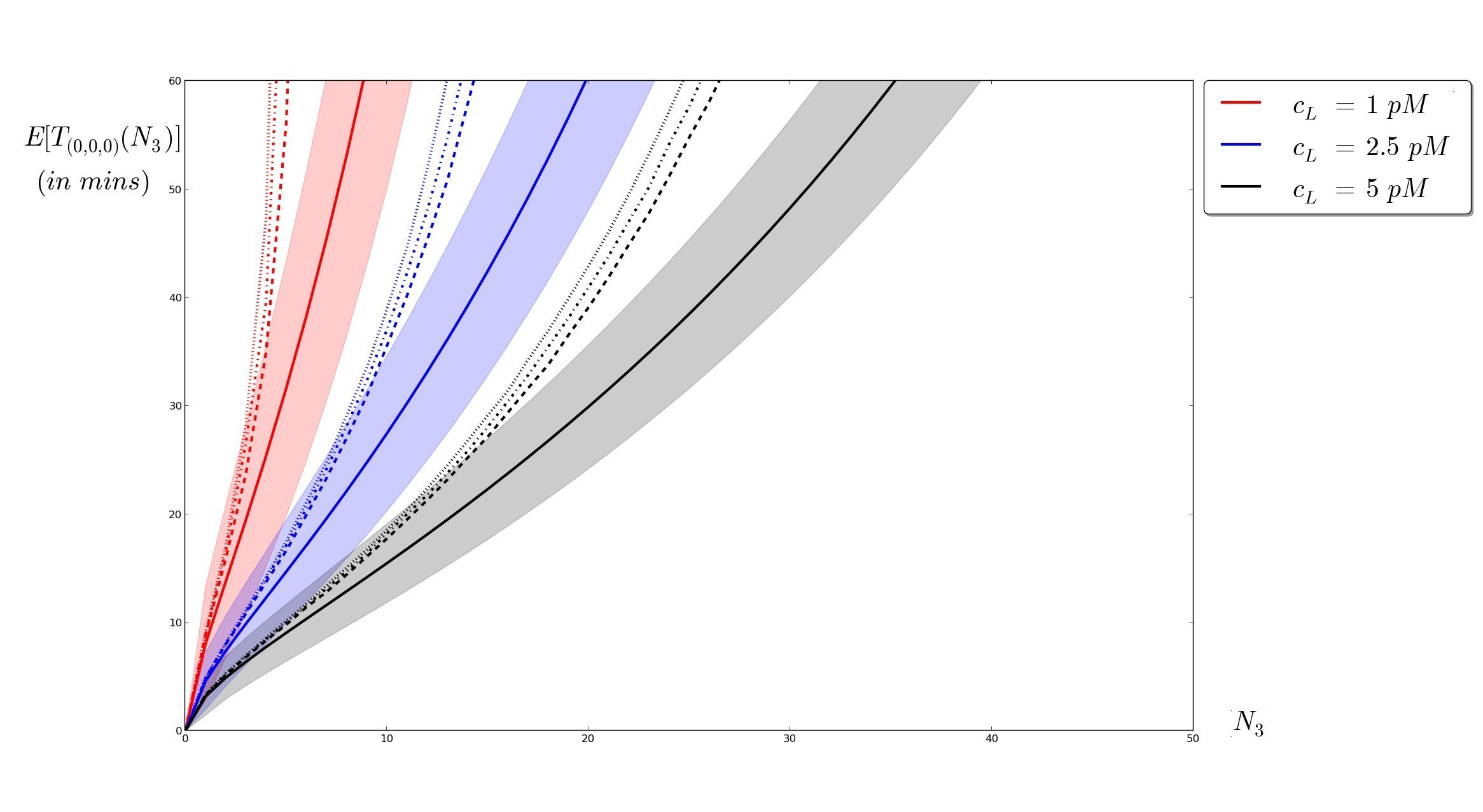}\newline
    \caption{$E[T_{(0,0,0)}(N_3)]$ for (from {\it left} to {\it right}) 
     ligand concentrations $c_L\in\{1 pM,2.5 pM,5 pM\}$. {\it Solid} curves represent the descriptor
in the absence of VEGFR1. {\it Dashed}, {\it dash-dotted} and {\it dotted} curves represent the descriptor with $64$, $72$ and $80$ numbers of VEGFR1s
 in the cell area under study, respectively.}
    \label{fig:3}
\end{figure}

In Figure~\ref{fig:3}, an analogous analysis 
to that of Figure~\ref{fig:1} is carried out by means of adapted versions of Algorithm~1 and Algorithm~2 in
Appendix~C. In this case, we require 
 the random variable $T_{(0,0,0)}(N_3)$
which describes the time to reach a total number,
 $N_3$, of phosphorylated $P_2$ complexes when delayed phosphorylation is assumed
 (introduced in Model~2 and Model~3.2).
The behaviour 
in Figure~\ref{fig:3} is similar to that observed in 
Figure~\ref{fig:1}, so that the consideration of delayed phosphorylation in the model does not seem to 
qualitatively affect the main features of the descriptor
under consideration. Therefore, 
the comments made in Figure~\ref{fig:1} about the descriptor $E[T_{(0,0)}(N_2)]$ for different
concentrations of ligand and VEGFR1 are also valid for Figure~\ref{fig:3},
but in this case for the descriptor $E[T_{(0,0,0)}(N_3)]$. However, the consideration of
phosphorylation as an independent reaction in the process clearly amounts to a delay in the time 
to reach the threshold $N_3$, when comparing it with
the time to reach the threshold $N_2$ in Figure~\ref{fig:1}:  every curve is displaced to the left. 
Finally, we have also computed 
the steady state
distribution: in the absence of VEGFR1 the mean number ${\hat m}_3$ of $P_2$ complexes in steady state 
is $18.25$, $45.51$ and $77.89$,
for ligand concentrations  $c_L=1pM$, $2.5 pM$ and $5 pM$, respectively.

\begin{figure}[h!]
    \centering
    \includegraphics[width=0.75\textwidth]{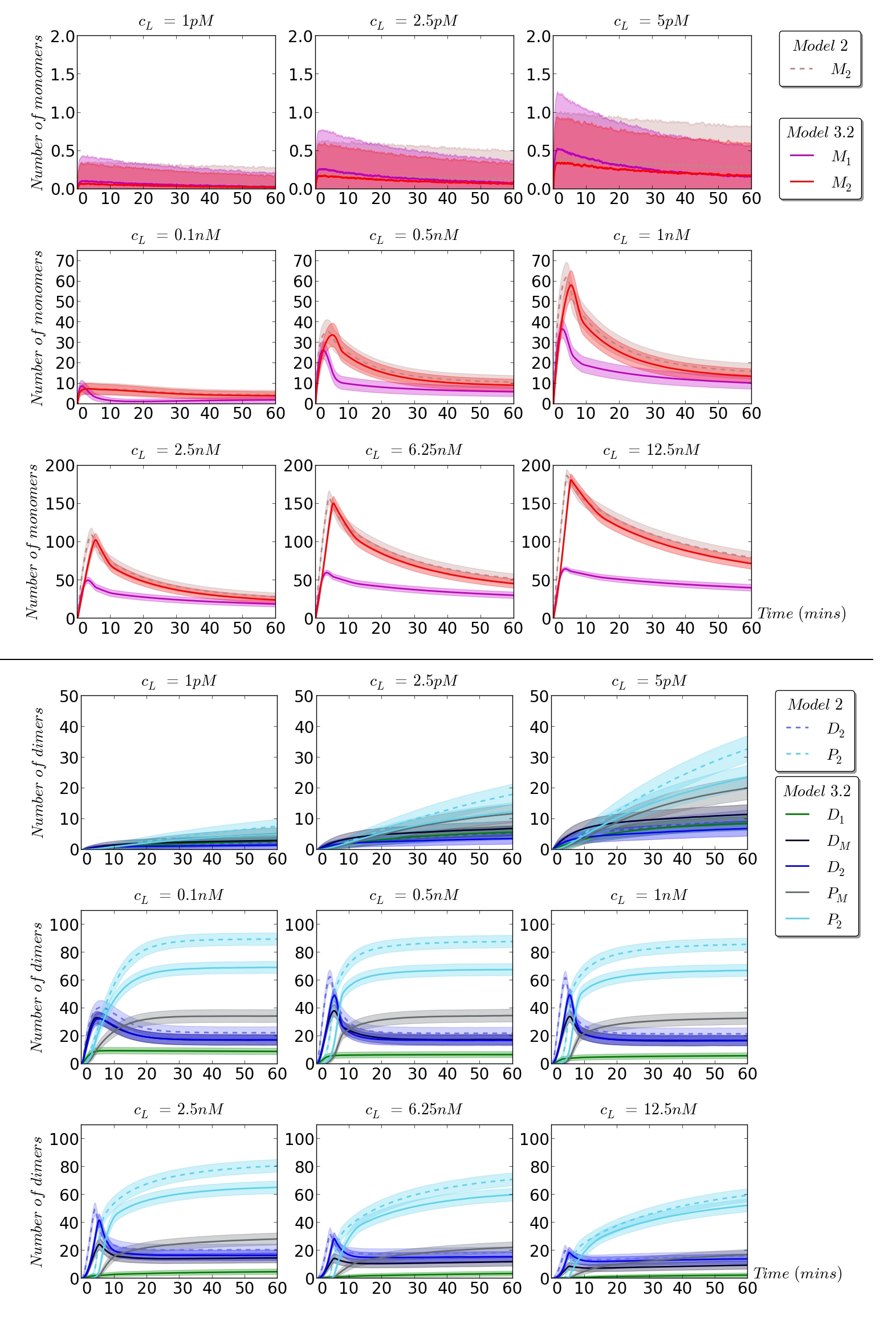}\newline
    \caption{Gillespie simulations of the process for different initial ligand concentrations $ c_L \in \{1pM, 2.5pM,$ $5pM, 0.1nM, 0.5nM, 1nM,2.5nM,6.25nM,12.5nM\}$.
{\it Dashed} lines correspond to Model~2 and {\it solid} lines correspond to Model~3.2. Time
course for monomers ({\it top}) and
dimers ({\it bottom}).}
    \label{simulationfigure}
\end{figure}

In Figure~\ref{simulationfigure}, we carry out Gillespie simulations to obtain the 
dynamics of the system in Model~3.2 for different ligand concentrations,
as before,
$ c_L \in \{1pM, 2.5pM,$ $5pM, 0.1nM, 0.5nM, 1nM,2.5nM,6.25nM,12.5nM\}$. For high ligand concentrations, phosphorylation events occur
within 10-20 minutes of ligand stimulation~\cite{Alarcon07,Ewan06,Tan13}. 
The number of non-active dimers ($D_1, D_M, D_2$) is, in general, lower than the number of
active dimers $P_M$ and $P_2$, in steady state. When enough ligand stimulation is given ($ c_L \in \{0.1nM, 0.5nM, 1nM,2.5nM,6.25nM,12.5nM\}$) the curves corresponding to dimers
$D_2$ and $D_M$ show a peak at early times, which is eventually lost
once these complexes become phosphorylated, as can be seen 
 in  the sudden increase for 
$P_2$ and $P_M$ complexes. Similar comments can be made regarding  monomer formation 
 (see Figure~\ref{simulationfigure}):  a peak 
 is seen during the first $5$ minutes, slightly before the dimeric peak. 
 This clearly indicates a {\it two-step} (monomer
and non-phosphorylated dimer) formation process, which is required for the subsequent creation of 
phosphorylated complexes on the cell surface. The optimum ligand concentration, $c_L$,
for phosphorylated dimers in steady state
 is approximately given by the
range $0.1nM-2.5nM$.  
As depticted in Figure~\ref{fig v-k instantaneous phospho}, for higher ligand concentrations
monomeric complexes are more likely to be formed than either non-phosphorylated
or phosphorylated dimeric complexes.
In this case, the peak for non-phosphorylated
dimers $D_2$ and $D_M$ is reduced, which is explained by the larger numbers
of  monomeric bound complexed formed  (see Figure~\ref{simulationfigure}).

When focusing on the number of dimers at $t = 60$ min, 
we observe an approximately $20 \%$ decrease for
the number of $P_2$ dimers in Model~2 with respect to Model~1, for small ligand concentrations $ c_L \in \{1pM, 2.5pM,5pM\}$. 
As $c_L$ grows,
the difference between the number of  $P_2$ dimers drops down to $16\%$. However, additional numerical results, not presented here, show that
the ratio between phosphorylated $P_2$ and non-phosphorylated $D_2$ dimers in Model~2 
does not change with ligand concentration, and it is approximately $4:1$.
Finally,
an additional aspect we are interested in, is the influence of competition on the number of dimers.
Our results indicate that 
there is a $35\%$ decrease in the number of $P_2$ complexes in  Model~3.1 with instantaneous phosphorylation for small ligand concentration,
$c_L = 1 pM$ with respect to Model~1. As the ligand concentration 
increases, this difference drops down to $22\%$, so that the competition between VEGFR1
and VEGFR2 is, again, reduced  when enough ligand stimulation is given. 
In the model with delayed phosphorylation, there is a $15\%$ decrease in
the number of  dimers $P_2$ in  Model~3.2 compared to Model~2, for small ligand concentration, $c_L = 1pM$.
As the concentration of ligand is increased, this difference also drops down to $1-2\%$.

\subsection{Sensitivity analysis}
\label{Subsect43}

Finally, the effect of the binding, dissociation and phosphorylation rates on the descriptors considered
 in this paper (for
phosphorylated $P_2$ complexes),
can be
estimated by means of the sensitivity analysis proposed in Section~\ref{Subsect24}. 
In Table~\ref{Tab2} we present the derivatives of the descriptors $E[T_{(0,0)}(N_2)]$,
$E[T_{(0,0,0)}(N_3)]$, $m_2$ and ${\hat m}_3$, when $N_2$ and $N_3$ are 
chosen to be $25\%$ of the total number of ligands $n_L$, and for different
concentrations of ligands $c_L$. 
As expected, the effect of each rate on any descriptor increases with increasing values of ligand concentration $c_L$.
However, we note that the rate $\alpha_+$ 
is the most influential one (it seems to have twice the
effect of any other rate), 
for all the descriptors. Thus, we conclude that 
 the formation of monomers seems to play here a more crucial
role for the formation of phosphorylated dimers than the phosphorylation or dimerisation rates themselves.

\begin{table}
\centering
{\scriptsize
\begin{tabular}{| c || c | c | c | c | c | c | c |}
\hline\noalign{\smallskip}
Partial Derivative & $c_L$ & $\alpha_+$ & $\alpha_-$ & $\beta_+$ & $\beta_-$ & $\gamma_+$ & $\gamma_-$ \\
\noalign{\smallskip}\hline\noalign{\smallskip}
$\frac{\partial E[T_{(0,0)}(N_2)]}{\partial\theta}$ & $1pM$ & $-8.40\times10^{7}$ & $6.85\times10^{2}$ & $-5.79\times10^{3}$ & $1.39\times10^{3}$ & $-$ & $-$
\\
 & $2.5pM$ & $-8.85\times10^{7}$ & $7.77\times10^{2}$ & $-6.50\times10^{3}$ & $2.01\times10^{3}$ & $-$ & $-$
\\
 & $5pM$ & $-9.26\times10^{7}$ & $9.00\times10^{2}$ & $-7.54\times10^{3}$ & $2.84\times10^{3}$ & $-$ & $-$
\\
\noalign{\smallskip}\hline\noalign{\smallskip}
$\frac{\partial E[T_{(0,0,0)}(N_3)]}{\partial\theta}$ & $1pM$ & $-8.85\times10^{7}$ & $6.79\times10^{2}$ & $-5.84\times10^{3}$ & $1.03\times10^{3}$ & $-2.34\times10^{3}$ & $3.66\times10^{3}$
\\
 & $2.5pM$ & $-9.96\times10^{7}$ & $8.13\times10^{2}$ & $-6.81\times10^{3}$ & $1.39\times10^{3}$ & $-3.08\times10^{3}$ & $6.27\times10^{3}$
\\
 & $5pM$ & $-1.10\times10^{8}$ & $9.99\times10^{2}$ & $-8.22\times10^{3}$ & $1.93\times10^{3}$ & $-3.75\times10^{3}$ & $8.85\times10^{3}$
\\
\noalign{\smallskip}\hline\noalign{\smallskip}
$\frac{\partial m_2}{\partial\theta}$ & $1pM$ & $2.09\times10^{6}$ & $-5.78\times10^{2}$ & $3.91\times10^{3}$ & $-1.08\times10^{4}$ & $-$ & $-$
\\
 & $2.5pM$ & $9.74\times10^{6}$ & $-2.70\times10^{3}$ & $1.77\times10^{4}$ & $-4.90\times10^{4}$ & $-$ & $-$
\\
 & $5pM$ & $2.41\times10^{7}$ & $-6.66\times10^{3}$ & $4.35\times10^{4}$ & $-1.20\times10^{5}$ & $-$ & $-$
\\
\noalign{\smallskip}\hline\noalign{\smallskip}
$\frac{\partial{\hat m}_3}{\partial\theta}$ & $1pM$ & $3.65\times10^{5}$ & $-1.01\times10^{2}$ & $6.83\times10^{2}$ & $-1.89\times10^{3}$ & $1.03\times10^{3}$ & $-4.11\times10^{3}$
\\
 & $2.5pM$ & $2.16\times10^{6}$ & $-5.97\times10^{2}$ & $3.90\times10^{3}$ & $-1.08\times10^{4}$ & $2.67\times10^{3}$ & $-1.07\times10^{4}$
\\
 & $5pM$ & $1.25\times10^{7}$ & $-3.47\times10^{3}$ & $2.21\times10^{4}$ & $-6.11\times10^{4}$ & $5.29\times10^{3}$ & $-2.12\times10^{4}$
\\
\noalign{\smallskip}\hline
\end{tabular}
\caption{Partial derivatives of the 
stochastic descriptors $E[T_{(0,0)}(N_2)]$ and $E[T_{(0,0,0)}(N_3)]$ (in $\frac{min}{s^{-1}}$)
 and descriptors $m_2$ and ${\hat m}_3$ (in $\frac{molecules}{s^{-1}}$),
with respect to each parameter $\theta_i\in\{\alpha_+,\alpha_-,\beta_+,\beta_-,\gamma_+,\gamma_-\}$
for different ligand concentrations $c_L\in\{1pM,2.5pM,5pM\}$.}
\label{Tab2}
}
\end{table}


\section{Discussion}
\label{Sect5}

In this paper, we have introduced
 different stochastic models to analyse the binding and phosphorylation dynamics of VEGF-A/VEGFR2 
 and
 VEGF-A/VEGFR1
 in
vascular endothelial cells. Model~1 in Section~\ref{Subsect21} and Model~2 in Section~\ref{Subsect22} 
consider the following 
processes:  ligand VEGF-A binds
receptor VEGFR2 in order to form monomers and dimers, which can eventually dissociate. Dimers become instantaneously phosphorylated in Model~1, while in
Model~2 phosphorylation is considered a
new and independent reaction. In these two models, matrix-analytic techniques have been applied to
study the time to reach a threshold of phosphorylated
dimers $P_2$ on the cell membrane, and the steady state distribution of the corresponding CTMCs. 
Moreover, the construction of Model~2, as an extension
of Model~1 in Section~\ref{Sect2} allows us, not only to analyse the role played by  phosphorylation 
events (in Section~\ref{Sect4}), but also to
show how different reactions may be incorporated while adapting the matrix-analytic approach.

We note that, although our arguments in Section~\ref{Subsect21}
and Section~\ref{Subsect22} might be adapted 
to include any number of reactions,
 the  computational efficiency of the algorithms would decrease with increasing number of reactions. 
 Thus, a balance between the complexity of the model and computational
considerations is required, and numerical approaches, such as moment-closure techniques or Gillespie simulations, 
may always prevail for
more complex models. In particular, Model~3.1 and  Model~3.2 in Section~\ref{Subsect23}, 
which incorporate the competition between VEGFR1 and VEGFR2 for ligand availability,
do not seem to be computationally effective when adapting procedures from Section~\ref{Subsect21}
and Section~\ref{Subsect22}. In light of this, Gillespie simulations have
been carried out in Section~\ref{Sect4} when dealing with these models.

\begin{figure}[h!]
\centering
\includegraphics[width=0.75\textwidth]{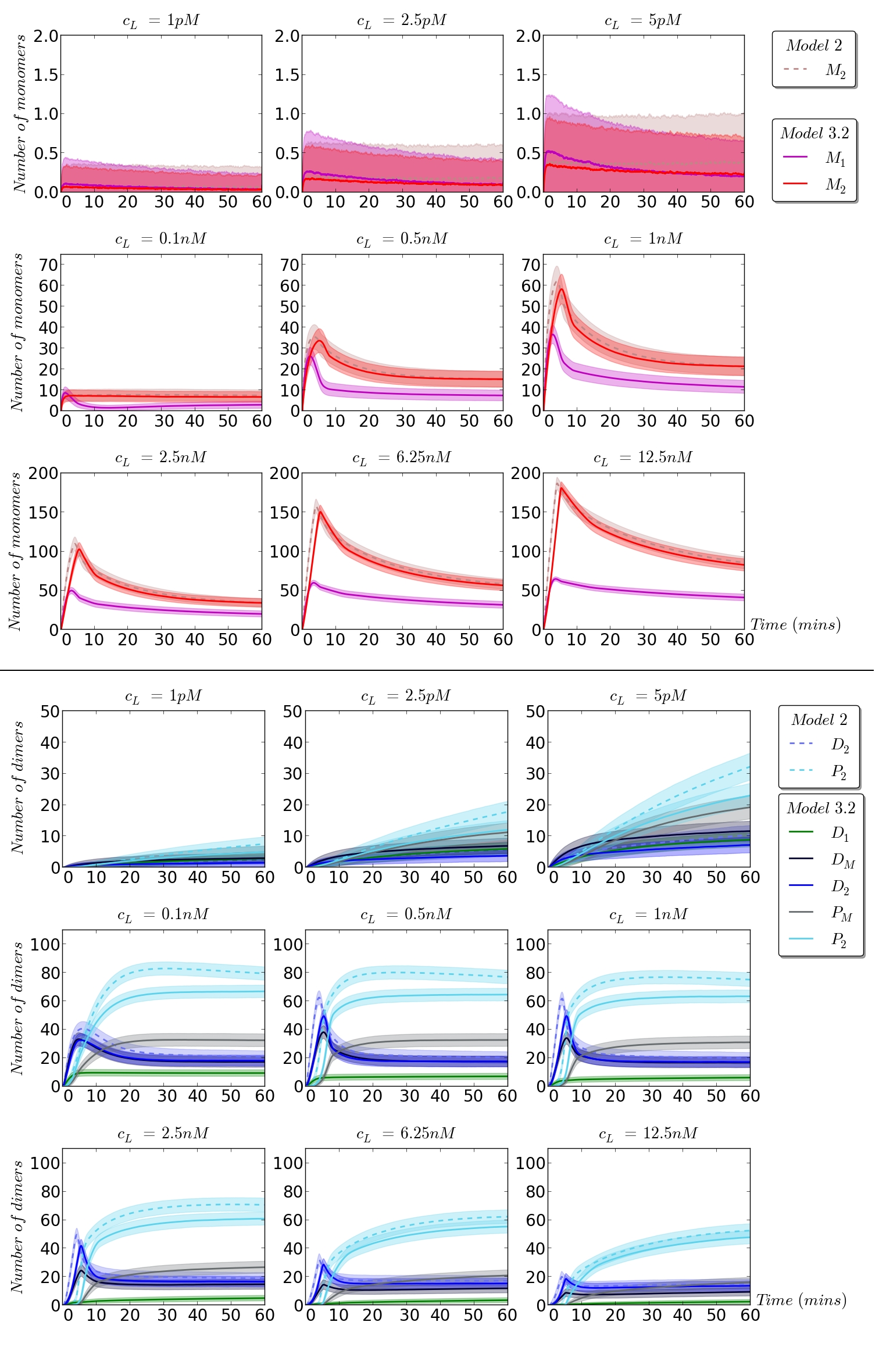}
\caption{
Generalisation of Model~2 and Model~3.2 to include ligand
dissociation of phosphorylated dimers. This figure is analogous to 
 Figure~\ref{simulationfigure}.}
\label{FigLast}
\end{figure}

A particular assumption in our models 
of Section~\ref{Sect2} is that dissociation of phosphorylated dimers requires
de-phosphorylation as a first step. 
For comparison, we introduce a more general model that allows ligand-dissociation of phosphorylated dimers. 
Three new reactions have been added to Model~2 and  Model~3.2 in order to consider the
dissociation of phosphorylated dimers:
 $P_2 \stackrel{\beta_{22-}}{\longrightarrow} R_2 + M_2$, $P_M \stackrel{\beta_{21-}}{\longrightarrow} R_2 + M_1$
and $P_M \stackrel{\beta_{12-}}{\longrightarrow}R_1 + M_2$. 
We have assumed that the previous ligand-dissociation reactions also imply de-phosphorylation of
the receptors, and thus, no phosphorylated 
monomeric receptors are allowed, which in turn, could bind ligand in a subsequent reaction.
This can be justified as de-phosphorylation is a fast process 
 (see Table~\ref{tab: biochem parameters}).
The results are shown in Figure~\ref{FigLast}, where the most significant difference with respect 
to Figure~\ref{simulationfigure}
is the slow decay that can be observed for
 the number of $P_2$ complexes after $20$ minutes under optimum ligand concentrations. This small effect,
which is more prominent in Model~2 than in Model~3.2, can be explained by the new de-phosphorylation pathway of
$P_2$ complexes (by ligand dissociation).
This clearly indicates a {\it two-step} (monomer
and non-phosphorylated dimer) formation process, which is required for the subsequent creation of 
phosphorylated dimers on the cell surface ($P_2$ and $P_M$ complexes).
 However, no other significant differences can be identified with the introduction of these three new reactions (see 
Figure~\ref{simulationfigure} and Figure~\ref{FigLast}). This suggests that the
 the additional 
reactions do not  qualitatively change
the dynamics or the steady state of the system, with or without VEGFR1.

Alternative approaches, such as moment-closure techniques, when analysing the master equation in 
Eq.~\eqref{eq: master R1-R2-L}, are available for our models.
 In particular, we have also analysed Model~3.1 and Model~3.2 
 making use of the
  van Kampen approximation, which is briefly
described in Appendix~D. 
The van Kampen approximation, as well as any other moment-closure method, 
aims to obtain the time evolution of the different order moments for the random variables of the CTMCs
under consideration. In particular, and as shown in Eq.~\eqref{eq: master R1-R2-L}, it is possible to obtain a system of
differential equations for the different order moments of the random variables considered. However, this system of
differential equations relates any order moment of a variable $X(t)$ to its immediately posterior order moment, which yields an infinite system of
differential equations. The main assumption of the van Kampen approximation 
is that  the distribution of the stochastic fluctuations around
a steady state follows a multi-variate normal distribution, which
allows to close the order moment hierarchy.
 Furthermore, one can also evaluate the
accuracy of this approximation by means of the covariance matrix of these random variables and the locus (contour) of constant probability~\cite{Papoulis02}.
This contour is the set of values of the random variables  for which their joint probability density function
 is greater or equal to some constant
$K$. Then, when we consider, for example, two given random variables of the process, 
the probability, ${\hat p}$, of a realisation of this bi-dimensional random vector to be in the contour of constant probability
(of constant value $K$) can be obtained from the $\chi$-square distribution by ${\hat p}=1-\exp(-K/2)$. 

\begin{figure}[h!]
\centering
\includegraphics[width = 15cm]{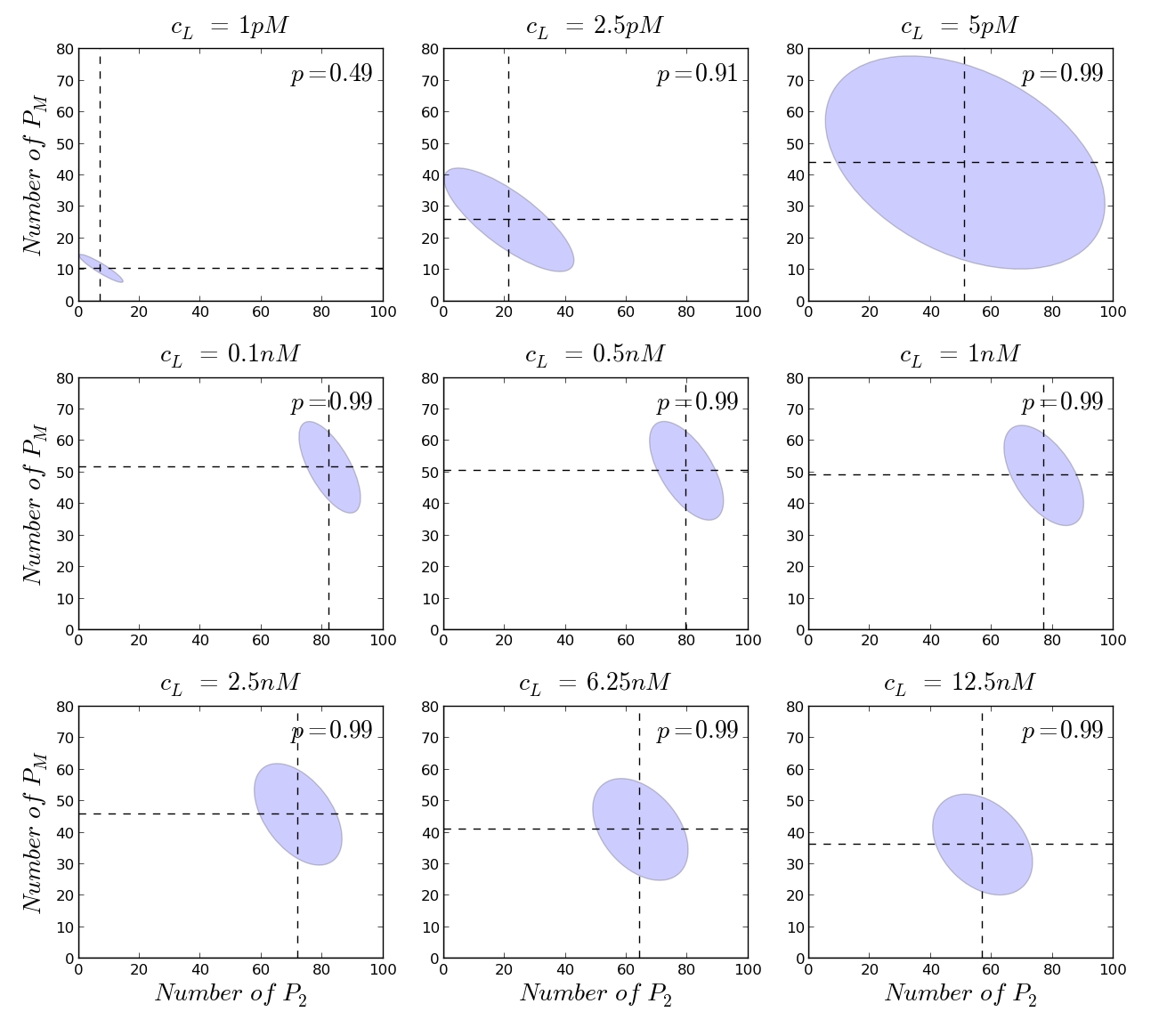}
\caption{Ellipse contours in Model~3.1, for initial ligand concentrations $ c_L \in \{1pM, 2.5pM,$ $5pM, 0.1nM, 0.5nM, 1nM,2.5nM,6.25nM,12.5nM\}$.
These ellipse contours correspond to $p=min\{0.99,{\hat p}\}$, where ${\hat p}$ is the maximum probability 
that yields a contour in the the first quadrant.}
\label{fig: ell1}
\end{figure}

As the random variables  in Models~3.1 and Model~3.2 represent numbers
of complexes for different molecular
  {\it species}, we should only focus on non-negative values for these random variables. 
  Therefore, given any two molecular species, we need to restrict the
confidence ellipses, corresponding to large enough values of ${\hat p}$, to be on the first quadrant. 
In Figure~\ref{fig: ell1} and for Model~3.1, we plot the
contours corresponding to the numbers $P_2$ of VEGFR2 homodimers and $P_M$ of heterodimers, in steady state, for ligand concentrations $ c_L \in \{1pM, 2.5pM,$ $5pM, 0.1nM, 0.5nM, 1nM,2.5nM,6.25nM,12.5nM\}$.
These contours correspond to a probability $p=min\{0.99,{\hat p}\}$, where ${\hat p}$ is the maximum probability
that yields a contour within the first
quadrant, so that $0.99$ is identified here with a proper performance of the
approximation method. When considering low concentrations, the probability $p$ increases until it reaches
its maximum value $p=0.99$ for $c_L=5pM$, which corresponds to a number of ligands 
which is of the same order than the total number of receptors $n_{R_2}$.
The performance of the van Kampen approximation clearly improves with increasing values of $c_L$, 
which is reflected in the smaller contours obtained amounting
to a total probability mass of $0.99$. These results are consistent with the fact that, under larger ligand concentrations, 
the behaviour of the stochastic
process can be properly approximated by its deterministic counterpart. On the other hand, under low ligand concentrations, the van Kampen approach can not
properly reflect the stochasticity in the system ($p=0.49$ for $c_L=1pM$ in Figure~\ref{fig: ell1}).

The numerical results presented  in Section~\ref{Sect4} 
have allowed us to quantify the effect of different ligand concentrations and different concentrations
of the competitor VEGFR1 on the dynamics of the process. Increasing ligand concentration
compensates the 
  competition effect caused by the
presence of VEGFR1, but at the same time too high ligand concentrations can result in 
saturated situations, where the phosphorylation of dimers is reduced and monomeric bound complexes are enhanced.
We also
note that the total number of VEGFR2s per cell varies according to Refs.~\cite{napione2012unraveling,Ewan06,ewan2006intrinsic}
and could be tenfold higher than used in our computations~\cite{Imoukhuede12}.
A larger number of VEGFR2s 
on the cell surface would, however, only quantitatively change our results: a reduced
VEGFR1 competition effect and a higher optimum ligand concentration. Finally,
the sensitivity analysis carried out for the descriptors enables us to show how the monomeric formation rate, 
$\alpha_+$,
plays a crucial role in these models, with an effect which is twice the effect of any other rate 
for any of the descriptors we have considered.


\bibliographystyle{unsrt}

\bibliography{vegfr-descriptor-22-junio-2016-submitted-arxiv}

\begin{thebibliography}{10}

\bibitem{Alarcon07}
T~Alarc{\'o}n and KM~Page.
\newblock Mathematical models of the {VEGF} receptor and its role in cancer
  therapy.
\newblock {\em Journal of The Royal Society Interface}, 4(13):283--304, 2007.

\bibitem{Olsson06}
AK~Olsson, A~Dimberg, J~Kreuger, and L~Claesson-Welsh.
\newblock {VEGF} receptor signalling? in control of vascular function.
\newblock {\em Nature Reviews Molecular Cell Biology}, 7(5):359--371, 2006.

\bibitem{Cross03}
MJ~Cross, J~Dixelius, T~Matsumoto, and L~Claesson-Welsh.
\newblock {VEGF}-receptor signal transduction.
\newblock {\em Trends in Biochemical Sciences}, 28(9):488--494, 2003.

\bibitem{Lauffenburger96}
DA~Lauffenburger and JJ~Linderman.
\newblock {\em Receptors: models for binding, trafficking, and signaling},
  volume 365.
\newblock Oxford University Press, New York, 1993.

\bibitem{Teis03}
D~Teis and LA~Huber.
\newblock The odd couple: signal transduction and endocytosis.
\newblock {\em Cellular and Molecular Life Sciences CMLS}, 60(10):2020--2033,
  2003.

\bibitem{Mac07}
F~Mac~Gabhann and AS~Popel.
\newblock Dimerization of {VEGF} receptors and implications for signal
  transduction: a computational study.
\newblock {\em Biophysical chemistry}, 128(2):125--139, 2007.

\bibitem{Starbuck90}
C~Starbuck, HS~Wiley, and DA~Lauffenburger.
\newblock Epidermal growth factor binding and trafficking dynamics in
  fibroblasts: relationship to cell proliferation.
\newblock {\em Chemical Engineering Science}, 45(8):2367--2373, 1990.

\bibitem{Tan13}
WH~Tan, AS~Popel, and F~Mac~Gabhann.
\newblock Computational model of {VEGFR2} pathway to erk activation and
  modulation through receptor trafficking.
\newblock {\em Cellular signalling}, 25(12):2496--2510, 2013.

\bibitem{Alarcon06}
T~Alarc{\'o}n and KM~Page.
\newblock Stochastic models of receptor oligomerization by bivalent ligand.
\newblock {\em Journal of The Royal Society Interface}, 3(9):545--559, 2006.

\bibitem{Mac05a}
F~Mac~Gabhann, MT~Yang, and AS~Popel.
\newblock Monte carlo simulations of {VEGF} binding to cell surface receptors
  in vitro.
\newblock {\em Biochimica et Biophysica Acta (BBA)-Molecular Cell Research},
  1746(2):95--107, 2005.

\bibitem{Neuts94}
MF~Neuts.
\newblock {\em Matrix-geometric solutions in stochastic models: an algorithmic
  approach, second edition}.
\newblock Courier Dover Publications, 1994.

\bibitem{Latouche99}
G~Latouche and V~Ramaswami.
\newblock {\em Introduction to matrix analytic methods in stochastic
  modelling}.
\newblock ASA-SIAM, Philadelphia, 1999.

\bibitem{GomezCorral12a}
A~G{\'o}mez-Corral and M~L{\'o}pez~Garc{\'i}a.
\newblock Extinction times and size of the surviving species in a two-species
  competition process.
\newblock {\em Journal of Mathematical Biology}, 64(1-2):255--289, 2012.

\bibitem{GomezCorral12b}
A~G{\'o}mez-Corral and M~L{\'o}pez~Garc{\'i}a.
\newblock On the number of births and deaths during an extinction cycle, and
  the survival of a certain individual in a competition process.
\newblock {\em Computers \& Mathematics with Applications}, 64(3):236--259,
  2012.

\bibitem{Van92}
NG~Van~Kampen.
\newblock {\em Stochastic processes in physics and chemistry}, volume~1.
\newblock Elsevier, 1992.

\bibitem{Grunewald10}
FS~Gr{\"u}newald, AE~Prota, A~Giese, and K~Ballmer-Hofer.
\newblock Structure--function analysis of {VEGF} receptor activation and the
  role of coreceptors in angiogenic signaling.
\newblock {\em Biochimica et Biophysica Acta (BBA)-Proteins and Proteomics},
  1804(3):567--580, 2010.

\bibitem{ruch2007structure}
C~Ruch, G~Skiniotis, MO~Steinmetz, T~Walz, and K~Ballmer-Hofer.
\newblock Structure of a {VEGF--VEGF} receptor complex determined by electron
  microscopy.
\newblock {\em Nature Structural \& Molecular Biology}, 14(3):249--250, 2007.

\bibitem{Kulkarni95}
VG~Kulkarni.
\newblock {\em Modeling and analysis of stochastic systems}.
\newblock Chapman \& Hall Texts in Statistical Science Series, London, 1996.

\bibitem{Gillespie77}
DT~Gillespie.
\newblock Exact stochastic simulation of coupled chemical reactions.
\newblock {\em The Journal of Physical Chemistry}, 81(25):2340--2361, 1977.

\bibitem{Gillespie09}
CS~Gillespie.
\newblock Moment-closure approximations for mass-action models.
\newblock {\em IET systems biology}, 3(1):52--58, 2009.

\bibitem{Hespanha08}
J~Hespanha.
\newblock Moment closure for biochemical networks.
\newblock In {\em Communications, Control and Signal Processing, 2008. ISCCSP
  2008. 3rd International Symposium on}, pages 142--147. IEEE, 2008.

\bibitem{Caswell11}
H~Caswell.
\newblock Perturbation analysis of continuous-time absorbing markov chains.
\newblock {\em Numerical Linear Algebra with Applications}, 18(6):901--917,
  2011.

\bibitem{Kut07}
C~Kut, F~Mac~Gabhann, and AS~Popel.
\newblock Where is {VEGF} in the body? a meta-analysis of {VEGF} distribution
  in cancer.
\newblock {\em British Journal of Cancer}, 97(7):978--985, 2007.

\bibitem{Abbate92}
J~Abate and W~Whitt.
\newblock Numerical inversion of probability generating functions.
\newblock {\em Operations Research Letters}, 12(4):245--251, 1992.

\bibitem{mac2004model}
F~Mac~Gabhann and AS~Popel.
\newblock Model of competitive binding of vascular endothelial growth factor
  and placental growth factor to {VEGF} receptors on endothelial cells.
\newblock {\em American Journal of Physiology-Heart and Circulatory
  Physiology}, 286(1):H153--H164, 2004.

\bibitem{park1993vascular}
John~E Park, Gilbert-A Keller, and Napoleone Ferrara.
\newblock The vascular endothelial growth factor ({VEGF}) isoforms:
  differential deposition into the subepithelial extracellular matrix and
  bioactivity of extracellular matrix-bound vegf.
\newblock {\em Molecular biology of the cell}, 4(12):1317--1326, 1993.

\bibitem{casaletto2012spatial}
JB~Casaletto and AI~McClatchey.
\newblock Spatial regulation of receptor tyrosine kinases in development and
  cancer.
\newblock {\em Nature Reviews Cancer}, 12(6):387--400, 2012.

\bibitem{Imoukhuede11}
PI~Imoukhuede and AS~Popel.
\newblock Quantification and cell-to-cell variation of vascular endothelial
  growth factor receptors.
\newblock {\em Experimental Cell Research}, 317(7):955--965, 2011.

\bibitem{Imoukhuede12}
PI~Imoukhuede and AS~Popel.
\newblock Expression of {VEGF} receptors on endothelial cells in mouse skeletal
  muscle.
\newblock {\em PloS ONE}, 7(9):e44791, 2012.

\bibitem{Magnus85}
JR~Magnus and H~Neudecker.
\newblock Matrix differential calculus with applications to simple, hadamard,
  and kronecker products.
\newblock {\em Journal of Mathematical Psychology}, 29(4):474--492, 1985.

\bibitem{Magnus88}
H~Neudecker and JR~Magnus.
\newblock Matrix differential calculus with applications in statistics and
  econometrics, 1988.

\bibitem{Weisz73}
PB~Weisz.
\newblock Diffusion and chemical transformation. an interdisciplinary
  excursion.
\newblock {\em Science}, 179(4072):433--440, 1973.

\bibitem{Ewan06}
LC~Ewan, HM~Jopling, H~Jia, S~Mittar, A~Bagherzadeh, GJ~Howell, JH~Walker,
  IC~Zachary, and S~Ponnambalam.
\newblock Intrinsic tyrosine kinase activity is required for vascular
  endothelial growth factor receptor 2 ubiquitination, sorting and degradation
  in endothelial cells.
\newblock {\em Traffic}, 7(9):1270--1282, 2006.

\bibitem{Mittar09}
S~Mittar, C~Ulyatt, GJ~Howell, AF~Bruns, I~Zachary, JH~Walker, and
  S~Ponnambalam.
\newblock {VEGFR1} receptor tyrosine kinase localization to the golgi apparatus
  is calcium-dependent.
\newblock {\em Experimental Cell Research}, 315(5):877--889, 2009.

\bibitem{Almqvist04}
N~Almqvist, R~Bhatia, G~Primbs, N~Desai, S~Banerjee, and R~Lal.
\newblock Elasticity and adhesion force mapping reveals real-time clustering of
  growth factor receptors and associated changes in local cellular rheological
  properties.
\newblock {\em Biophysical Journal}, 86(3):1753--1762, 2004.

\bibitem{bikfalvi1991interaction}
A~Bikfalvi, C~Sauzeau, H~Moukadiri, J~Maclouf, N~Busso, M~Bryckaert, J~Plouet,
  and G~Tobelem.
\newblock Interaction of vasculotropin/vascular endothelial cell growth factor
  with human umbilical vein endothelial cells: binding, internalization,
  degradation, and biological effects.
\newblock {\em Journal of Cellular Physiology}, 149(1):50--59, 1991.

\bibitem{ewan2006intrinsic}
LC~Ewan, HM~Jopling, H~Jia, S~Mittar, A~Bagherzadeh, GJ~Howell, JH~Walker,
  IC~Zachary, and S~Ponnambalam.
\newblock Intrinsic tyrosine kinase activity is required for vascular
  endothelial growth factor receptor 2 ubiquitination, sorting and degradation
  in endothelial cells.
\newblock {\em Traffic}, 7(9):1270--1282, 2006.

\bibitem{huang1998expression}
X~Huang, C~Gottstein, RA~Brekken, and PE~Thorpe.
\newblock Expression of soluble {VEGF} receptor 2 and characterization of its
  binding by surface plasmon resonance.
\newblock {\em Biochemical and Biophysical Research Communications},
  252(3):643--648, 1998.

\bibitem{waltenberger1994different}
J~Waltenberger, L~Claesson-Welsh, A~Siegbahn, M~Shibuya, and CH~Heldin.
\newblock Different signal transduction properties of {KDR} and flt1, two
  receptors for vascular endothelial growth factor.
\newblock {\em Journal of Biological Chemistry}, 269(43):26988--26995, 1994.

\bibitem{Mac05b}
F~Mac~Gabhann and AS~Popel.
\newblock Differential binding of {VEGF} isoforms to {VEGF} receptor 2 in the
  presence of neuropilin-1: a computational model.
\newblock {\em American Journal of Physiology-Heart and Circulatory
  Physiology}, 288(6):H2851--H2860, 2005.

\bibitem{Linderman89}
JJ~Linderman and DA~Lauffenburger.
\newblock {\em Receptor/ligand sorting along the endocytic pathway}.
\newblock Oxford University Press, New York, 1989.

\bibitem{Aird07}
WC~Aird.
\newblock Phenotypic heterogeneity of the endothelium i. structure, function,
  and mechanisms.
\newblock {\em Circulation Research}, 100(2):158--173, 2007.

\bibitem{Papoulis02}
A~Papoulis and SU~Pillai.
\newblock {\em Probability, random variables, and stochastic processes}.
\newblock Tata McGraw-Hill Education, 2002.

\bibitem{napione2012unraveling}
Lucia Napione, Simona Pavan, Andrea Veglio, Andrea Picco, Guido Boffetta,
  Antonio Celani, Giorgio Seano, Luca Primo, Andrea Gamba, and Federico
  Bussolino.
\newblock Unraveling the influence of endothelial cell density on {VEGF-A}
  signaling.
\newblock {\em Blood}, 119(23):5599--5607, 2012.

\end{thebibliography}


\newpage
 
\appendix{\par\noindent \Large \bf Appendix~A. Notation}

\vspace{0.2cm}

In this appendix we set some standard notation we use throughout the paper. 
First of all, $\delta_{i,j}$ represents  Kronecker's delta, that is,
\begin{eqnarray*}
 \delta_{i,j} &=& \left\{\begin{array}{ll}
1,& \hbox{if $i=j$},\\
0,& \hbox{otherwise}.
\end{array}\right.
\end{eqnarray*}
Given a set ${\cal S}$, $\#$ represents its cardinality. 
Regarding matrix notation, matrices and vectors are always given in bold, where ${\bf 0}_p$ (${\bf e}_q$) 
represents a column vector of
zeros (ones) with dimension $p$ ($q$). The symbol $^{T}$ represents the transposition operator and, for a matrix ${\bf A}(\theta)$, 
we use the calculus
notation
\begin{eqnarray*}
 {\bf A}^{(l)}(0) &=& \left.\frac{d^l}{d\theta^l}{\bf A}(\theta)\right|_{\theta=0}.
\end{eqnarray*}

\par\noindent Finally, when a matrix depends on different parameters, ${\bf A}(\alpha,\theta)$, its first order partial derivatives 
with respect to each parameter are given by ${\bf A}^{(\alpha)}(\alpha,\theta)$
and ${\bf A}^{(\theta)}(\alpha,\theta)$, respectively.


\newpage

\appendix{\par\noindent \Large \bf Appendix~B. \large Matrices introduced in Section~\ref{Sect2}}

\section*{\normalsize B.1. Matrices ${\bf A}_{k,k'}$ in~\eqref{Eqn3}}

Matrices ${\bf A}_{k,k'}$  in~\eqref{Eqn3} contain the infinitesimal transition rates 
for those transitions from states in level $L(k)$ to states in level
$L(k')$, for $k'\in\{k-1,k,k+1\}$, and are obtained from~\eqref{Eqn2} as follows:
\begin{itemize}
 \item For $1\leq k\leq n_L$,
\begin{eqnarray*}
\left({\bf A}_{k,k-1}\right)_{ij} &=& \left\{\begin{array}{ll}
2\beta_{-}k, & \hbox{if $j=i+1$},\\
0, & \hbox{otherwise,}
\end{array}\right.
\end{eqnarray*}
where $0\leq i\leq n_L-k$, $0\leq j\leq n_L-k+1$.
 \item For $0\leq k\leq n_L$,
\begin{eqnarray*}
\left({\bf A}_{k,k}\right)_{ij} &=& \left\{\begin{array}{ll}
2\alpha_{+}(n_{R_2}-i-2k)(n_L-i-k), \quad\quad\quad\quad\quad\quad\quad\quad\quad\quad\quad\quad\quad \ \hbox{\it if $j=i+1$},\\
\alpha_{-}i, \quad\quad\quad\quad\quad\quad\quad\quad\quad\quad\quad\quad\quad\quad\quad\quad\quad\quad\quad\quad\quad\quad\quad\quad\quad \hbox{\it if $j=i-1$},\\
-\left(2\alpha_{+}(n_{R_2}-i-2k)(n_L-i-k)+\alpha_{-}i+2\beta_{-}k+\beta_{+}i(n_{R_2}-i-2k)\right),\\
\quad\quad\quad\quad\quad\quad\quad\quad\quad\quad\quad\quad\quad\quad\quad\quad\quad\quad\quad\quad\quad\quad\quad\quad\quad\quad\quad \hbox{\it if $j=i$},\\
0,\quad\quad\quad\quad\quad\quad\quad\quad\quad\quad\quad\quad\quad\quad\quad\quad\quad\quad\quad\quad\quad\quad\quad\quad\quad\quad \hbox{\it otherwise,}
\end{array}\right.
\end{eqnarray*}
where $0\leq i\leq n_L-k$, $0\leq j\leq n_L-k$.
  \item For $0\leq k\leq n_L-1$,
\begin{eqnarray*}
\left({\bf A}_{k,k+1}\right)_{ij} &=& \left\{\begin{array}{ll}
\beta_{+}i(n_{R_2}-i-2k), & \hbox{if $j=i-1$},\\
0, & \hbox{otherwise,}
\end{array}\right.
\end{eqnarray*}
where $0\leq i\leq n_L-k$, $0\leq j\leq n_L-k-1$.
\end{itemize}

\section*{\normalsize B.2. Sub-matrices ${\bf A}_{k,k'}(z)$ and sub-vector ${\bf a}_{N_2-1}(z)$ in~\eqref{Eqn5}}

Sub-matrices ${\bf A}_{k,k'}(z)$ and sub-vector ${\bf a}_{N_2-1}(z)$ in~\eqref{Eqn5} are given by:

\begin{itemize}
 \item $\left({\bf a}_{N_2-1}(z)\right)_i ~=~ \frac{\beta_{+}i(n_{R_2}-i-2(N_2-1))}{z+A_{(i,N_2-1)}}$, for $0\leq i\leq n_L-N_2+1$.
 \item For $1\leq k\leq n_L$,
\begin{eqnarray*}
\left({\bf A}_{k,k-1}(z)\right)_{ij} &=& \left\{\begin{array}{ll}
\frac{2\beta_{-}k}{z+A_{(i,k)}}, & \hbox{if $j=i+1$},\\
0, & \hbox{otherwise,}
\end{array}\right.
\end{eqnarray*}
where $0\leq i\leq n_L-k$, $0\leq j\leq n_L-k+1$.
 \item For $0\leq k\leq n_L$,
\begin{eqnarray*}
\left({\bf A}_{k,k}(z)\right)_{ij} &=& \left\{\begin{array}{ll}
\frac{2\alpha_{+}(n_{R_2}-i-2k)(n_L-i-k)}{z+A_{(i,k)}}, & \hbox{if $j=i+1$},\\
\frac{\alpha_{-}i}{z+A_{(i,k)}}, & \hbox{if $j=i-1$},\\
0, & \hbox{otherwise,}
\end{array}\right.
\end{eqnarray*}
where $0\leq i\leq n_L-k$, $0\leq j\leq n_L-k$.
  \item For $0\leq k\leq n_L-1$,
\begin{eqnarray*}
\left({\bf A}_{k,k+1}(z)\right)_{ij} &=& \left\{\begin{array}{ll}
\frac{\beta_{+}i(n_{R_2}-i-2k)}{z+A_{(i,k)}}, & \hbox{if $j=i-1$},\\
0, & \hbox{otherwise,}
\end{array}\right.
\end{eqnarray*}
where $0\leq i\leq n_L-k$, $0\leq j\leq n_L-k-1$.
\end{itemize}

\section*{\normalsize B.3. Derivative matrices ${\bf A}^{N_2,(p)}(0)$ and ${\bf a}^{N_2,(p)}(0)$ in~\eqref{Eqn7}}

Matrices ${\bf A}^{N_2,(p)}(0)$ and ${\bf a}^{N_2,(p)}(0)$ in~\eqref{Eqn7} are given by

\begin{eqnarray*}
 {\bf A}^{N_2,(p)}(0) &=& \left(\begin{array}{ccccccc}
{\bf A}_{0,0}^{(p)}(0) & {\bf A}_{0,1}^{(p)}(0) & {\bf 0}_{J(0)\times J(2)} & \dots & {\bf 0}_{J(0)\times J(N_2-2)} & {\bf 0}_{J(0)\times J(N_2-1)} \\
{\bf A}_{1,0}^{(p)}(0) & {\bf A}_{1,1}^{(p)}(0) & {\bf A}_{1,2}^{(p)}(0) & \dots & {\bf 0}_{J(1)\times J(N_2-2)} & {\bf 0}_{J(1)\times J(N_2-1)} \\
{\bf 0}_{J(2)\times J(0)} & {\bf A}^{(p)}_{2,1}(0) & {\bf A}^{(p)}_{2,2}(0) & \dots & {\bf 0}_{J(2)\times J(N_2-2)} & {\bf 0}_{J(2)\times J(N_2-1)} \\
\vdots & \vdots & \vdots & \ddots & \vdots & \vdots \\
{\bf 0}_{J(N_2-2)\times J(0)} & {\bf 0}_{J(N_2-2)\times J(1)} & {\bf 0}_{J(N_2-2)\times J(2)} & \dots & {\bf A}_{N_2-2,N_2-2}^{(p)}(0) & {\bf A}_{N_2-2,N_2-1}^{(p)}(0) \\
{\bf 0}_{J(N_2-1)\times J(0)} & {\bf 0}_{J(N_2-1)\times J(1)} & {\bf 0}_{J(N_2-1)\times J(2)} & \dots & {\bf A}_{N_2-1,N_2-2}^{(p)}(0) & {\bf A}_{N_2-1,N_2-1}^{(p)}(0)
\end{array}\right),\\
{\bf a}^{N_2,(p)}(0) &=& \left(\begin{array}{c}
{\bf 0}_{J(0)}\\
{\bf 0}_{J(1)}\\
\vdots\\
{\bf 0}_{J(N_2-2)}\\
{\bf a}^{(p)}_{N_2-1}(0)
\end{array}\right),
\end{eqnarray*}
where expressions for ${\bf a}^{(p)}_{N_2-1}(0)$ and ${\bf A}_{k,k'}^{(p)}(0)$ are as follows:

\begin{itemize}
 \item $\left({\bf a}^{(p)}_{N_2-1}(0)\right)_i ~=~ (-1)^pp!\frac{\beta_{+}i(n_{R_2}-i-2(N_2-1))}{A_{(i,N_2-1)}^{p+1}}$, for $0\leq i\leq n_L-N_2+1$, $p\geq1$.
 \item For $1\leq k\leq n_L$, $p\geq1$,
\begin{eqnarray*}
\left({\bf A}_{k,k-1}^{(p)}(0)\right)_{ij} &=& \left\{\begin{array}{ll}
(-1)^pp!\frac{2\beta_{-}k}{A_{(i,k)}^{p+1}}, & \hbox{if $j=i+1$},\\
0, & \hbox{otherwise,}
\end{array}\right.
\end{eqnarray*}
where $0\leq i\leq n_L-k$, $0\leq j\leq n_L-k+1$.
 \item For $0\leq k\leq n_L$, $p\geq1$,
\begin{eqnarray*}
\left({\bf A}_{k,k}^{(p)}(0)\right)_{ij} &=& \left\{\begin{array}{ll}
(-1)^pp!\frac{2\alpha_{+}(n_{R_2}-i-2k)(n_L-i-k)}{A_{(i,k)}^{p+1}}, & \hbox{if $j=i+1$},\\
(-1)^pp!\frac{\alpha_{-}i}{A_{(i,k)}^{p+1}}, & \hbox{if $j=i-1$},\\
0, & \hbox{otherwise,}
\end{array}\right.
\end{eqnarray*}
where $0\leq i\leq n_L-k$, $0\leq j\leq n_L-k$.
  \item For $0\leq k\leq n_L-1$, $p\geq1$,
\begin{eqnarray*}
\left({\bf A}_{k,k+1}^{(p)}(0)\right)_{ij} &=& \left\{\begin{array}{ll}
(-1)^pp!\frac{\beta_{+}i(n_{R_2}-i-2k)}{A_{(i,k)}^{p+1}}, & \hbox{if $j=i-1$},\\
0, & \hbox{otherwise,}
\end{array}\right.
\end{eqnarray*}
where $0\leq i\leq n_L-k$, $0\leq j\leq n_L-k-1$.
\end{itemize}

\section*{\normalsize B.4. Matrices ${\bf \hat A}_{k,k'}$ in Model~2}

For $0\leq k\leq n_L$
\begin{eqnarray*}
 {\bf \hat A}_{k,k} &=& \left(\begin{array}{cccccc}
{\bf B}_{0,0}^{k,k} & {\bf B}_{0,1}^{k,k} & {\bf 0} & \dots & {\bf 0} & {\bf 0} \\
{\bf B}_{1,0}^{k,k} & {\bf B}_{1,1}^{k,k} & {\bf B}_{1,2}^{k,k} & \dots & {\bf 0} & {\bf 0} \\
{\bf 0} & {\bf B}_{2,1}^{k,k} & {\bf B}_{2,2}^{k,k} & \dots & {\bf 0} & {\bf 0} \\
\vdots & \vdots & \vdots & \ddots & \vdots & \vdots \\
{\bf 0} & {\bf 0} & {\bf 0} & \dots & {\bf B}_{n_L-k-1,n_L-k-1}^{k,k} & {\bf B}_{n_L-k-1,n_L-k}^{k,k} \\
{\bf 0} & {\bf 0} & {\bf 0} & \dots & {\bf B}_{n_L-k,n_L-k-1}^{k,k} & {\bf B}_{n_L-k,n_L-k}^{k,k}
                   \end{array}\right),
\end{eqnarray*}
for $0\leq k\leq n_L-1$,
\begin{eqnarray*}
  {\bf \hat A}_{k,k+1} &=& \left(\begin{array}{cccccc}
{\bf 0} & {\bf 0} & {\bf 0} & \dots & {\bf 0} & {\bf 0} \\
{\bf B}_{1,0}^{k,k+1} & {\bf 0} & {\bf 0} & \dots & {\bf 0} & {\bf 0} \\
{\bf 0} & {\bf B}_{2,1}^{k,k+1} & {\bf 0} & \dots & {\bf 0} & {\bf 0} \\
\vdots & \vdots & \vdots & \ddots & \vdots & \vdots \\
{\bf 0} & {\bf 0} & {\bf 0} & \dots & {\bf 0} & {\bf 0} \\
{\bf 0} & {\bf 0} & {\bf 0} & \dots & {\bf B}_{n_L-k,n_L-k-1}^{k,k+1} & {\bf 0}
                   \end{array}\right),
\end{eqnarray*}
and, for $1\leq k\leq n_L$,
\begin{eqnarray*}
 {\bf \hat A}_{k,k-1} &=& \left(\begin{array}{cccccc}
{\bf 0} & {\bf B}_{0,1}^{k,k-1} & {\bf 0} & \dots & {\bf 0} & {\bf 0} \\
{\bf 0} & {\bf 0} & {\bf B}_{1,2}^{k,k-1} & \dots & {\bf 0} & {\bf 0} \\
{\bf 0} & {\bf 0} & {\bf 0} & \dots & {\bf 0} & {\bf 0} \\
\vdots & \vdots & \vdots & \ddots & \vdots & \vdots \\
{\bf 0} & {\bf 0} & {\bf 0} & \dots & {\bf B}_{n_L-k-1,n_L-k}^{k,k-1} & {\bf 0} \\
{\bf 0} & {\bf 0} & {\bf 0} & \dots & {\bf 0} & {\bf B}_{n_L-k,n_L-k+1}^{k,k-1}
                   \end{array}\right).
\end{eqnarray*}

We note that, although we are omitting  the dimensions of the matrices ${\bf 0}$ for the ease of notation,
the dimension of each matrix ${\bf 0}$, representing transitions from states in sub-level $l(k;r)$ to states in sub-level $l(k';r')$,
 is
$J(k;r)\times J(k';r')$. The expressions for the matrices ${\bf B}_{r,r'}^{k,k'}$ are given as follows:

\begin{itemize}
 \item For $0\leq r\leq n_L-k$, $0\leq k\leq n_L$,
\begin{eqnarray*}
\left({\bf B}_{r,r}^{k,k}\right)_{ij} &=& \left\{\begin{array}{ll}
\alpha_{-}i, & \hbox{if $j=i-1$},\\
-A_{(i,r,k)}, & \hbox{if $j=i$},\\
2\alpha_{+}(n_{R_2}-i-2r-2k)(n_L-i-r-k), & \hbox{if $j=i+1$},\\
0, & \hbox{otherwise,}
\end{array}\right.
\end{eqnarray*}
where $0\leq i\leq n_L-r-k$, $0\leq j\leq n_L-r-k$, and, from now on, $A_{(i,r,k)}=2\alpha_{+}(n_{R_2}-i-2r-2k)(n_L-i-r-k)+\alpha_{-}i+\beta_{+}i(n_{R_2}-i-2r-2k)+2\beta_{-}r+\gamma_{+}r+\gamma_{-}k$.
\item For $0\leq r\leq n_L-k-1$, $0\leq k\leq n_L$,
\begin{eqnarray*}
\left({\bf B}_{r,r+1}^{k,k}\right)_{ij} &=& \left\{\begin{array}{ll}
\beta_{+}i(n_{R_2}-i-2r-2k), & \hbox{if $j=i-1$},\\
0, & \hbox{otherwise,}
\end{array}\right.
\end{eqnarray*}
where $0\leq i\leq n_L-r-k$, $0\leq j\leq n_L-r-k-1$.
\item For $1\leq r\leq n_L-k$, $0\leq k\leq n_L$,
\begin{eqnarray*}
\left({\bf B}_{r,r-1}^{k,k}\right)_{ij} &=& \left\{\begin{array}{ll}
2\beta_{-}r, & \hbox{if $j=i+1$},\\
0, & \hbox{otherwise,}
\end{array}\right.
\end{eqnarray*}
where $0\leq i\leq n_L-r-k$, $0\leq j\leq n_L-r-k+1$.
\item For $1\leq r\leq n_L-k$, $0\leq k\leq n_L-1$,
\begin{eqnarray*}
\left({\bf B}_{r,r-1}^{k,k+1}\right)_{ij} &=& \left\{\begin{array}{ll}
\gamma_{+}r, & \hbox{if $j=i$},\\
0, & \hbox{otherwise,}
\end{array}\right.
\end{eqnarray*}
where $0\leq i\leq n_L-r-k$, $0\leq j\leq n_L-r-k$.
\item For $0\leq r\leq n_L-k$, $1\leq k\leq n_L$,
\begin{eqnarray*}
\left({\bf B}_{r,r+1}^{k,k-1}\right)_{ij} &=& \left\{\begin{array}{ll}
\gamma_{-}k, & \hbox{if $j=i$},\\
0, & \hbox{otherwise,}
\end{array}\right.
\end{eqnarray*}
where $0\leq i\leq n_L-r-k$, $0\leq j\leq n_L-r-k$.
\end{itemize}


\newpage

\appendix{\par\noindent \Large \bf Appendix~C. \large Algorithms}

\vspace{0.25cm}

\par \noindent{\bf Algorithm~1} (to obtain 
the Laplace-Stieltjes transforms ${\bf g}^{N_2}(z)$ and the $l$-th order moments ${\bf m}^{N_2,(l)}$)
\begin{description}
  \item \bf PART 1
  \item ${\bf H}^{N_2}_{0}(z) ~=~ {\bf I}_{J(0)}-{\bf A}_{0,0}(z)$;
  \item \it For $k=1,\dots,N_2-1$:
  \item $~$\hspace{0.5cm} ${\bf H}^{N_2}_{k}(z) ~=~ {\bf I}_{J(k)}-{\bf A}_{k,k}(z)-{\bf A}_{k,k-1}(z){\bf H}^{N_2}_{k-1}(z)^{-1}{\bf A}_{k-1,k}(z)$;
  \item ${\bf g}_{N_2-1}^{N_2}(z) ~=~ {\bf H}^{N_2}_{N_2-1}(z)^{-1}{\bf a}_{N_2-1}(z)$;
  \item ${\bf m}^{N_2,(0)}_{N_2-1} ~=~ {\bf g}^{N_2}_{N_2-1}(0)$;
  \item \it For $k=N_2-2,\dots,1,0$:
  \item $~$\hspace{0.5cm} ${\bf g}^{N_2}_{k}(z) ~=~ {\bf H}^{N_2}_{k}(z)^{-1}{\bf A}_{k,k+1}(z){\bf g}^{N_2}_{k+1}(z)$;
  \item $~$\hspace{0.5cm} ${\bf m}^{N_2,(0)}_{k} ~=~ {\bf g}^{N_2}_{k}(0)$;
  \item \bf PART 2
  \item ${\bf m}^{N_2,(0)}_{N_2-1} ~=~ {\bf g}^{N_2}_{N_2-1}(0)$;
  \item \it For $k=N_2-2,\dots,1,0$:
  \item $~$\hspace{0.5cm} ${\bf m}^{N_2,(0)}_{k} ~=~ {\bf g}^{N_2}_{k}(0)$;
  \item For $p=1,\dots,l$:
  \item $~$\hspace{0.5cm} ${\bf P}^{N_2,(p)}_{0} ~=~ \sum\limits_{k=1}^{p}\binom{p}{k}(-1)^k\left({\bf A}_{0,0}^{(k)}(0){\bf m}^{N_2,(p-k)}_{0}+{\bf A}_{0,1}^{(k)}(0){\bf m}^{N_2,(p-k)}_{1}\right)$;
  \item $~$\hspace{0.5cm} For $j=1,\dots,N_2-1$:
  \item $~$\hspace{1.25cm} ${\bf P}^{N_2,(p)}_{j} ~=~ {\bf A}_{j,j-1}(0){\bf H}^{N_2}_{j-1}(0)^{-1}{\bf P}^{N_2,(p)}_{j-1}+\sum\limits_{k=1}^{p}\binom{p}{k}(-1)^k\left({\bf A}_{j,j-1}^{(k)}(0){\bf m}^{N_2,(p-k)}_{j-1}\right.$
  \item $~$\hspace{3cm} $\left.+{\bf A}_{j,j}^{(k)}(0){\bf m}^{N_2,(p-k)}_{j}+(1-\delta_{j,N_2-1}){\bf A}_{j,j+1}^{(k)}(0){\bf m}^{N_2,(p-k)}_{j+1}\right)$;
  \item $~$\hspace{0.5cm} ${\bf m}^{N_2,(p)}_{N_2-1} ~=~ {\bf H}^{N_2}_{N_2-1}(0)^{-1}\left({\bf P}^{N_2,(p)}_{N_2-1}+(-1)^p{\bf a}^{(p)}_{N_2-1}(0)\right)$;
  \item $~$\hspace{0.5cm} For $j=N_2-2,\dots,1,0$:
  \item $~$\hspace{1.25cm} ${\bf m}^{N_2,(p)}_{j} ~=~ {\bf H}^{N_2}_{j}(0)^{-1}\left({\bf P}^{N_2,(p)}_{j}+{\bf A}_{j,j+1}(0){\bf m}^{N_2,(p)}_{j+1}\right)$;
\end{description}

\vspace{0.25cm}

\par \noindent{\bf Algorithm~2} (to obtain the stationary distribution ${\boldsymbol \pi}$)
\begin{description}
  \item ${\bf H}_{0} ~=~ {\bf A}_{0,0}$;
  \item For $k=1,\dots,n_L-1$:
  \item $~$\hspace{0.5cm} ${\bf H}_{k} ~=~ {\bf A}_{k,k}-{\bf A}_{k,k-1}{\bf H}_{k-1}^{-1}{\bf A}_{k-1,k}$;
  \item ${\boldsymbol \pi}^{*}_{n_L} ~=~ 1$;
  \item For $k=n_L-1,\dots,0$:
  \item $~$\hspace{0.5cm} ${\boldsymbol \pi}^{*}_{k} ~=~ -{\boldsymbol \pi}^{*}_{k+1}{\bf A}_{k+1,k}{\bf H}_{k}^{-1}$;
  \item For $k=0,\dots,n_L$:
  \item $~$\hspace{0.5cm} ${\boldsymbol \pi}_{k} ~=~ \frac{1}{\sum\limits_{j=0}^{n_L}{\boldsymbol \pi}^{*}_{j}{\bf e}_{J(r)}}{\boldsymbol \pi}^{*}_{k}$;
\end{description}

\vspace{0.25cm}

\par \noindent{\bf Algorithm~1S} (to obtain the derivative of the $r$-th order moments $E[T_{(n_1,n_2)}(N_2)^{r}]$ with respect\\
$~$\hspace{2.5cm} $\theta_i\in\{\alpha_{+},\alpha_{-},\beta_{+},\beta_{-}\}$)
\begin{description}
  \item ${\bf H}^{N_2,(\theta_i)}_{0}(0) ~=~ -{\bf A}_{0,0}^{(\theta_i)}(0)$;
  \item \it For $k=1,\dots,N_2-1$:
  \item $~$\hspace{0.5cm} ${\bf H}^{N_2,(\theta_i)}_{k}(0) ~=~ -{\bf A}_{k,k}^{(\theta_i)}(0)-\left({\bf A}_{k,k-1}^{(\theta_i)}(0){\bf H}^{N_2}_{k-1}(0)^{-1}{\bf A}_{k-1,k}(0)-{\bf A}_{k,k-1}(0){\bf H}^{N_2}_{k-1}(0)^{-1}\right.$
  \item $~$\hspace{3cm} $\left.\times{\bf H}^{N_2,(\theta_i)}_{k-1}(0){\bf H}^{N_2}_{k-1}(0)^{-1}{\bf A}_{k-1,k}(0)+{\bf A}_{k,k-1}(0){\bf H}^{N_2}_{k-1}(0)^{-1}{\bf A}^{(\theta_i)}_{k-1,k}(0)\right)$;
  \item ${\bf m}^{N_2,(0,\theta_i)}_{N_2-1} ~=~ -{\bf H}^{N_2}_{N_2-1}(0)^{-1}{\bf H}^{N_2,(\theta_i)}_{N_2-1}(0){\bf H}^{N_2}_{N_2-1}(0)^{-1}{\bf a}_{N_2-1}(0)+{\bf H}^{N_2}_{N_2-1}(z)^{-1}{\bf a}^{,(\theta_i)}_{N_2-1}(z)$;
  \item \it For $k=N_2-2,\dots,1,0$:
  \item $~$\hspace{0.5cm} ${\bf m}^{N_2,(0,\theta_i)}_{k} ~=~ -{\bf H}^{N_2}_{k}(0)^{-1}{\bf H}^{N_2,(\theta_i)}_{k}(0){\bf H}^{N_2}_{k}(0)^{-1}{\bf A}_{k,k+1}(0){\bf m}^{N_2,(0)}_{k+1}+{\bf H}^{N_2}_{k}(0)^{-1}{\bf A}^{(\theta_i)}_{k,k+1}(0)$
  \item $~$\hspace{2.5cm} $\times{\bf m}^{N_2,(0)}_{k+1}+{\bf H}^{N_2}_{k}(0)^{-1}{\bf A}_{k,k+1}(0){\bf m}^{N_2,(0,\theta_i)}_{k+1}$;
  \item For $j=1,\dots,r$:
  \item $~$\hspace{0.5cm} ${\bf P}^{N_2,(j,\theta_i)}_{0} ~=~ \sum\limits_{k=1}^{j}\binom{j}{k}(-1)^k\left({\bf A}_{0,0}^{(k,\theta_i)}(0){\bf m}^{N_2,(j-k)}_{0}+{\bf A}_{0,0}^{(k)}(0){\bf m}^{N_2,(j-k,\theta_i)}_{0}+{\bf A}_{0,1}^{(k,\theta_i)}(0){\bf m}^{N_2,(j-k)}_{1}\right.$
  \item $~$\hspace{2.5cm} $\left.+{\bf A}_{0,1}^{(k)}(0){\bf m}^{N_2,(j-k,\theta_i)}_{1}\right)$;
  \item $~$\hspace{0.5cm} For $p=1,\dots,N_2-1$:
  \item $~$\hspace{1.25cm} ${\bf P}^{N_2,(j,\theta_i)}_{p} ~=~ {\bf A}^{(\theta_i)}_{p,p-1}(0){\bf H}^{N_2}_{p-1}(0)^{-1}{\bf P}^{N_2,(j)}_{p-1}-{\bf A}_{p,p-1}(0){\bf H}^{N_2}_{p-1}(0)^{-1}{\bf H}^{N_2,(\theta_i)}_{p-1}(0)$
  \item $~$\hspace{3.5cm} $\times{\bf H}^{N_2}_{p-1}(0)^{-1}{\bf P}^{N_2,(j)}_{p-1}+{\bf A}_{p,p-1}(0){\bf H}^{N_2}_{p-1}(0)^{-1}{\bf P}^{N_2,(j,\theta_i)}_{p-1}+\sum\limits_{k=1}^{j}\binom{i}{k}(-1)^k$
  \item $~$\hspace{3.5cm} $\times\left({\bf A}_{p,p-1}^{(k,\theta_i)}(0){\bf m}^{N_2,(j-k)}_{p-1}+{\bf A}_{p,p-1}^{(k)}(0){\bf m}^{N_2,(j-k,\theta_i)}_{p-1}+{\bf A}_{p,p}^{(k,\theta_i)}(0){\bf m}^{N_2,(j-k)}_{p}\right.$
  \item $~$\hspace{3.5cm} $+{\bf A}_{p,p}^{(k)}(0){\bf m}^{N_2,(j-k,\theta_i)}_{p}+(1-\delta_{p,N_2-1})\left({\bf A}_{p,p+1}^{(k,\theta_i)}(0){\bf m}^{N_2,(j-k)}_{p+1}+{\bf A}_{p,p+1}^{(k)}(0)\right.$
  \item $~$\hspace{3.5cm} $\left.\left.\times{\bf m}^{N_2,(j-k,\theta_i)}_{p+1}\right)\right)$;
  \item $~$\hspace{0.5cm} ${\bf m}^{N_2,(j,\theta_i)}_{N_2-1} ~=~ -{\bf H}^{N_2}_{N_2-1}(0)^{-1}{\bf H}^{N_2,(\theta_i)}_{N_2-1}(0){\bf H}^{N_2}_{N_2-1}(0)^{-1}\left({\bf P}^{N_2,(j)}_{N_2-1}+(-1)^j{\bf a}^{,(j)}_{N_2-1}(0)\right)$
  \item $~$\hspace{2.75cm} $+{\bf H}^{N_2}_{N_2-1}(0)^{-1}\left({\bf P}^{N_2,(j,\theta_i)}_{N_2-1}+(-1)^j{\bf a}^{,(j,\theta_i)}_{N_2-1}(0)\right)$;
  \item $~$\hspace{0.5cm} For $p=N_2-2,\dots,1,0$:
  \item $~$\hspace{1.25cm} ${\bf m}^{N_2,(j,\theta_i)}_{p} ~=~ -{\bf H}^{N_2}_{p}(0)^{-1}{\bf H}^{N_2,(\theta_i)}_{p}(0){\bf H}^{N_2}_{p}(0)^{-1}\left({\bf P}^{N_2,(j)}_{p}+{\bf A}_{p,p+1}(0){\bf m}^{N_2,(j)}_{p+1}\right)+{\bf H}^{N_2}_{p}(0)^{-1}$
  \item $~$\hspace{3.5cm} $\times\left({\bf P}^{N_2,(j,\theta_i)}_{p}+{\bf A}_{p,p+1}^{(\theta_i)}(0){\bf m}_{p+1}^{N_2,(j)}+{\bf A}_{p,p+1}(0){\bf m}^{N_2,(j,\theta_i)}_{p+1}\right)$;
\end{description}
We point out that ${\bf m}_{k}^{N_2,(r,\theta_i)}$ and ${\bf A}_{k,k'}^{(r,\theta_i)}(0)$ in Algorithm 1S, which corresponds to the model with instantaneous
phosphorylation, represent the derivatives of ${\bf m}_{k}^{N_2,(r)}$ and ${\bf A}_{k,k'}^{(r)}(0)$, respectively, with respect $\theta_i$, for
$\theta_i\in\{\alpha_{+},\alpha_{-},\beta_{+},\beta_{-}\}$.

\vspace{0.25cm}

\par \noindent{\bf Algorithm~2S} (to obtain the derivative of $m_1$ and $m_2$ with respect $\theta_i\in\{\alpha_{+},\alpha_{-},\beta_{+},\beta_{-}\}$)
\begin{description}
  \item ${\bf H}^{(\theta_i)}_{0} ~=~ {\bf A}^{(\theta_i)}_{0,0}$;
  \item For $k=1,\dots,n_L-1$:
  \item $~$\hspace{0.5cm} ${\bf H}^{(\theta_i)}_{k} ~=~ {\bf A}^{(\theta_i)}_{k,k}-\left({\bf A}^{(\theta_i)}_{k,k-1}{\bf H}_{k-1}^{-1}{\bf A}_{k-1,k}-{\bf A}_{k,k-1}{\bf H}_{k-1}^{-1}{\bf H}_{k-1}^{(\theta_i)}{\bf H}_{k-1}^{-1}{\bf A}_{k-1,k}+{\bf A}_{k,k-1}{\bf H}_{k-1}^{-1}{\bf A}_{k-1,k}^{(\theta_i)}\right)$;
  \item ${\boldsymbol \pi}^{*,(\theta_i)}_{n_L} ~=~ 0$;
  \item For $k=n_L-1,\dots,1,0$:
  \item $~$\hspace{0.5cm} ${\boldsymbol \pi}^{*,(\theta_i)}_{k} ~=~ -\left({\boldsymbol \pi}^{*,(\theta_i)}_{k+1}{\bf A}_{k+1,k}{\bf H}_{k}^{-1}+{\boldsymbol \pi}^{*}_{k+1}{\bf A}^{(\theta_i)}_{k+1,k}{\bf H}_{k}^{-1}-{\boldsymbol \pi}^{*}_{k+1}{\bf A}_{k+1,k}{\bf H}_{k}^{-1}{\bf H}^{(\theta_i)}_{k}{\bf H}_{k}^{-1}\right)$;
  \item For $k=0,\dots,n_L$:
  \item $~$\hspace{0.5cm} ${\boldsymbol \pi}^{(\theta_i)}_{k} ~=~ \frac{1}{{\boldsymbol \pi^{*}}{\bf e}_{\#{\cal S}}}\left({\boldsymbol \pi}_k^{*,(\theta_i)}-{\boldsymbol \pi}_k{\boldsymbol \pi^{*,(\theta_i)}}{\bf e}_{\#{\cal S}}\right)$;
  \item $m_1^{(\theta_i)} ~=~ \sum\limits_{k=0}^{n_L}k\left(\sum\limits_{j=0}^{n_L}({\boldsymbol \pi}_{j}^{(\theta_i)})_k\right)$;
  \item $m_2^{(\theta_i)} ~=~ \sum\limits_{k=0}^{n_L}k\left({\boldsymbol \pi}_{k}^{(\theta_i)}{\bf e}_{J(k)}\right)$;
\end{description}

\newpage

\appendix{\par\noindent \Large \bf Appendix~D. \large van Kampen approximation}

\vspace{0.25cm}

We have made use of  the system-size expansion technique, based on the power series approximation proposed by van 
Kampen~\cite{Van92}, to 
study~\eqref{eq: master R1-R2-L}. 
In this Appendix, and for completeness, we 
derive the equations for the moments, which will be used for numerical
simulations.

We first identify the expansion parameter $\Omega$, with $\Omega \gg 1$. In our study $\Omega$ represents
the volume of the system, so that 
 fluctuations are of  order  $\Omega^{\frac{1}{2}}$. We write any discrete variable
  $n_i$  as a sum of
``mean'' number of particles of type $i$ ($x_i$ - macroscopic concentration) and fluctuations $\xi_i$ of order $\Omega^{\frac{1}{2}}$:
$ n_i = \Omega x_i + \Omega^{\frac{1}{2}} \xi_i. $ This transformation of variables implies
\[
{ P_{\bf n}}(t) \longrightarrow \Psi({\bf \xi},t), \ \ \
\Omega^{\frac{1}{2}} \dfrac{\partial}{\partial n_i} P_{\bf n}(t) = \dfrac{\partial}{\partial \xi_i} \Psi(\xi, t).
\]
Therefore the left-hand-side of the master equation~\eqref{eq: master R1-R2-L} might be rewritten as
\begin{equation}
\dfrac{dP_{\bf n}(t)}{dt} = \dfrac{\partial \Psi }{\partial t} + \sum_i \dfrac{\partial \Psi}{\partial \xi_i} \dfrac{d \xi_i}{dt}.
\label{eq: master VK mod1 L-R}
\end{equation}
Our choice of transformation implies
\[
\dfrac{d\xi_i}{dt} = -\Omega^{\frac{1}{2}} \dfrac{dx_i}{dt},
\]
so that we can write
\begin{equation}
\dfrac{dP_{\bf n}(t)}{dt} = \dfrac{\partial \Psi}{\partial t} - \Omega^{\frac{1}{2}} \sum_i \dfrac{\partial \Psi}{\partial \xi_i} \dfrac{dx_i}{dt}.
\label{eq: master VK mod 0.1 L-R}
\end{equation}
The system of ODES for the mean variables $x_i$ is obtained from the term
{\small $\sum_i \dfrac{\partial \Psi}{\partial \xi_i} \dfrac{dx_i}{dt}$} of equation~\eqref{eq: master VK mod 0.1 L-R} and the system 
of ODEs for the moments of the fluctuations
is obtained  from  the term {$\dfrac{\partial \Psi}{\partial t}$} of the
equation~\eqref{eq: master VK mod 0.1 L-R}. 
If we collect all terms of order
$\Omega^{0}$, we obtain the following Fokker-Planck equation
\begin{equation*}
\frac{\partial \Psi}{\partial t} = -\sum_{i,j} A_{ij} \frac{\partial}{\partial \xi_i} (\xi_j \Psi) +\frac{1}{2} 
\sum_{i,j} B_{ij} \frac{\partial^2 \Psi}{\partial \xi_i \partial \xi_j}.
\label{FPE R-L}
\end{equation*}
From the Fokker-Plank equation it is possible to obtain equations for the mean, the variance and the covariance of
the fluctuations
\begin{equation*}
\dfrac{d \mathbb{E}(\xi_i)}{dt} = \sum_j A_{ij} \mathbb{E} (\xi_j),
\label{eq: moments VK EX}
\end{equation*}
\begin{equation*}
\dfrac{d \mathbb{E}(\xi_i \xi_j)}{dt} = \sum_k A_{ik} \mathbb{E}(\xi_k \xi_j) + \sum_k A_{jk} \mathbb{E} (\xi_i \xi_k) + B_{ij}.
\label{eq: moments VK EX2}
\end{equation*}

\paragraph{van Kampen approximation for Model~3.1}

In the case of Model~3.1 the deterministic variables are given by
\[
(M_1, M_2, D_1, P_2, P_M) \sim (x_1, x_2, x_3, x_4, x_5).
\]
The following parameters have been rescaled:
\[
\begin{array}{c}
n_L = N_L \Omega, \ \ \ n_{R_1} = N_{R_1} \Omega, \ \ \ n_{R_2} = N_{R_2} \Omega, \ \ \ \alpha_{1+} = a_{1+} \Omega^{-1}, \ \ \ \alpha_{2+} = a_{2+} \Omega^{-1}, \\
\\
\beta_{11+} = b_{11+} \Omega^{-1}, \ \ \ \beta_{12+} = b_{12+} \Omega^{-1} \ \ \ \beta_{21+} = b_{21+} \Omega^{-1}, \ \ \ \beta_{22+} = b_{22+} \Omega^{-1}.
\end{array}
\]
The deterministic equations of  Model~3.1 are given by

\begin{eqnarray}
\dfrac{dx_1}{dt}  &= & 2a_{1+}(N_{R_1}-x_1-2x_3-x_5)(N_L-x_1-x_2-x_3-x_4-x_5) 
\nonumber
\\
& - &b_{11+}x_1(N_{R_1}-x_1-2x_3-x_5)-b_{12+}x_1(N_{R_2}-x_2-2x_4-x_5)
\nonumber
\\
&-&\alpha_{1-}x_1+2\beta_{11-}x_3+\beta_{12-}x_5,
\nonumber
\\
\dfrac{dx_2}{dt} &= & 2a_{2+}(N_{R_2}-x_2-2x_4-x_5)(N_L-x_1-x_2-x_3-x_4-x_5)
\nonumber
\\
&-&b_{22+}x_2(N_{R_2}-x_2-2x_4-x_5) - b_{21+}x_2(N_{R_1}-x_1-2x_3-x_5) 
\nonumber
\\
& -&\alpha_{2-}x_2 + 2\beta_{22-}x_4+\beta_{21-}x_5,
\nonumber
\\
\dfrac{dx_3}{dt} &= & -2\beta_{11-}x_3+b_{11+}x_1(N_{R_1}-x_1-2x_3-x_5),
\nonumber
\\
\dfrac{dx_4}{dt} &= & b_{22+}x_2(N_{R_2}-x_2-2x_4-x_5)-2\beta_{22-}x_4,
\nonumber
\\
\dfrac{dx_5}{dt} &= & b_{12+}x_1(N_{R_2}-x_2-2x_4-x_5)+b_{21+}x_2(N_{R_1}-x_1-2x_3-x_5)-\beta_{12-}x_5-\beta_{21-}x_5.
\nonumber
\label{ODE: R1 R1 L}
\end{eqnarray}
We will denote by $(x_1^*,x_2^*,x_3^*,x_4^*,x_5^*)$
the unique stable steady state solution of Eqs.~\eqref{ODE: R1 R1 L}.

The coefficients $A_{ij}$ and $B_{ij}=B_{ji}$
 for  the competition model with immediate phosphorylation (Model~3.1) are
$A_{ij}=0$ and $B_{ij}=0$,
 except for  the following
\begin{eqnarray}
A_{11} &= &  -2a_{1+}\{(N_L -x_1^*-x_2^*-x_3^*-x_4^*-x_5^*)+(N_{R_1}-x_1^*-2x_3^*-x_5^*)\} 
\nonumber
\\
& -&b_{11+}(N_{R_1}-x_1^*-2x_3^*-x_5^*) - b_{12+}(N_{R_2}-x_2^*-2x_4^*-x_5^*) + b_{11+}x_1^* - \alpha_{1-},
\nonumber
\\
A_{12} &= & b_{12+}x_1^* -2a_{1+}(N_{R_1}-x_1^*-2x_3^*-x_5^*),
\nonumber
\\
A_{13} &= & -2a_{1+}\{2(N_L -x_1^*-x_2^*-x_3^*-x_4^*-x_5^*)+(N_{R_1}-x_1^*-2x_3^*-x_5^*)\} + 2b_{11+}x_1^*+2\beta_{11-},
\nonumber
\\
A_{14} &= & 2b_{12+}x_1 - 2a_{1+}(N_{R_1}-x_1^*-2x_3^*-x_5^*),
\nonumber
\\
A_{15} &= & b_{11+}x_1^* + b_{12+}x_1^* + \beta_{12-} -2a_{1+}\{(N_L -x_1^*-x_2^*-x_3^*-x_4^*-x_5^*)+(N_{R_1}-x_1^*-2x_3^*-x_5^*)\},
\nonumber
\\ 
A_{21} &= & b_{21+}x_2^* - 2a_{2+}(N_{R_2}-x_2^*-2x_4^*-x_5^*),
\nonumber
\\
A_{22} &= & -2a_{2+}\{(N_L -x_1^*-x_2^*-x_3^*-x_4^*-x_5^*)+(N_{R_2}-x_2^*-2x_4^*-x_5^*)\} 
\nonumber
\\
& -&b_{21+}(N_{R_1}-x_1^*-2x_3^*-x_5^*)-b_{22+}(N_{R_2}-x_2^*-2x_4^*-x_5^*) + b_{22+}x_2^*-\alpha_{2-},
\nonumber
\\
A_{23} &= & 2b_{21+}x_2^* - 2a_{2+}(N_{R_2}-x_2^*-2x_4^*-x_5^*),
\nonumber
\\
A_{24} &= &  -2a_{2+}\{2(N_L -x_1^*-x_2^*-x_3^*-x_4^*-x_5^*)+(N_{R_2}-x_2^*-2x_4^*-x_5^*)\} + 2b_{22+}x_2^*+2\beta_{22-},
\nonumber
\\
A_{25} &= & -2a_{2+}\{(N_L -x_1^*-x_2^*-x_3^*-x_4^*-x_5^*)+(N_{R_2}-x_2^*-2x_4^*-x_5^*)\} + b_{21+}x_2^* + b_{22+}x_2^* + \beta_{21-},
\nonumber
\\
A_{31} &= & b_{11+}(N_{R_1}-x_1^*-2x_3^*-x_5^*)-b_{11+}x_1^*,
\nonumber
\\
A_{33} &= & -2b_{11+}x_1^* - 2\beta_{11-}, 
\nonumber
\\
A_{35} &= & -b_{11+}x_1^*,
\nonumber
\\
A_{42} &= & b_{22+}(N_{R_2}-x_2^*-2x_4^*-x_5^*) - b_{22+}x_2^*,
\nonumber
\\
A_{44} &= & -2b_{22+} x_2^* - 2\beta_{22-},
\nonumber
\\
A_{45} &= & -b_{22+}x_2^*,
\nonumber
\\
A_{51} &= & -b_{21+} x_2^* + b_{12+}(N_{R_2}-x_2^*-2x_4^*-x_5^*),
\nonumber
\\
A_{52} &= & -b_{12+} x_1^* + b_{21+} (N_{R_1}-x_1^*-2x_3^*-x_5^*),
\nonumber
\\
A_{53} &= & -2b_{21+}x_2^*,
\nonumber
\\
A_{54} &= & -2b_{12+}x_1^*,
\nonumber
\\
A_{55} &= & -b_{12+}x_1^* -b_{21+}x_2^* - \beta_{12-}-\beta_{21-},
\nonumber
\\
B_{11} &= & 2a_{1+}(N_L -x_1^*-x_2^*-x_3^*-x_4^*-x_5^*)(N_{R_1}-x_1^*-2x_3^*-x_5^*)  + b_{11+}x_1^*(N_{R_1}-x_1^*-2x_3^*-x_5^*) 
\nonumber
\\
& + &b_{12+}x_1^*(N_{R_2}-x_2^*-2x_4^*-x_5^*) + \alpha_{1-}x_1^*+2\beta_{11-}x_3^*+ \beta_{12-}x_5^*,
\nonumber
\\
B_{13} &= &-2b_{11+}x_1^*(N_{R_1}-x_1^*-2x_3^*-x_5^*) - 4\beta_{11-}x_3^*,
\nonumber
\\
B_{15} &= & -2b_{12+}x_1^*(N_{R_2}-x_2^*-2x_4^*-x_5^*) - 2\beta_{12-}x_5^*,
\nonumber
\\
B_{22} &= & 2a_{2+}(N_L -x_1^*-x_2^*-x_3^*-x_4^*-x_5^*)(N_{R_2}-x_2^*-2x_4^*-x_5^*) +  b_{21+}x_2^*(N_{R_1}-x_1^*-2x_3^*-x_5^*)
\nonumber
\\
& +& b_{22+}x_2^*(N_{R_2}-x_2^*-2x_4^*-x_5^*) + \alpha_{2-}x_2^*+\beta_{21-}x_5^*+2\beta_{22-}x_4^*,
\nonumber
\\
B_{24} &= & -2b_{22+}x_2^*(N_{R_2}-x_2^*-2x_4^*-x_5^*)-4\beta_{22-}x_4^*,
\nonumber
\\
B_{25} &= & -2b_{21+}x_2^*(N_{R_1}-x_1^*-2x_3^*-x_5^*) -2\beta_{21-}x_5^*,
\nonumber
\\
B_{33} &= & b_{11+}x_1^*(N_{R_1}-x_1^*-2x_3^*-x_5^*)+2\beta_{11-}x_3^*,
\nonumber
\\
B_{44} &= & b_{22+}x_2^*(N_{R_2}-x_2^*-2x_4^*-x_5^*)+2\beta_{22-}x_4,
\nonumber
\\
B_{55} &= & b_{12+}x_1^*(N_{R_2}-x_2^*-2x_4^*-x_5^*) + b_{21+}x_2^*(N_{R_1}-x_1^*-2x_3^*-x_5^*) + \beta_{12-}x_5^*+\beta_{21-}x_5^*.
\nonumber
\end{eqnarray}

\paragraph{van Kampen approximation for Model~3.2}

In the case of Model~3.2 the deterministic variables are given by
\[
(M_1, M_2, D_1, D_2, D_M, P_2, P_M) \sim (x_1, x_2, x_3, x_4, x_5,x_6,x_7),
\]
\\
and the deterministic equations can be written as
\begin{eqnarray}
\dfrac{dx_1}{dt} &= & 2a_{1+}(N_{R_1}-x_1-2x_3-x_5)(N_L-x_1-x_2-x_3-x_4-x_5)
\nonumber
 \\
& -&b_{11+}x_1(N_{R_1}-x_1-2x_3-x_5)-b_{12+}x_1(N_{R_2}-x_2-2x_4-x_5)
\nonumber
\\
&-&\alpha_{1-}x_1+2\beta_{11-}x_3+\beta_{12-}x_5,
\nonumber
\\
\dfrac{dx_2}{dt} &= & 2a_{2+}(N_{R_2}-x_2-2x_4-x_5)(N_L-x_1-x_2-x_3-x_4-x_5)
\nonumber
\\
&-&b_{22+}x_2(N_{R_2}-x_2-2x_4-x_5) - b_{21+}x_2(N_{R_1}-x_1-2x_3-x_5)
\nonumber
 \\
& -&\alpha_{2-}x_2 + 2\beta_{22-}x_4+\beta_{21-}x_5,
\nonumber
\\
\dfrac{dx_3}{dt} &= & -2\beta_{11-}x_3+b_{11+}x_1(N_{R_1}-x_1-2x_3-x_5),
\nonumber
\\
\dfrac{dx_4}{dt} &= & b_{22+}x_2(N_{R_2}-x_2-2x_4-x_5)-2\beta_{22-}x_4 - \gamma_{22+}x_4+\gamma_{22-}x_6,
\nonumber
\\
\dfrac{dx_5}{dt} &= & b_{12+}x_1(N_{R_2}-x_2-2x_4-x_5)+b_{21+}x_2(N_{R_1}-x_1-2x_3-x_5)-\beta_{12-}x_5-\beta_{21-}x_5
\nonumber
\\
& -&\gamma_{12+}x_5 + \gamma_{12-}x_7,
\nonumber
\\
\dfrac{dx_6}{dt} &= & \gamma_{22+} x_4 - \gamma_{22-}x_6,
\nonumber
\\
\dfrac{dx_7}{dt} &= & \gamma_{12+} x_5 - \gamma_{12-}x_7.
\nonumber
\label{ODE: model 3.2}
\end{eqnarray}
For Model~3.2
(competition model with delayed phosphorylation),
the coefficients $A_{ij}$ and $B_{ij}=B_{ji}$ are the same as in Model~3.1, except for the following
\begin{eqnarray}
A_{44} &= & -2b_{22+} x_2^* - 2\beta_{22-} -\gamma_{22+},
\nonumber
\\
A_{55} &= & -b_{12+}x_1^* -b_{21+}x_2^* - \beta_{12-}-\beta_{21-} - \gamma_{12+},
\nonumber
\\
A_{46} &=& \gamma_{22-},
\nonumber
\\
 A_{64} &=& \gamma_{22+},  
 \nonumber
\\
 A_{66} &=& - \gamma_{22-}, 
\nonumber
\\ 
 A_{57} &=& \gamma_{12-}, 
\nonumber
\\
 A_{75} &=& \gamma_{12+}, 
\nonumber
\\ 
  A_{77} &=& -\gamma_{12-},
\nonumber
\\
B_{44} &= & b_{22+}x_2^*(N_{R_2}-x_2^*-2x_4^*-x_5^*)+2\beta_{22-}x_4 +\gamma_{22-}x_6^*+\gamma_{22+}x_4^*,
\nonumber
\\
B_{55} &= & b_{12+}x_1^*(N_{R_2}-x_2^*-2x_4^*-x_5^*) + b_{21+}x_2^*(N_{R_1}-x_1^*-2x_3^*-x_5^*) + \beta_{12-}x_5^*
+\beta_{21-}x_5^* +\gamma_{12-} x_7^*+  \gamma_{12+} x_5^*
,
\nonumber
\\
B_{46} &=&  -2\gamma_{22-}x_6^* - 2\gamma_{22+} x_4^*, 
\nonumber
\\
B_{57} &=&  -2\gamma_{12-}x_7^* - 2\gamma_{12+} x_5^*, 
\nonumber
\\
B_{66} &=&  \gamma_{22-} x_6^* + \gamma_{22+}x_4^*, 
\nonumber
\\
B_{77} &=&  \gamma_{12-} x_7^* + \gamma_{12+} x_5^*.
\nonumber
\end{eqnarray}

\end{document}